\documentclass[aps,prl,reprint,superscriptaddress,twocolumn,nobibnotes]{revtex4-2}
\pdfoutput=1

\usepackage{graphicx,amsmath,amssymb,dcolumn,physics,color}

\begin{document}
\title{The energy cost and optimal design of networks for biological discrimination} 
\author{Qiwei Yu}
\affiliation{Center for Theoretical Biological Physics, Rice University, Houston, TX 77005}
\affiliation{Lewis-Sigler Institute for Integrative Genomics, Princeton University, Princeton, NJ 08544}

\author{Anatoly B.~Kolomeisky} 
\affiliation{Center for Theoretical Biological Physics, Rice University, Houston, TX 77005}
\affiliation{Department of Chemistry, Rice University, Houston, TX 77005}
\affiliation{Department of Chemical and Biomolecular Engineering, Rice University, Houston, TX 77005}
\affiliation{Department of Physics and Astronomy, Rice University, Houston, TX 77005}

\author{Oleg A.~Igoshin}
\email[Corresponding author: ]{igoshin@rice.edu}
\affiliation{Center for Theoretical Biological Physics, Rice University, Houston, TX 77005}
\affiliation{Department of Chemistry, Rice University, Houston, TX 77005}
\affiliation{Department of Bioengineering, Rice University, Houston, TX 77005}
\affiliation{Department of Biosciences, Rice University, Houston, TX 77005}
\date{\today}

\begin{abstract}
	Many biological processes discriminate between correct and incorrect substrates through the kinetic proofreading mechanism which enables lower error at the cost of higher energy dissipation.
	Elucidating physicochemical constraints for global minimization of dissipation and error is important for understanding enzyme evolution.	Here, we identify theoretically a fundamental error-cost bound which tightly constrains the performance of proofreading networks under any parameter variations preserving the rate discrimination between substrates.
	The bound is kinetically controlled, i.e. completely determined by the difference between the transition state energies on the underlying free energy landscape.
	The importance of the bound is analyzed for three biological processes. 
	DNA replication by T7 DNA polymerase is shown to be nearly optimized, i.e. its kinetic parameters place it in the immediate proximity of the error-cost bound. 
	The isoleucyl-tRNA synthetase (IleRS) of \textit{E. coli} also operates close to the bound, but further optimization is prevented by the need for reaction speed.
	In contrast, \textit{E. coli} ribosome operates in a high-dissipation regime, potentially in order to speed up protein production.
	Together, these findings establish a fundamental error-dissipation relation in biological proofreading networks and provide a theoretical framework for studying error-dissipation trade-off in other systems with biological discrimination.
\end{abstract}

\maketitle

\section{Introduction}

The remarkable fidelity in cellular information processing, including DNA replication~\citep{Kunkel2000}, transcription~\citep{Sydow2009}, and translation~\citep{Rodnina2001,Zaher2009}, is realized through a nonequilibrium error-reduction mechanism called kinetic proofreading~\citep{Hopfield1974,Ninio1975}. The proofreading process is dissipative as it introduces an extra energy cost in exchange for improved discrimination against the formation of incorrect products~\citep{Bennett1979}. Besides error and energy dissipation, the reaction speed constitutes another important property of the proofreading system, shown to be optimized in processes such as replication and translation~\citep{Banerjee2017PNAS,Mallory2019,Wong1991,Gromadski2004}.
The interplay among speed, accuracy, and energy dissipation in systems involving kinetic proofreading (KPR) has been studied in different contexts~\citep{Savageau1979Biochemistry,Freter1980,Savageau1981,Ehrenberg1980,Blomberg1981,Murugan2012,Murugan2014,Wong2018,Sartori2013,Sartori2015,Banerjee2017PNAS,Mallory2019,Yu2020,Mallory2020PNAS,Mallory2020JPCB,Hartich2015,Galstyan2020}, providing insights to both general KPR networks and specific biological systems achieving discrimination through KPR.

Nonetheless, a fundamental difference distinguishes speed from error and dissipation. In a nonequilibrium steady state, the magnitude of a probability flux is generally affected by the energy levels of both barriers (maxima) and discrete states (minima) of the free energy landscape, but the ratio of fluxes only depends on the barriers (maxima)~\cite{Mallory2020PNAS}. 
In the KPR network, speed is characterized by the magnitude of the product-forming flux, whereas error and dissipation (per product formed) can be expressed as flux ratios.
Therefore, speed depends on both minima and maxima, while error and dissipation are determined only by energy maxima. 
The variation of energy barriers could create a fundamental constraint (trade-off) between error and dissipation, but speed is decoupled from this trade-off since it can be varied independently by perturbing energy minima. 
Elucidating this fundamental error-dissipation trade-off is of great importance to the mechanistic understanding of KPR.

Besides the theoretical motivation, the need to quantitatively understand experimentally characterized KPR systems also necessitates the investigation of the error-dissipation bound.
Specifically, the trade-offs between speed, error, dissipation, and noise in KPR systems have been studied locally, i.e. by examining the change in these characteristic properties due to the variation of a certain rate constant~\citep{Banerjee2017PNAS,Mallory2019}. However, different reaction steps may have different priorities in the optimization of characteristic properties.
In the KPR network of tRNA\textsuperscript{Ile} aminoacylation, for instance, the amino acid activation step optimizes speed, but the amino acid transfer step optimizes dissipation~\citep{Yu2020}.
A more global approach which examines the effect of simultaneously varying multiple rate constants might be better suited to understand the evolutionary principle and consequence in the placement of the rate constants.
This approach reveals a global error-dissipation constraint which illustrates the importance of minimizing the energy cost in tRNA\textsuperscript{Ile} aminoacylation~\citep{Yu2020} and coronavirus genome replication~\citep{Mallory2021}. The error-dissipation constraint defines a manifold along which the decrease in error will lead to increase in energy cost and vice versa. In this sense, it is reminiscent of the concept of the Pareto front in phenotype space due to natural selection~\citep{Shoval2012}.
However, much remains unknown about this constraint including its physical origin and biological importance.
Previously, a general matrix method was developed to study the relation between error and energy cost under different constraints~\citep{Savageau1981}.
Furthermore, multiple proofreading regimes where accuracy depends on binding energy difference in distinct fashions were discovered~\citep{Murugan2012,Murugan2014}. More recently, scaling analysis was employed to obtain an asymptotic energy-accuracy-speed relation~\citep{Wong2018}.
The thermodynamic uncertainty relation also imposes a lower bound on the energy dissipation rate~\citep{Pineros2020}.
To unify these relations and apply them to understanding biological proofreading systems~\citep{Zaher2009,Gromadski2004,Johnson1993}, it is crucial to develop a general method to obtain the explicit relationship between minimal error and dissipation for biologically relevant models.

In this work, we seek to address these challenges by developing a unified understanding of the fundamental error-dissipation trade-off in general KPR networks. 
To this end, a theoretical framework is developed with a three-pronged approach unifying the perspectives of chemical kinetics, reaction fluxes, and free energy landscape. The KPR process is described by the steady-state of a chemical reaction network governed by chemical master equations (CME), which can be explicitly transformed into a flux-based formalism enabling the derivation of the exact error-dissipation bound.
The bound strictly encapsulates all possible systems, and it can only be approached in the presence of strong nonequilibrium driving in the proofreading cycle and with the fine-tuning of certain flux-splitting ratios.
From the free energy landscape perspective, the bound is only determined by the difference of energy barriers between cognate and noncognate networks, indicating that the trade-off is under kinetic rather than thermodynamic control~\citep{Mallory2020JPCB}.

The general theoretical framework developed here could be utilized to identify the error-dissipation bound in a large class of KPR networks.
We first illustrate its usage in well-recognized models such as Hopfield's scheme and multi-stage proofreading networks with dissociation-based rate discrimination. The methodology's impact, however, is not limited to simplified systems. It is applicable to complex models with arbitrary discrimination factors and multiple intermediates or proofreading pathways. The error-dissipation bound in these systems reveal important physical and biological insights. To demonstrate this, we study three examples whose reaction networks were previously characterized: DNA replication by T7 DNA polymerase, aminoacyl-tRNA selection by \textit{E. coli} ribosome~\citep{Banerjee2017PNAS}, and aminoacylation by \textit{E. coli} isoleucyl-tRNA synthetase (IleRS)~\citep{Yu2020}.
The global parameter sampling confirms that the error-dissipation bound is valid in these systems and that the nonequilibrium driving provided by hydrolyzing energy-rich molecules in the futile cycles is indeed sufficiently large, allowing for the bound to be closely approached.
By comparing the native systems with the optimal ones that sit on the error-dissipation bound, we search for general constraints and principles in these biological discrimination systems.

\section{Results}

\subsection{Theoretical Formalism}
\subsubsection{Error and cost in Hopfield's kinetic proofreading scheme}
\begin{figure}[tb]
	\centering
	\includegraphics[width=\linewidth]{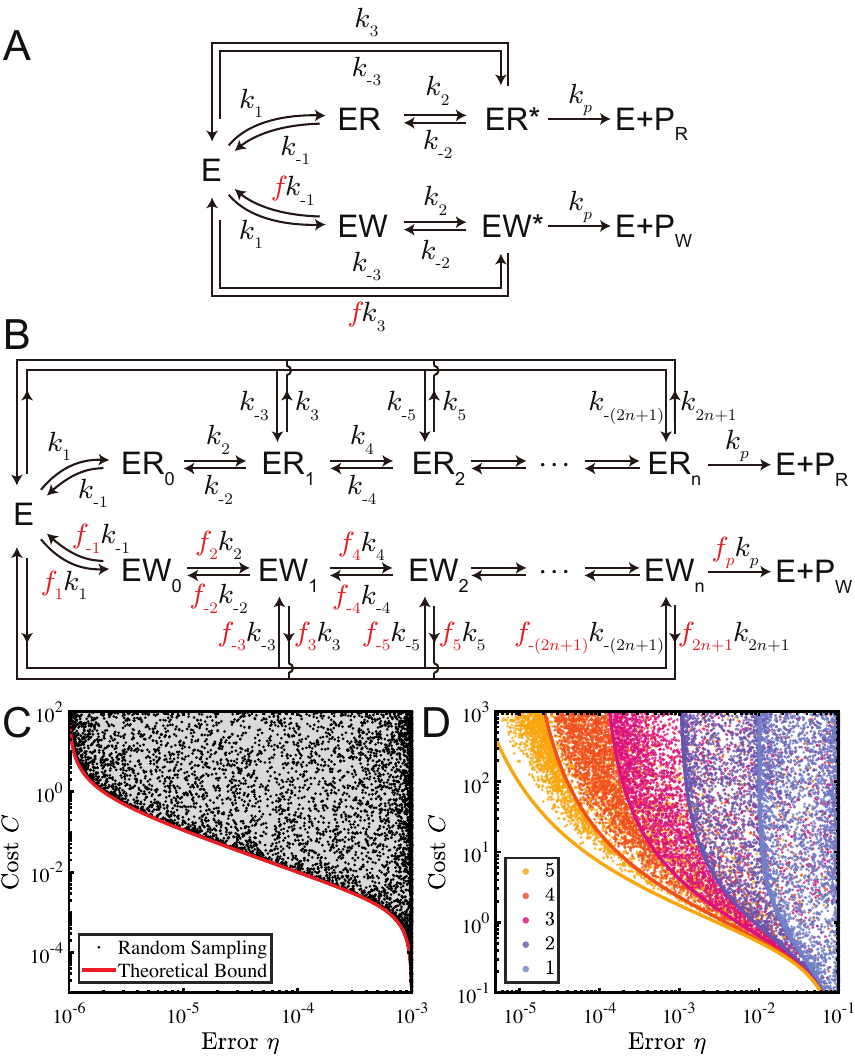}
	\caption{Proofreading schemes and the error-cost trade-off.
		(A) The proofreading scheme proposed by Hopfield~\citep{Hopfield1974} with one proofreading pathway and dissociation-based discrimination.
		(B) A generalized proofreading scheme with $n$ proofreading pathways and discrimination factors in all reaction steps.
		The discrimination factors $f$ and $f_{i}$ are marked in red, with $i$ labeling the reactions.
		(C) The error-cost relation in the Hopfield scheme (panel A) with $f{}=1000$.
		{$N=2\times10^4$ points are shown.}
		Red line: theoretical bound in Eq.~\ref{Eq:HopfieldSchemeBoundExpression}.
		(D) The error-cost relation in the $n$-stage dissociation-based-discrimination scheme with $f{}=10$ and $n=1,2,3,4,5$. Solid lines of the corresponding color indicate the theoretical bound in Eq.~\ref{Eq:nstageHopfieldBoundExpression}.
		{The points are generated with a biased sampling method that prefers points with low error and cost~\cite{Estrada2016}. $N=3\times10^4$ points are shown in total.}
		The thermodynamic constraint is $\ln\gamma=20$ for C and $\ln\gamma=30$ for D.
	}
	\label{fig:BasicSchemeAndSampling}
\end{figure}

In the classic proofreading scheme proposed by Hopfield~\citep{Hopfield1974}, the free enzyme E can either bind to the correct substrate (R) forming the cognate complex ER or bind to the incorrect substrate (W) forming EW.
The complex then enters an intermediate state ER\textsuperscript{*} or EW\textsuperscript{*}, where it can either generate a product P\textsubscript{R/W} or undergo proofreading, i.e. resetting without generating any product.
Both proofreading and product formation return the enzyme to the unbound state E.
This reaction scheme is shown in Fig.~\ref{fig:BasicSchemeAndSampling}A. All reactions are pseudo-first-order as fixed concentrations of the substrates and products are maintained. The networks for right and wrong substrates are identical in structure but differ in reaction rates (highlighted in red). In Hopfield's scheme, such difference only exists in dissociation steps, where the rate for the wrong substrate is $f{}$-fold larger than the rate for the right substrate.

The state of the enzyme at any time $t$ is characterized by a probability distribution vector $\mathbf{P}(t)=\left[P_\mathrm{E}, P_\mathrm{ER}, P_\mathrm{EW}, P_\mathrm{ER^{*}}, P_\mathrm{EW^{*}}\right]^{T}$, where $P_\mathrm{A}$ denotes the probability of staying in state A.
The probabilities are normalized by $\mathbf{1}^{T}\cdot\mathbf{P}=1$.
The time evolution of the probability distribution is governed by the chemical master equation (CME):
\begin{equation}
	\dv{\mathbf{P}}{t} = \mathbf{K} \cdot \mathbf{P},
\end{equation}
with the transition matrix $\mathbf{K}$ given by
\begin{equation}
	K_{j, i}=\left\{\begin{array}{ll}
		k_{i, j},                  & \text { for } j \neq i \\
		-\sum_{i \neq m} k_{i, m}, & \text { for } j=i
	\end{array}\right.
\end{equation}
$k_{i,j}$ denotes the rate of transition from state $i$ to state $j$. Specifically, we study the properties of the system at steady state, which satisfies  $\mathbf{K}\cdot\mathbf{P}=\boldsymbol{0}$ and $\mathbf{1}^{T}\cdot\mathbf{P}=1$.

The steady-state properties of the reaction network can be quantified with two key (dimensionless) properties: error $\eta$ and proofreading energy cost $C$.
Error is defined as the rate of forming the incorrect product $P$\textsubscript{W} divided by the rate of forming the correct product $P$\textsubscript{R}:
\begin{equation}
	\eta = \frac{J_\mathrm{W}}{J_\mathrm{R}},
\end{equation}
where $J_\mathrm{R} = k_p P_\mathrm{ER^{*}}$ and $J_\mathrm{W} = k_p P_\mathrm{EW^{*}}$ are the probability fluxes of forming a correct/incorrect product, respectively. Another important property of such a nonequilibrium reaction system is the free energy dissipation, which could be quantified by the total energy dissipation per correct product formed~\citep{Hill1977,Qian2006, Mallory2019}:
\begin{equation}
	\sigma = \sigma_0 + C (1+\eta) \Delta \mu_\text{futile},
	\label{Eq:EntropyProduction}
\end{equation}
where $\sigma_0=\sigma_\mathrm{R} + \eta \sigma_\mathrm{W}$ is the (fixed) energy cost of making the products and $\Delta \mu_\text{futile}$ is the chemical potential difference for the (futile) proofreading cycles, usually corresponding to the hydrolysis of energy-rich molecules such as nucleotide triphosphate (NTP). 
The cost $C$ is the number of futile hydrolysis reactions per any product formed, calculated by taking the ratio of the total (futile) proofreading flux to the total product formation flux (including both cognate and noncognate products)~\citep{Freter1980,Banerjee2017PNAS}
\begin{equation}
	C = \frac{J_\text{futile}}{J_\mathrm{R}+J_\mathrm{W}},
\end{equation}
For example, the futile flux in the Hopfield scheme is
\begin{equation}
	J_\text{futile} = \qty(k_3P_\mathrm{ER^{*}}-k_{-3}P_\mathrm{E}) + \qty(k_3f{} P_\mathrm{EW^{*}}-k_{-3}P_\mathrm{E}).
\end{equation}
In this study, we consider $\sigma_0$ and $\Delta \mu_\mathrm{futile}$ as constants since they are usually fixed by constraints external to the enzyme, such as the chemical potential of substrates, products, and other molecules involved in the futile cycle. The chemical potential $\Delta\mu_\mathrm{futile}$ is related to the thermodynamic drive of the futile cycle, 
$\gamma=e^{\Delta\mu_\text{futile}}=\qty(k_1k_2k_3)/\qty(k_{-1}k_{-2}k_{-3})$~\cite{Qian2006,Hill1977}. Hence, Eq.~\ref{Eq:EntropyProduction} indicates that the cost $C$ is a measure of the true (physical) dissipation rate, and the interplay between accuracy and energy dissipation of the proofreading network can be studied by directly investigating the relation between dimensionless numbers $C$ and $\eta$.
Before analysing their relation, however, we generalize the definitions to other proofreading networks.

\subsubsection{Generalizing proofreading schemes}
The reaction scheme in Fig.~\ref{fig:BasicSchemeAndSampling}A suffers from a few limitations.
First, the difference in reaction rates is only present in dissociation steps, while experiment data suggests that disparity in rate constants can exist in any step of the reaction scheme~\citep{Johnson1993,Gromadski2004,Zaher2009,Ling2009,Cvetesic2012,Dulic2014,Cvetesic2015}.
Second, the Hopfield scheme allows for only one proofreading pathway that resets the enzyme, yet many biological systems, such as isoleucyl-tRNA synthetase, involve multiple proofreading pathways~\citep{Ling2009}.
To address these limitations, we study the interplay between error and cost in a generalized scheme which has $n$ proofreading pathways and allows for rate discrimination in all reactions (Fig.~\ref{fig:BasicSchemeAndSampling}B).
All proofreading pathways reset the enzyme to the initial empty state thereby resulting in dissipative cycles.
The discrimination factors are highlighted in red with $f_i$ denoting the ratio of rates in step ${i}$.
Although the network is structurally similar to the McKeithan network~\citep{McKeithan1995}, the additional proofreading stages here do not involve multiple phosphorylation and therefore dissipate the same amount of free energy. Thus the energy dissipation rate is still given by Eq.~\ref{Eq:EntropyProduction}, with the error $\eta$ defined as the ratio of the flux forming the incorrect product to that forming the correct product, and the cost $C$  defined as the ratio of the total proofreading flux to the total product formation flux.

In particular, we first illustrate our methodology using direct generalization of the Hopfield scheme in which $n$ proofreading pathways coexist, but the rate discrimination is still limited to dissociation steps with the same factor $f$. The generalized scheme, which we name the ``$n$-stage scheme with dissociation-based discrimination'' ($n$-stage DBD), has the same network structure as Fig.~\ref{fig:BasicSchemeAndSampling}B with discrimination only in a subset of reactions:
\begin{equation}
	f_{-1}=f_3=f_5=\cdots =f_{2n+1}=f>1,
	\label{Eq:DBD discrimination factors}
\end{equation}
All the other reactions carry no discrimination factor.
Rich theoretical insights obtained from studying this scheme would be extended to networks which allow for different discrimination factor in all reactions, especially those describing real biological proofreading processes.

\subsubsection{Parameter Sampling}
The parameter sampling (perturbation) is performed by varying the rate constants $\{k\}$ with fixed discrimination factors $\{ f\}$, which is the ratio of a rate constant in the noncognate network to the rate of the corresponding reaction in the cognate network. For one-stage proofreading systems, the rates $\{k\}$ are sampled from a log-uniform distribution. For multi-stage systems, a sampling method biased towards points with low error and low cost is used~\cite{Estrada2016}. 
Fixing the discrimination factors is equivalent to maintaining the same energy barrier differences between cognate and noncognate reactions, thus exerting the same level of kinetic control on substrate discrimination as the original (unperturbed) system.
Assuming that the cognate and noncognate energy barriers correspond the enzyme in the same conformational state interacting with the respective substrates, perturbation to the enzyme structure would introduce variation to both barriers by the same amount, thus maintaining the same discrimination factor.
This is motivated by the commonly used Linear Free Energy Relationship which we assume between the cognate and noncognate reactions~\citep{Kirsch1972,Hammett1970}.

\subsection{Minimal energy cost in Hopfield's proofreading scheme}
Fig.~\ref{fig:BasicSchemeAndSampling}C depicts the relation between error and energy cost in the prototypical Hopfield model (Fig.~\ref{fig:BasicSchemeAndSampling}A) with rate constants $k$ sampled from {a log-uniform distribution over the range $[10^{-5},10^5]$} and the discrimination factor $f$ kept constant.
Each black point represents the error $\eta$ and cost $C$ of a distinct combination of rate constants $\{ k \}$.
We find all of the points constrained above a boundary (red line),  which effectively defines a Pareto front where no further improvement in error is possible without compromising the cost.
Along the boundary, cost decreases monotonically with error, ranging from zero at $\eta_\mathrm{eq}=f{}^{-1}$ to infinity at $\eta_\mathrm{min} = f{}^{-2}$.
This is consistent with the previous finding that dissipative proofreading is only necessary if the error needs to be reduced below the equilibrium value $f{}^{-1}$, but the error can never be suppressed below $f{}^{-2}$~\citep{Hopfield1974,Ninio1975}.
We focus on the exact relation between error and cost in the dissipative proofreading regime $\eta\in(\eta_\mathrm{min},\eta_\mathrm{eq})$, where the proofreading mechanism becomes necessary.

\begin{figure}[tb]
	\centering
	\includegraphics[width=\linewidth]{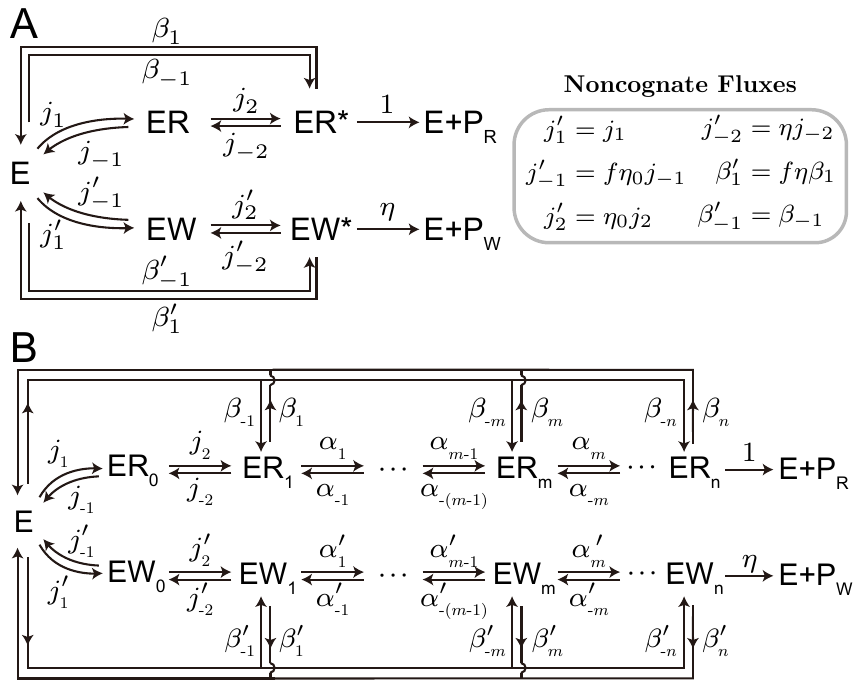}
	\caption{
		The flux-based formalism for the original Hopfield scheme (A) and the n-stage proofreading scheme (B). The expressions for noncognate fluxes in (B) are given in \textit{SI Appendix, Section II}.
	}
	\label{fig:HopfieldFluxFormalism}
\end{figure}

To aid the mathematical analysis of the error-cost bound, we introduce a flux-based formalism which changes the primary variables of the chemical master equation (CME) from the probability of each states ($\{P_A\}$, where $A$ labels all possible enzyme states) to the probability fluxes normalized by the correct-product-forming flux.
Fig.~\ref{fig:HopfieldFluxFormalism}A illustrates the flux-based formalism of the Hopfield scheme.
In the correct half of the network, the normalized fluxes are given by $j_i = J_i/J_\mathrm{R}$ ($i=\pm1,\pm2$) and $\beta_{\pm 1} = J_{\pm 3}/J_\mathrm{R}$, where $J_i$ is the probability flux of reaction step $i$ and $J_\mathrm{R}=k_p P_\mathrm{{R}}$ is the probability flux of forming the right product.
To quantify the normalized probability fluxes in the incorrect network, we define an additional error rate $\eta_0$ as the ratio of the forward fluxes from EW (ER) to EW\textsuperscript{${*}$} (ER\textsuperscript{${*}$}):
\begin{equation}
	\eta_0 = \frac{j_2'}{j_2} =  \frac{k_2 P_\mathrm{EW}}{k_2 P_\mathrm{ER}} =  \frac{ P_\mathrm{EW}}{ P_\mathrm{ER}}.
\end{equation}
We refer to $\eta_0$ as the zeroth-stage error rate as it is the error measured before the first proofreading step.
The resulting expressions for the normalized fluxes in the incorrect network are presented in Fig.~\ref{fig:HopfieldFluxFormalism}A with derivation detailed in \textit{SI Appendix, Section I}. In terms of the normalized fluxes, the energy cost is given by
\begin{equation}
	C =\frac{\qty(1+\eta f{})\beta_1 - 2\beta_{-1}}{1+\eta}.
\end{equation}

The steady-state condition $\mathbf{K}\cdot\mathbf{P}=\boldsymbol{0}$ in the CME translates to a set of stationary conditions in the flux formalism, stipulating that each state must have equal (normalized) fluxes entering and leaving it.
The stationary conditions impose four independent constraints on variables $\qty{j,\beta, \eta}$, leaving three degrees of freedom.
We choose $\eta_0$, $j_{-2}$, and $\beta_{-1}$ as free variables with respect to which the cost is minimized.
Moreover, $\eta_0$ is bounded by the equilibrium error rate, namely $\eta_0>\eta_\mathrm{eq}=f{}^{-1}$, where the minimum is only achieved in the limit of fast binding and unbinding between the free enzyme and the substrate.
We discover that the minimum cost is achieved when $\eta_0\to f{}^{-1}$ and $j_{-2}, \beta_{-1}\to 0$ (see \textit{SI Appendix, Section I} for detailed derivation):
\begin{equation}
	C_\mathrm{min}(\eta, f{}) = \frac{1-\eta^2f{}^2}{\qty(1+\eta)\qty(\eta f{}^2-1)},
	\label{Eq:HopfieldSchemeBoundExpression}
\end{equation}
which exactly bounds all data points found in numeric sampling (Fig.~\ref{fig:BasicSchemeAndSampling}C, red line).
Notably, as $\eta$ decreases within the range $\eta\in(f{}^{-2},f{}^{-1})$, the cost $C_\mathrm{min}$ increases monotonically and exhibits a divergence at the minimum error.
This can be compared with the error-cost bound in multi-stage proofreading schemes, where the cost diverges much faster toward a smaller error minimum.

The conditions for minimizing the cost reveal how the probability fluxes should be arranged for the scheme to be energetically optimal without impairing the accuracy.
The first condition, $\eta_0=f{}^{-1}$, indicates that the reactions between E, ER, and EW are in fast equilibrium. Hence, the ratio of probabilities $P_\mathrm{EW}$ and $P_\mathrm{ER}$ is determined by the ratio of their respective association constants with the enzyme, which is $f{}^{-1}$.
In many biochemical systems, the first step corresponds to the binding between enzyme and substrates, which is indeed in fast equilibrium compared to the subsequent catalytic reactions.
The second condition, $j_{-2}\to 0$, indicates that this reverse flux only increases the energy dissipation.
To better understand it, let us imagine a perturbation redirecting $j_{-2}$ to forming the correct product and $j'_{-2}$ to forming the incorrect product. This perturbation will not change the error rate since $j'_{-2}/j_{-2}=\eta$.
The stationary conditions also remain unaffected since the product forming fluxes return to the free enzyme state, which is in fast equilibrium with states ER and EW.
In this way, however, we have generated more products without increasing the futile fluxes and thereby reduced the cost. Therefore, vanishing fluxes $j_{-2}$ and $j_{-2}'$ is always energetically favorable.
The third condition $\beta_{-1}\to 0$ deals with the reverse proofreading fluxes. Although these fluxes seem to reduce the dissipation, they also significantly increase the error: going directly from E to ER\textsuperscript{*}/EW\textsuperscript{*} introduces error $\beta_{-1}'/\beta_{-1}=1$, which is always higher than the error $j_{2}'/j_{2}=\eta_0$ from going through intermediate states ER/EW.
The condition of vanishing $\beta_{-1}$ indicates that the reduction in dissipation due to reverse proofreading is outweighed by the increase in forward proofreading fluxes needed to mitigate the increase in error.

To summarize, the optimal proofreading system consists of three independent steps:
first, the error is reduced to $\eta_0 = \frac{ P_\mathrm{EW}}{ P_\mathrm{ER}}\to f{}^{-1}$ through the fast equilibrium in the binding and unbinding between the enzyme E and the substrate R/W;
second, the complex undergoes an activation step (from ER/EW to ER\textsuperscript{${*}$}/EW\textsuperscript{${*}$}) which is almost irreversible;
third, the error is reduced from $\eta_0$ to $\eta$ with a proofreading mechanism that is also almost irreversible.
Note that the activation and proofreading steps cannot be strictly irreversible due to the thermodynamic constraint $\gamma = (k_1k_2k_3)/(k_{-1}k_{-2}k_{-3})$. However, this constraint only increase the minimum cost by a small correction term of the order $\gamma^{-1/2}$ (see \textit{SI Appendix, Section I}), which is usually negligible in real networks since $\gamma\gg 1$ (for instance, $\gamma = e^{20}$ in the DNA replication network).

The key factor that characterizes the intensity of proofreading is the partition ratio of proofreading over product formation
\begin{equation}
	\beta_1 = \frac{J_3}{J_R} = \frac{1-\eta f{}}{\eta f{}^2-1},
\end{equation}
Under the optimal setting, the ratio decreases monotonically with $\eta$, indicating one-to-one monotonic correspondence between the optimal proofreading intensity and the desired accuracy. Networks that are non-optimal always have a larger partition ratio compared to the optimal network with the same error rate.
The significance of this ratio will be further illustrated in networks with multiple proofreading pathways.

\subsection{Minimum energy cost in multi-stage schemes with dissociation-based discrimination}
The natural generalization of the Hopfield scheme is to have multiple proofreading pathways while still localizing the discrimination to dissociation steps.
Hence, we study the so-called $n$-stage dissociation-based-discrimination (DBD) scheme, with network structure shown in Fig.~\ref{fig:BasicSchemeAndSampling}B and discrimination factors given by Eq.~\ref{Eq:DBD discrimination factors}.
By preferentially sampling systems with low error and proofreading cost~\citep{Estrada2016}, we identify error-cost boundaries for systems with different number of proofreading pathways $n$ (Fig.~\ref{fig:BasicSchemeAndSampling}D). For each system, the minimum dissipation starts from zero at $\eta_\mathrm{eq}=f{}^{-1}$ and increases as the error decreases before diverging to infinity at $\eta_\mathrm{min} = f{}^{-n-1}$. The same minimum error has also been obtained with a graph theoretic approach~\citep{Cetiner2020}.

\begin{figure*}[tb]
	\centering
	\includegraphics[width=0.8\linewidth]{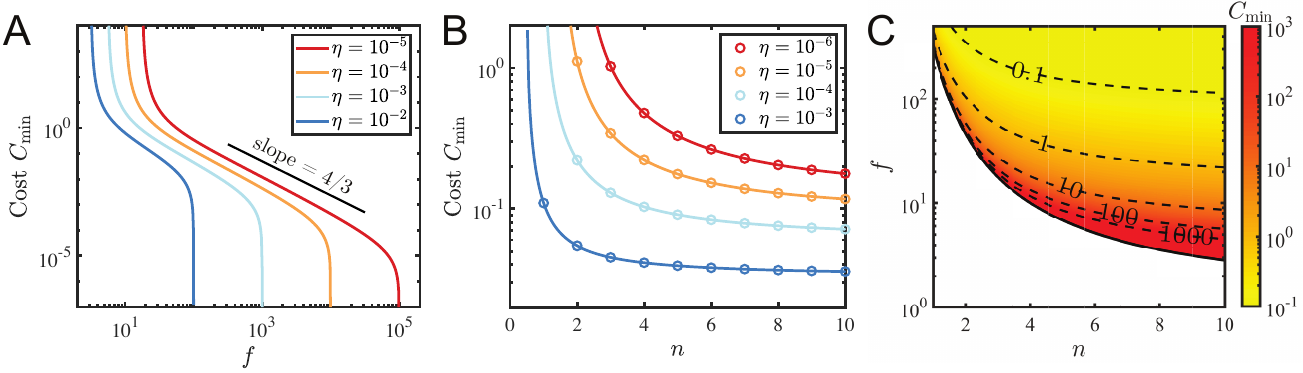}
	\caption{Energy cost in the $n$-stage dissociation-based-discrimination scheme can be reduced by increasing $n$ or $f$.
	(A) For a given error rate, the minimum cost decreases with the discrimination factor $f$. Here $n=3$ and $\eta$ is specified in the legend.
	(B) For a given error rate, the minimum cost decreases with the number of proofreading pathways $n$. Here $f=100$ and $\eta$ is specified in the legend.
	(C) Heatmap of $C_\mathrm{min}$ as a function of $f$ and $n$ at a fixed error $\eta=10^{-5}$. The area below the black line $\eta = f^{-(n+1)}$ is strictly inaccessible.
	}
	\label{fig:HopfieldNumericAnalysis}
\end{figure*}

Next, we analyze the error-cost bound in the $n$-stage DBD scheme with the flux-based formalism (Fig.~\ref{fig:HopfieldFluxFormalism}B).
For each stage $m$ ($m=1,2,\dots,n$), we define $\alpha_{\pm m}$ as the normalized forward/backward fluxes and $\beta_{\pm m}$ as the normalized proofreading/reverse proofreading fluxes.
The $m$-stage error $\eta_m$ is defined as the ratio of the forward fluxes going from EW\textsubscript{m} (ER\textsubscript{m}) to EW\textsubscript{m+1} (ER\textsubscript{m+1}). The final error is $\eta=\eta_n$. These $\eta$ characterize how the error is sequentially reduced from $\eta_0=\eta_\mathrm{eq}$ to $\eta_n=\eta$ through $n$ proofreading steps.
Similar to the case of the original Hopfield scheme, the normalized fluxes in the noncognate network can be expressed in terms of $\alpha$, $\beta$, $\eta$, and $f{}$ (see \textit{SI Appendix, Section II}). Through mathematical induction, we found that the proofreading cost in a $n$-stage DBD scheme $C_n$ is bounded by:
\begin{equation}
	C_n \geq \frac{(1+f{}^{-1})(f{}-1)^n}{1+\eta} \prod_{m=1}^{n} \frac{\eta_{m}}{\eta_{m} f{}-\eta_{m-1}} - 1,
	\label{Eq:nstageHopfieldBoundFunctionofEta}
\end{equation}
whose equality condition is $\alpha_{-i}, \beta_{-i}\to 0$ for $i=1,2,\dots, n$, $j_{-2}\to 0$, and $\eta_0=f{}^{-1}$.
Effectively, the system is optimal when the binding step is in fast equilibrium and the chemical reactions towards product formation or proofreading are nearly irreversible.
Complete irreversibility is precluded by the thermodynamic constraint, which increases the minimum cost by a correction term of the order $\gamma^{-1/(n+1)}$ (see \textit{SI Appendix, Section II}).
The intermediate error rates $\{\eta_m\}$ characterize the distribution of proofreading burden between different reaction stages, which can be further optimized.
We introduce $\lambda_m=\eta_{m-1}/\eta_{m}>1$ to quantify the increase in accuracy at stage $m$. These ratios are constrained by $\prod_{m=1}^{n}\lambda_m=\eta_0/\eta=\qty(f{}\eta)^{-1}$. The product under the $\Pi$ notation in Eq.~\ref{Eq:nstageHopfieldBoundFunctionofEta} is thus given by $\prod_{m=1}^{n}\qty(f{}-\lambda_m)^{-1}$.
Due to symmetry, the optimal system has equal $\lambda$ at all stages, i.e. $\lambda_m = (\eta_0/\eta_n)^{1/n} = \qty(f{}\eta)^{-1/n}$, which leads to the expression for minimum energy cost: (also see \textit{SI Appendix, Section II})
\begin{equation}
	C_{n,\mathrm{\ min}}(\eta,f{}) = \frac{(f{}-1)^n \qty(f+1) }{\qty(\qty(\eta f{})^{1/n}f{}-1)^n}\frac{\eta}{1+\eta}- 1.
	\label{Eq:nstageHopfieldBoundExpression}
\end{equation}
Indeed, this boundary constraints all points on the error-cost plane for each $n$ (solid lines, Fig.~\ref{fig:BasicSchemeAndSampling}D).
In the optimal scheme, the reduction of error is carried out sequentially with each proofreading step reducing the error by a factor of $\eta_{m}/\eta_{m-1} = \qty(f{}\eta)^{1/n}$. This implies that the burden of correcting errors is evenly distributed across $n$ proofreading pathways without any preference to early or late pathways.
Notably, the minimum cost exhibits $n$-th order divergence in the limit of minimum error [i.e. $C_\mathrm{min} \propto (\eta-\eta_\mathrm{min})^{-n}$], which is much stronger than the first-order divergence in the original Hopfield scheme.

The error-cost trade-off in the $n$-stage DBD scheme is controlled by two parameters: the discrimination factor $f$ and the number of proofreading pathways $n$.
Increasing either $f$ or $n$ reduces the overall minimum error $\eta_\mathrm{min} = f^{-n-1}$ as well as the minimum energy cost at any given error rate (Eq.~\ref{Eq:nstageHopfieldBoundExpression}).
They correspond to two error-correcting strategies: enhancing the discriminating capability of each individual proofreading pathways or redistributing the burden of error correction to additional pathways.
In real biochemical systems, increasing $n$ often requires the enzyme to have multiple reaction/proofreading domains, and the discrimination factor $f$ is determined by the free energy landscape of the underlying biochemical reactions.
Here we analyze the kinetic effects and defer the implication in specific biochemical contexts to the Discussion section.
Fig.~\ref{fig:HopfieldNumericAnalysis}A studies the relation between the minimum cost and the discrimination factor $f$.
For any fixed $n$ and $\eta$, the minimum cost diverges to infinity when $f$ is so small that the system operates near the minimum error and approaches zero when $f$ tends to $\eta^{-1}$ allowing the system to operate in the equilibrium discrimination regime. In the intermediate error range, however, we find the minimum cost decreasing with $f$ following a power law with exponent $\qty(1+n^{-1})$ (see \textit{SI Appendix, Section II} for derivation). Increasing $f$ in this range results in a non-diminishing return in the decrease of cost.
In contrast, although increasing $n$ also reduces the minimum cost, the benefit becomes marginal when $n$ is large, and  the minimum cost never decreases to zero even when $n$ tends to infinity (see Fig.~\ref{fig:HopfieldNumericAnalysis}B).
These two effects can be summarized in Fig.~\ref{fig:HopfieldNumericAnalysis}C which depicts how the minimum cost depends on $n$ and $f$ for a given error rate. The orientation of the (dashed) cost contours demonstrates that the cost decreases with both $n$ and $f$ but varies more rapidly along the $f$ direction.
The solid black line shows the error minimum, where the cost diverges.

\subsection{The optimal scheme is kinetically controlled}
The intermediate error rates $\eta_m$ in the optimal $n$-stage DBD system form a geometric series, where each proofreading stages reduces the error by the same factor.
This is achieved by a specific combination of rate constants. In particular, each intermediate state in the optimal scheme has the equal ratio of partition between proofreading (resetting to E) and moving forward to the next intermediate state along the product formation pathway:
\begin{equation}
	a_m=\frac{\beta_m}{\alpha_m} = \frac{1-\qty(\eta f{})^{1/n}}{f{}\qty(\eta f{})^{1/n}-1} = \frac{k_{2m+1}}{k_{2m+2}}.
	\label{Eq:nstageHopfieldAlphaBetaRatio}
\end{equation}

\begin{figure}[tb]
	\centering
	\includegraphics[width=\linewidth]{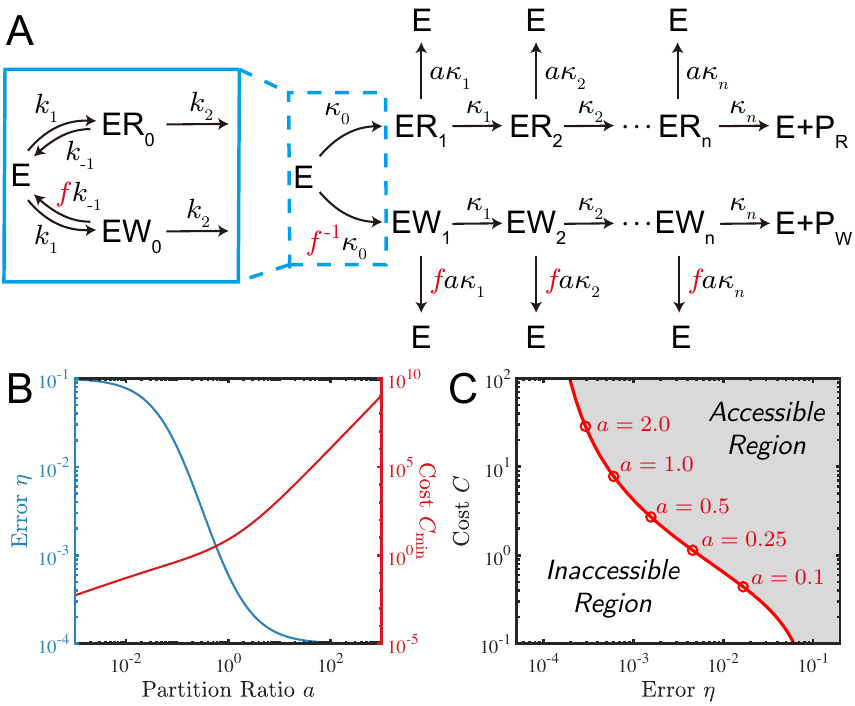}
	\caption{
		The kinetics of optimal proofreading network with only dissociation-based discrimination is captured by a simple kinetic model.
		(A) minimal kinetic model for the $n$-stage proofreading scheme.
		(B) the partition ratio $a$ serves as a controlling parameter simultaneously tuning error $\eta$ and cost $C$.
		Parameters are $n=3$ and $f=10$.
		(C) varying $a$ in the minimal model corresponds to moving the system along the optimal front in the error-cost trade-off, which separates the accessible and inaccessible regions.
	}
	\label{fig:MinimalScheme}
\end{figure}

In fact, this partition ratio reveals how features of the free energy landscape shape the fundamental trade-off between accuracy and energy dissipation.
To illustrate this, we construct a simple kinetic model (Fig.~\ref{fig:MinimalScheme}A) which elucidates the flux dynamics in optimal multi-stage proofreading networks.
The $f$-fold discrimination due to the fast equilibrium in the binding steps is captured by the two reactions in the blue dashed box, which create a $f{}^{-1}$-fold difference in the production of EW\textsubscript{1} compared to ER\textsubscript{1}.
All the subsequent intermediate states are assigned with a reaction rate $\kappa_i$ (or equivalently a time scale $\tau_i=\kappa_i^{-1}$) and a partition ratio $a$ which is identical for all proofreading pathways. We also assign $\kappa_0$ to the initial steps.
By allowing for completely irreversible reactions, this model only calculates the leading order term of cost in the large $\gamma$ limit.
The steady-state probability distribution of this model is obtained by directly solving the CME (see \textit{SI Appendix, Section III}), which reveals that both error $\eta$ and cost $C$ are completely determined by only $f$ and $a$. They are independent of all the $\kappa_i$ ($i=0,1,\dots,n$).
As illustrated in Fig.~\ref{fig:MinimalScheme}B, increasing $a$ continuously from zero to infinity drives the system from the equilibrium discrimination regime with no proofreading cost ($\eta_\mathrm{eq}=f{}^{-1}$ and $C=0$) toward the nonequilibrium limit with highest accuracy and diverging cost ($\eta\to\eta_\mathrm{min} = f{}^{-n-1}$ and $C \to + \infty$).
Varying $\kappa$ has no effect on either error or cost.
We also find that the intermediate error rates $\eta_m$ forms a geometric series the same way as seen in the $n$-stage DBD scheme, and eliminating $a$ from the expression of $\eta$ and $C$ recovers the error-cost bound in Eq.~\ref{Eq:nstageHopfieldBoundExpression}.
Therefore, the system sits on the optimal error-cost boundary as long as all proofreading stages share the same partition ratio $a$.
Moreover, the ratio $a$ functions as a tuning parameter which only moves the system along the Pareto front of the error-cost trade-off (i.e., theoretical bounds in Fig.~\ref{fig:BasicSchemeAndSampling}CD).
This relation is depicted in Fig.~\ref{fig:MinimalScheme}C, where each value of $a$ corresponds to the optimal scheme for a different error rate and all the systems obtained in the previous parameter sampling fall within the grey accessible region above the optimal bound.

It is not a coincidence that both error and cost are determined by the partition ratio $a$ but not rates $\kappa_i$. In fact, the deeper explanation lies in the different features of the underlying free energy landscape captured by these rates.
Previous work has shown that any quantities that can be expressed as ratios of stationary fluxes, including both $\eta$ and $C$, are invariant against perturbation of the energy level of discrete states (minima on the free energy landscape) and only affected by perturbation of the energy barriers (maxima)~\citep{Mallory2020PNAS}. Expressing $\kappa_m$ and $a$ in terms of the energy levels~\citep{Laidler1987}, we find:
\begin{equation}
	\kappa_m\propto e^{\epsilon_m-\epsilon_{m,m+1}^\dagger},\quad a\kappa_m\propto e^{\epsilon_m-\epsilon_{m,p}^\dagger},
\end{equation}
where $\epsilon_m$ is the energy of state ER\textsubscript{m}, $\epsilon_{m,m+1}^\dagger$ is the free energy barrier between ER\textsubscript{m} and ER\textsubscript{m+1}, and $\epsilon_{m,p}^\dagger$ is the free energy barrier in the proofreading step (from ER\textsubscript{m} to E).
Hence, $a$ is associated with the difference in energy barriers $\qty(\epsilon_{m,m+1}^\dagger-\epsilon_{m,p}^\dagger)$ which is key to the kinetic control of stationary flux ratios~\citep{Mallory2020PNAS}. In contrast, perturbing $\kappa_m$ is equivalent to varying the energy level of ER\textsubscript{m}, which is irrelevant to any ratio of stationary fluxes, such as $\eta$ and $C$.
These results reinforce the argument that KPR is kinetically controlled~\citep{Mallory2020JPCB} and highlights the importance of the partition ratio in both investigating natural biological proofreading systems and engineering synthetic biological systems with high selectivity.

\begin{figure}[tb]
	\centering
	\includegraphics[width=\linewidth]{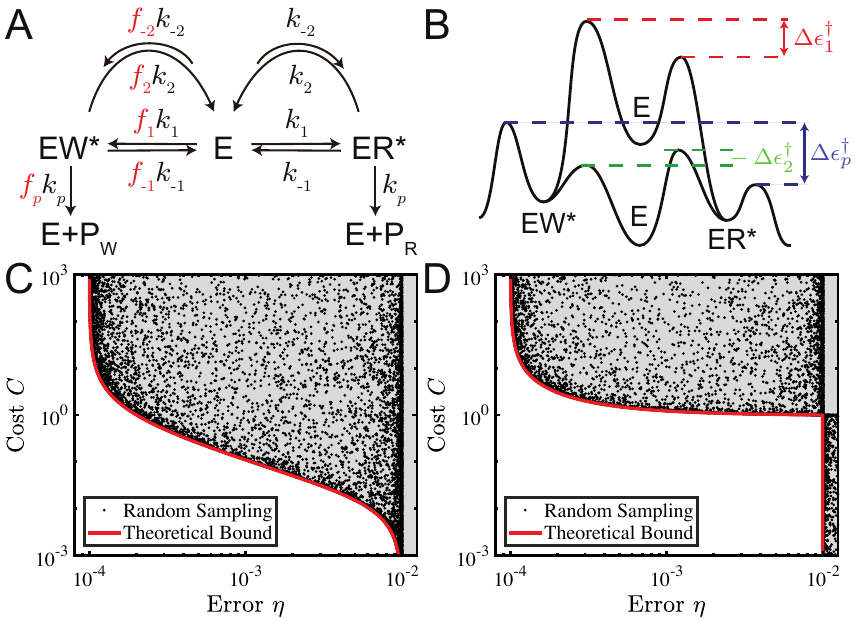}
	\caption{Michaelis-Menten reaction scheme with dissipative resetting.
	(A) The kinetic scheme. The discrimination factors $f$ are highlighted in red.
	(B) The free energy landscape of the kinetic scheme.
	(C, D) Error and energy cost of sampled networks (black dots) with two different set of discrimination factors.
	Red line: theoretical boundary given by Eq.~\ref{Eq:MMresettingMinCost}.
	(C) $f_1=10^{-2}$, $f_{-1}=1$, $f_{2}=10^2$, $f_p = 1$, $f_{-2}=1$;
	(D) $f_1=f_p=1$, $f_{-1} = 10^2$, $f_{2} = 10^4$, $f_{-2}=10^2$.
	The chemical potential for the futile cycle is set to $\Delta \mu = 20 k_BT$ ($\ln \gamma = 20$).  {$N=2\times 10^4$ points are shown in (C) and (D).}
	}
	\label{fig:MMresettingScheme}
\end{figure}

\subsection{Networks with arbitrary discrimination factors}
To elucidate the biological implications of the error-dissipation trade-off studied so far, we need to generalize it to networks which allow for disparity in the rate constants in all reaction steps (Fig.~\ref{fig:BasicSchemeAndSampling}B).
To illustrate the idea for generalization, we first study the classic Michaelis-Menten scheme with the addition of a resetting reaction (Fig.~\ref{fig:MMresettingScheme}A, henceforth named MM-with-proofreading), which could be regarded as a basic building block that makes up more complex proofreading networks (e.g. with multiple proofreading pathways).
The resetting cycle is driven by the cycle chemical potential difference $\Delta \mu= k_BT \ln \qty[k_1k_2/\qty(k_{-1}k_{-2})]$, which allows for increasing accuracy at the cost of energy dissipation.
The disparity in reaction rates is quantified by the discrimination factors $f_{i}$ (highlighted in red; $i$ labels the reaction steps), which obey the thermodynamic constraint $f_1f_2=f_{-1}f_{-2}$. Following the previous methodology, we define error $\eta$ as the ratio of the product formation fluxes and cost $C$ as the sum of the futile fluxes normalized by the total product formation flux.

Before studying the accuracy-energy trade-off, we first identify the parameter regime where dissipative proofreading becomes relevant. The resetting only improves the accuracy when it dissociates EW\textsuperscript{*} more readily than ER\textsuperscript{*}. Hence, $f_2$ should be sufficiently large for proofreading to be effective. In fact, we find proofreading meaningful only when $f_2$ is greater than both $f_{-1}$ and $f_p$, which is explained as following.
First, the dissipative resetting is only useful when it creates more bias than the non-dissipative dissociation step $k_{-1}$, which requires $f_2>f_{-1}$; otherwise utilizing only the equilibrium discrimination would be more accurate and less dissipative.
Second, the proofreading mechanism should proportionally dissociate more wrong complexes compared to the right ones, which requires the noncognate network to have a larger proofreading-to-product-formation partition ratio, namely $k_{2,W}/k_{p,W}>k_{2,R}/k_{p,R}$. This is equivalent to stipulating $f_2>f_p$.
With these two conditions, we find the system capable of achieving the minimum error $\eta_\mathrm{min} = f_1f_p/f_2$, which is lower than the minimum error of equilibrium discrimination $\eta_\mathrm{eq}=\min\qty(f_1, f_1f_p/f_{-1})$, calculated in the absence of proofreading ($k_{\pm 2}=0$).
Therefore, the trade-off between accuracy and energy dissipation is analyzed in the error range $\eta\in(\eta_\mathrm{min},\eta_\mathrm{eq})$, which can only be realized with proofreading.
With the flux-based method, it can be shown that for a given error rate, the energy cost $C$ is minimized when the reverse reaction rates $k_{-1, -2}$ become vanishing (see \textit{SI Appendix, Section IV}). The minimum cost reads:
\begin{equation}
	C_\mathrm{min} = \frac{(f_1-\eta)\qty(1+\eta\frac{f_2}{f_p})}{\qty(1+\eta)\qty(\eta\frac{f_2}{f_p}-f_1)}.
	\label{Eq:MMresettingMinCost}
\end{equation}

To examine this bound numerically, we sample all reaction rates in the MM-with-proofreading scheme with fixed discrimination factors and chemical potential. Fig.~\ref{fig:MMresettingScheme}CD presents the sampling results for two different sets of discrimination factors.
Consistent with theory, all sampled systems reside above the boundary given by Eq.~\ref{Eq:MMresettingMinCost} (red lines).
The cost diverges to infinity as the error approaches minimum.
Towards the $\eta\to\eta_\mathrm{eq}^-$ limit, however, the minimum cost approaches zero in the case of Fig.~\ref{fig:MMresettingScheme}C but converges to a positive value in Fig.~\ref{fig:MMresettingScheme}D.
As zero proofreading cost is expected in the equilibrium regime ($\eta>\eta_\mathrm{eq}$), the discontinuity in Fig.~\ref{fig:MMresettingScheme}D indicates a sudden change of the optimal rate configuration during the transition from the nonequilibrium regime to the equilibrium regime.
In the absence of proofreading, the minimum error is achieved by making either $k_1$ or $k_p$ rate-limiting, which leads to error rates of $\eta=f_1$ or $\eta=f_1f_p/f_{-1}$, respectively. $\eta_\mathrm{eq}$ corresponds to the more accurate one of these two configurations.
In the optimal scheme in the nonequilibrium regime, nonetheless, the rate $k_{-1}$ is always vanishing as seen in the cost minimization condition derived above.
Hence, there is a discontinuous regime change if $f_p<f_{-1}$, where the $k_{\pm 1}$ step is in fast equilibrium on the equilibrium side and almost irreversibly driven forward on the nonequilibrium side, resulting in the discontinuous error-cost relation as seen in Fig.~\ref{fig:MMresettingScheme}D. If $f_p\geq f_{-1}$, however, $k_{-1}$ is vanishing on both sides of $\eta_\mathrm{eq}$, and the error-cost bound is continuous at $\eta_\mathrm{eq}$, which is the case in Fig.~\ref{fig:MMresettingScheme}C.
This regime change is a feature present only in networks whose rate discrimination is not limited to dissociation steps.
Other than the discontinuity at $\eta_\mathrm{eq}$, the error-cost bound has the same quantitative profile as the original Hopfield scheme.

It is remarkable that out of all the discrimination factors, the error-cost relation (Eq.~\ref{Eq:MMresettingMinCost}) is only controlled by $f_1$ and the ratio $f_2/f_p$.
This can be understood from the perspective of free energy landscape (Fig.~\ref{fig:MMresettingScheme}B), where previous analysis has shown that both error and cost are determined solely by the energy barriers.
We denote the energy barriers by $\epsilon^{\dagger}_{\lambda, \mathrm{R}/\mathrm{W}}$, where $\lambda\in\qty{1,2,p}$ labels the reaction and R/W distinguish the right/wrong half of the network.
The discrimination factors are associated with the difference of energy barriers for cognate and noncognate substrates:
\begin{equation}
	f_1 \propto e^{-\Delta\epsilon_1^\dagger},
	\quad \frac{f_1f_2}{f_{-1}} \propto e^{-\Delta\epsilon_2^\dagger},
	\quad \frac{f_1f_p}{f_{-1}} \propto e^{-\Delta\epsilon_p^\dagger},
\end{equation}
where $\Delta\epsilon^\dagger_\lambda = \epsilon^\dagger_{\lambda,\mathrm{W}} - \epsilon^\dagger_{\lambda,\mathrm{R}}$ denotes the difference of energy barriers. Hence, the ratio $f_2/f_p$, which equals to the partition ratio in the noncognate network divided by the partition ratio in cognate network, is proportional to $e^{\Delta\epsilon_p^\dagger-\Delta\epsilon_2^\dagger}$.
In other words, the fundamental error-dissipation trade-off is governed only by the energy barrier differences $\Delta\epsilon_1^\dagger$ and $\qty(\Delta\epsilon_p^\dagger-\Delta\epsilon_2^\dagger)$.
Moreover, the error-cost bound in more complex biological proofreading networks can be readily derived by identifying the equivalents of $f_1$ and $f_2/f_p$ since they already capture all the relevant features of the energy landscape.
Next, we apply this technique to three real proofreading systems with parameters provided in previous studies~\citep{Banerjee2017PNAS,Mallory2019,Yu2020} and discuss biological implications.

\subsection{Error-cost trade-off in real biological networks}
To illustrate the robustness of the bound, we apply our theoretical framework to three real biological proofreading networks, where numerical sampling confirms that the bound formulated with our theory tightly constrains the error and cost of all systems sampled. The position of the native systems compared to the bound reveals the relative importance in the evolutionary optimization of functionalities including speed, accuracy, and dissipation in KPR systems.

\begin{figure}[tb]
	\centering
	\includegraphics[width=\linewidth]{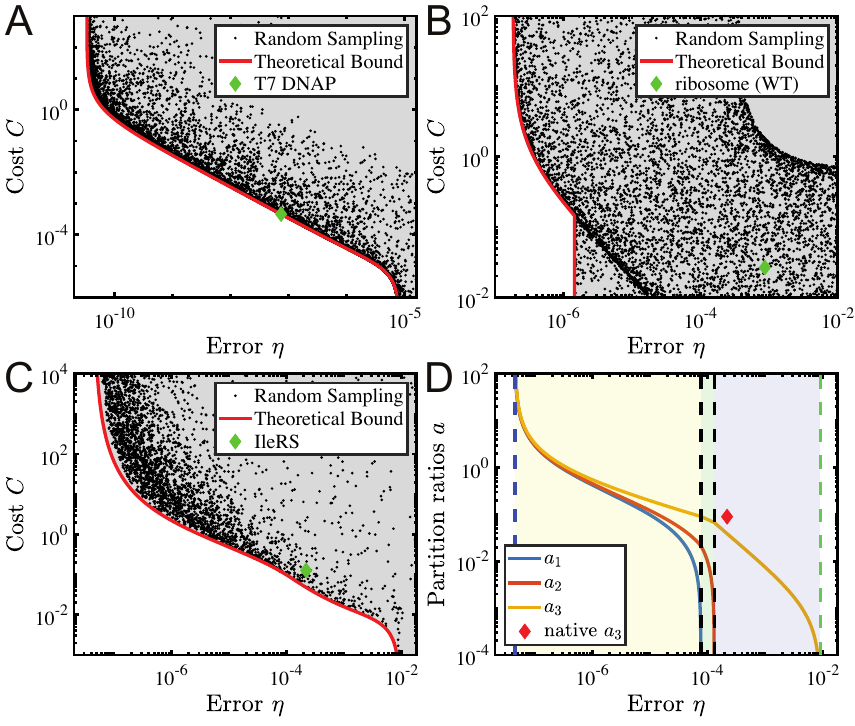}
	\caption{
		Results in biological networks involving proofreading (see \textit{SI Appendix, Section V} for details of the reaction schemes).
		(A) The error-cost relation in the T7 DNA polymerase network due to sampling all parameters.
		(B) The error-cost relation in the protein synthesis network.
		(C) The error-cost relation in the isoleucyl-tRNA synthetase (IleRS) network.
		All rate constants sampled obey their original thermodynamic constraints.
		(D) The optimal (solid lines) and native (red diamond) partition ratios in the IleRS network.
		The blue dashed line indicates the minimum error, and the green dashed line indicates minimum error for discrimination without dissipation.
		The black dashed lines show thresholds where $a_1$ and $a_2$ vanish.
		The yellow, green, and blue shades indicate phases of $n=3,2,1$.
	}
	\label{fig:RealisticNetworkResults}
\end{figure}

We start with DNA replication by T7 DNA polymerase, which employs a one-cycle proofreading mechanism~\citep{Wong1991,Banerjee2017PNAS}.
Its reaction network is similar to the MM-with-proofreading scheme with an additional intermediate state in the proofreading step.
Since the proofreading step is irreversible at the bound, the presence of this intermediate state does not affect the validity of the error-cost bound given in Eq.~\ref{Eq:MMresettingMinCost} (see \textit{SI Appendix, Section V} for details).
Fig.~\ref{fig:RealisticNetworkResults}A presents the result of sampling all rate constants while fixing the discrimination factors. Indeed, all the sampled systems fall exactly above the theoretical bound outlined by the red line.
Notably, the native system (green diamond) resides close to the boundary.
The energy cost $C$ for the native system is only 4.3\% larger than the minimum possible cost at the native error rate, which means that only a small portion of futile hydrolysis is excessive.
The energy efficiency of the DNA polymerase could also be assessed by the proofreading-to-product-formation partition ratio $a=k_2/k_p=8\times 10^{-4}$, which is close to its minimum possible value $a_\mathrm{min}=4.5\times 10^{-4}$ for the optimal system at the native error rate.
In analogy to the original Hopfield scheme, the DNAP network approaches the error-cost bound by driving both polymerization and proofreading irreversibly forward.
Indeed, we find the forward/backward polymerization rate ratio to be $k_1/k_{-1}=250 \gg 1$, and the rate ratio in proofreading to be $(k_2k_3)/(k_{-2}k_{-3}) = 2 \times 10^6 \gg 1$.
These reactions are driven forward by the nonequilibrium driving $\Delta \mu=20k_BT$ in the futile cycle, which comes from hydrolyzing one dNTP molecule.
Going below the native error rate, the system has the potential of reducing the error by more than two orders of magnitude (from $10^{-8}$ to $10^{-10}$). However, this will lead to about one futile hydrolysis per product formed ($C \approx 1$), effectively doubling the total energy dissipation of DNA replication.
Hence, it is possible that further increase in replication accuracy is prevented in order to avoid potentially disadvantageous excessive energy dissipation.

Next, we consider the aminiacyl-tRNA (aa-tRNA) selection process by ribosome during translation~\citep{Zaher2010,Banerjee2017PNAS}.
This network also has only one proofreading pathway, and the error-cost bound is fully captured by Eq.~\ref{Eq:MMresettingMinCost} with $f_1$ replaced by the minimum error achievable in steps prior to proofreading and $f_2$ replaced by the discrimination in the proofreading step in this network (see \textit{SI Appendix, Section V} for details). The error-cost bound is also obtained for a more detailed ribosome model~\citep{Wohlgemuth2011}, which demonstrates that taking additional intermediate states into account does not change the form of the bound (see \textit{SI Appendix, Section V} for details).
As shown in Fig.~\ref{fig:RealisticNetworkResults}B, the bound (red line) exactly encapsulates all sampled systems on the error-cost plane.
The vertical part of the bound indicates the minimum error for equilibrium discrimination $\eta_\mathrm{eq}$, above which proofreading becomes unnecessary and the minimum cost is zero.
Unexpectedly, the native system resides within this equilibrium regime, seemingly suggesting that the proofreading mechanism is redundant.
The error-cost trade-off has also been analyzed for the error-prone and hyperaccurate mutants of the ribosome~\citep{Zaher2010}, whose native error rates are also larger than $\eta_\mathrm{eq}$ (see \textit{SI Appendix, Section V}).

To understand what prevents the ribosome from realizing the theoretical possibility of maintaining the native error rate without proofreading, we remove both forward and backward proofreading reactions ($k_{\pm 2}=0$) and examine adjustments in  the other rate constants required to achieve the native error rate. Theoretically, the error is minimized when the product formation step is rate-limiting. It requires a time-scale separation where $k_p$, which corresponds to the accommodation of the aa-tRNA into the A site of the large subunit and the subsequent peptidyl-transfer, is much smaller than the rate constants of all the preceding reactions. Experiment measurements, however, found $k_p$ comparable to the rates of preceding reactions~\citep{Zaher2010}. It is three-fold smaller than the GTP hydrolysis rate and five-fold smaller than the binding rate of the ternary complex containing tRNA, EF-Tu, and GTP. Restoration of the native error rate would require reducing $k_p$ and/or increasing other rate constants.
On the one hand, reducing $k_p$ directly slows down the speed of protein synthesis and eventually the speed of cell growth, especially since translation is suggested as a rate-governing process in bacterial growth~\citep{Belliveau2021}.
If all the other rate constants remain invariant, $k_p$ needs to be decreased $700$-fold to recover the native error rate. This leads to significant decrease in the growth rate, which would seem evolutionarily detrimental.
On the other hand, amplification of the rates of the preceding reactions faces physical limitations. For example, the rate of ternary complex binding is already close to its upper limit which corresponds to diffusion-limited reaction~\citep{Klumpp2013}, rendering further rate increase impossible.
Therefore, it would seem that the condition to maintain the native error rate without proofreading could not be fulfilled without sacrificing the overall rate of protein synthesis and bacterial growth.
The analysis above indicates that reaction speed becomes an important factor when considering real proofreading networks, where the low-cost equilibrium discrimination regime permissible in the theory could be kinetically prohibited.
Therefore, the proofreading mechanism is still necessary in the native system, contributing to a 20-fold increase in the translation fidelity~\citep{Zaher2010}.
The above analysis indicates that the minimum dissipation for protein translation is limited by the speed constraints rather than the error-cost trade-off.

To extend our analysis to multi-stage proofreading networks, we study the reaction network for Isoleucyl-tRNA synthetase (IleRS) in \textit{E. coli}~\citep{Yu2020}.
The enzyme pairs tRNA\textsuperscript{Ile} with the cognate amino acid isoleucine (Ile) by discriminating it against a chemically similar amino acid, valine (Val)~\citep{Ling2009}.
The network has the structure of Fig.~\ref{fig:BasicSchemeAndSampling}B with $n=3$ proofreading stages.
The error-cost bound could be derived by generalizing the bound in MM-with-proofreading scheme with the mathematical induction method used in the $n$-stage DBD scheme (see \textit{SI Appendix, Section V}).
Fig.~\ref{fig:RealisticNetworkResults}C presents the error-cost relation due to the rates sampling, demonstrating that all systems sampled fall above the theoretical error-cost bound.
The native system falls within the non-equilibrium discrimination regime (i.e. $\eta<\eta_\mathrm{eq}$). Similar to T7 DNA polymerase, the enzyme resides close to the boundary, whose cost is 2.6-fold of the minimum cost required to maintain the same error rate.
In terms of the energy dissipated per Ile-tRNA\textsuperscript{Ile} formed ($\sigma$), the dissipation of the native system is only slightly (less than $20\%$) larger than the optimal system.
This finding reaffirms that IleRS is energetically efficient~\citep{Yu2020}.

The reason why the dissipation could not be further reduced could be explained by analyzing the optimal schemes corresponding to the bound.
In contrast to the $n$-stage DBD scheme where all proofreading pathways are equally leveraged, the IleRS network has three different proofreading regimes characterized by different number of ``effective'' proofreading pathways with nonzero proofreading fluxes.
To  understand this, we calculate the optimal partition ratios $a_{1,2,3}$ as a function of error $\eta$ (see Fig.~\ref{fig:RealisticNetworkResults}D). The full error range $\eta\in\qty(\eta_\mathrm{min},\eta_\mathrm{eq})$ can be categorized into three phases (represented by different shades in Fig.~\ref{fig:RealisticNetworkResults}D) by the number of nonzero partition ratios.
When the error is sufficiently small, all three stages need to be functional (yellow phase, $n=3$).
Due to the different discrimination factor, however, the partition ratios are different among the three stages.
The post-transfer proofreading pathway, which has the most discrimination, has the largest partition ratio $a_3$.
Conversely, the first pre-transfer proofreading pathway has the smallest partition ratio $a_1$.
These partition ratios decrease as the error is increased until a threshold (left black dashed line) is reached where $a_1\to 0$.
Further increase in error leads to a negative $a_1$, which is prohibited since the rate constants are always positive. Negative $a_1$ is also thermodynamically impossible since all proofreading pathways have the same nonequilibrium driving $\gamma$ which precludes coexistence of proofreading and anti-proofreading in different pathways (otherwise there will be cyclic flux on a reaction loop with no driving, such as the loop ER\textsubscript{1}--E--ER\textsubscript{2} in Fig.~\ref{fig:BasicSchemeAndSampling}B).
Therefore, the first pre-transfer proofreading pathway is turned off (namely $a_1=0$) for error larger than this threshold.
Similarly, there is a second threshold (right black dashed line) where $a_2\to 0$, turning off the second pre-transfer proofreading pathway.
The optimal system effectively operates in a two-stage proofreading regime for error rates between the two thresholds (green phase, $n=2$) and in a one-stage regime for error larger than the second threshold (blue phase, $n=1$).
The native system (red diamond) resides in the one-stage regime where the optimal scheme utilizes only the post-transfer proofreading mechanism.
The native partition ratios $a_1$ and $a_2$ are indeed negligibly small, while the native $a_3$ is about two-fold of its optimal value, accounting for the increased cost.
The increased $a_3$ can in fact be attributed to steps before proofreading, where the actual error rate is about twice of the minimum error $\eta_\mathrm{eq}$.
Reducing this error rate requires decreasing the rate of amino acid activation, which will decrease the speed of product formation.
This is consistent with the previous trade-off analysis on the activation rate, which indicated that the reaction optimizes speed over energy dissipation~\citep{Yu2020}.
Therefore, similar to the case of translation, the native IleRS system's deviation from the optimal bound could be explained by the speed requirement.
Moreover, the small partition ratios in pre-transfer editing ($a_{1,2}$) could be a result of selective pressure to reduce the energy dissipation in aminoacylation.
It is possible that after the early emergence of the CP1 editing domain which is responsible for post-transfer editing~\citep{Bullwinkle2014}, the pre-transfer editing activity ($a_{1,2}$) evolved to decrease, allowing errors to be corrected more efficiently in post-transfer editing ($a_3$).
Taken together, the analyses of translation and aminoacylation seem to suggest that the \textit{E. coli} places high priority on optimizing the rate of protein synthesis and therefore growth rate, even at the expense of higher proofreading cost.

\section{Discussion}
\subsection{The error-dissipation trade-off is kinetically controlled}
We have analyzed the error and energy cost of kinetic proofreading in a large class of reaction networks whose dynamics are governed by the chemical master equation (CME).
In terms of methodology, we propose a formalism whereby the probability fluxes serve as the primary variables of interest.
The flux-based formalism complements the CME formalism and provides a useful mathematical device for understanding the flux kinetics in reaction networks, especially those with symmetric and branching structures.
Applying the flux-based formalism to models of biological proofreading could reveal important biological insights.
In terms of physical interpretation in the context of the free energy landscape, we demonstrate that both error $\eta$ and cost $C$ depend only on the energy barriers rather than the energy levels of discrete states (kinetic control).
More precisely, the energy barriers determine error and cost through the partition ratio $a$.
Having uniform partition ratios in different proofreading stages is necessary for optimizing the error-cost relation in the $n$-stage DBD scheme, and the magnitude of the partition ratio determines the extent to which error or dissipation is prioritized in their trade-off.
In the minimal multi-stage proofreading scheme (Fig.~\ref{fig:MinimalScheme}A), it is further demonstrated that the energy level of discrete states is irrelevant to both error and cost, and that the uniform partition ratio parameterizes the system's position along the error-cost bound (Fig.~\ref{fig:MinimalScheme}C).

These theoretical analyses suggest experimental characterization of reaction fluxes rather than the rate constants as an important way to understand the kinetics of networks involving proofreading or similar branching structures.
Moreover, any properties that involve only the ratio of stationary fluxes, such as error and cost studied here, are fully characterized with the knowledge of energy barriers, which are, in turn, fully captured by the partition ratio of fluxes without knowing all the reaction rates.

The significance of transition state energy is further elucidated in MM-with-proofreading scheme. It is shown that while the energy barriers ($\epsilon^\dagger_{\lambda,\text{R/W}}$) affect error and cost, the difference of energy barriers between the correct and incorrect networks ($\Delta\epsilon^\dagger_\lambda$) fully determines the fundamental error-cost bound.
More precisely, the energy barrier difference corresponds to the free energy difference between cognate and noncognate substrates interacting with the enzyme in the same conformational state, which would be invariant under perturbations to the enzyme structure if a Linear Free Energy Relationship is assumed.
The rate sampling is equivalent to perturbing the energy landscape without affecting the energy barrier difference, and the networks are optimized in the sense of tuning $\epsilon^\dagger$ with fixed $\Delta \epsilon^\dagger$.
While the energy difference of discrete states is well characterized, for example, by the ratio of association constants, it is more difficult to determine the difference of energy barriers both experimentally and computationally.
A system-specific molecular dynamic analysis of the transition state configuration might be useful to account for the barrier difference of $\Delta\epsilon^\dagger$, which is the key to understanding how a specific biological discrimination process is kinetically controlled. Importantly, although the expression of the error-dissipation bound is system-specific, the theory and the kinetic control picture are general and by no means limited to the models studied here.

\subsection{Constraints and strategies in real proofreading systems}
The error-cost bound obtained in this work has several implications.
First, although it is known that the minimum error always corresponds to infinite dissipation, the bound provides the complete quantitative description of how fast dissipation must increase and eventually diverge as error is decreased. It also helps explain why several biological systems capable of achieving very low error maintain a relatively higher error instead~\citep{Banerjee2017PNAS,Mallory2019,Yu2020}.
Second, the multi-stage proofreading schemes reveals two approaches of reducing the cost at a given error: increasing the discrimination capacity of each proofreading pathway ($f$) or the number of proofreading pathways ($n$).
In biological systems, however, the values of $f$ and $n$ are upper bounded by various constraints.
$f$ is constrained by the difference in transition state energies, and $n$ is limited since introducing new proofreading pathways requires the enzyme to have either additional conformational states or dedicated domains for proofreading, such as the editing site in aminoacyl-tRNA synthetases~\citep{Nureki1998,Ling2009}.
Depending on the biochemical structure as well as the functional purpose of the enzyme, one of the two constraints might be predominantly challenging to circumvent, resulting in the enzyme preferentially adopting the alternate strategy to reduce the energy cost of proofreading.
However, the analysis in the multi-stage DBD scheme demonstrates that increasing $n$ leads to a diminishing benefit of cost reduction compared to increasing $f$.
This suggests that proofreading with fewer pathways and larger discrimination factors is favorable to proofreading with numerous pathways and small discrimination factors, potentially accounting for the rarity of multi-stage proofreading in nature.
The method is also applied to models with multiple intermediate states, demonstrating these states do not change the form of the bound.

The general applicability of our theoretical framework is demonstrated in three real biological proofreading networks. Surprisingly, the three systems seem to operate in different regimes.
The DNA polymerase resides remarkably close to the error-dissipation bound, demonstrating high energy efficiency in DNA replication. In contrast, the ribosome is unable to optimize its energy dissipation, possibly as a consequence of maximizing speed.
IleRS operates relatively close to the error-dissipation bound, but further optimization also seems to be prohibited by speed requirements.
Therefore, although speed is in theory decoupled from the error-dissipation trade-off, it still plays an important role in the evolutionary trade-offs among characteristic properties due to realistic limitations to which biochemical reactions can be accelerated.
The generalization of our formalism in these complex models allows identification of key states or mechanisms, which will be important to characterize experimentally.
The theoretical optimal scheme requires the binding and unbinding reactions to be much faster than the subsequent reaction. In reality, however, such time scale separation is not always possible since the binding/unbinding rates are upper-bounded by processes such as diffusion and substrate recognition, and the rate of the subsequent reaction is lower-bounded by the minimum overall reaction speed. Future studies generalizing the theoretical framework to include speed should take into consideration how these restrictions on rate constants constrain the free energy landscape and thereby affect the optimal error-cost bound.

Finally, our work provides a general framework of analyzing the error-dissipation trade-off in biochemical reaction networks capable of achieving high fidelity with nonequilibrium proofreading.
Given the importance of partition ratios, the error-cost bound could be determined as soon as a few key discrimination factors are measured.
Further insights could be revealed by subjecting the energy barrier differences, which exert the key kinetic control on both error and dissipation, to a more detailed and specific molecular dynamics or experimental investigation.
It would also be important to see the implication of the error-cost bound for other biological systems involving proofreading.

\section*{Acknowledgments}
The authors thank D.~Thirumalai, J.~Gunawardena, F.~Wong, U.~Cetiner and K.~Banerjee for valuable comments on the manuscript. 
This work was supported by Welch Foundation Grant C-1995 (to OAI) and Center for Theoretical Biological Physics National Science Foundation (NSF) Grant PHY-2019745.  A.B.K. also acknowledges support from the Welch Foundation Grant C-1559 and from NSF Grants CHE-1664218 and MCB-1941106.

\bibliography{KPR_ref}

\begin{thebibliography}{48}%
\makeatletter
\providecommand \@ifxundefined [1]{%
 \@ifx{#1\undefined}
}%
\providecommand \@ifnum [1]{%
 \ifnum #1\expandafter \@firstoftwo
 \else \expandafter \@secondoftwo
 \fi
}%
\providecommand \@ifx [1]{%
 \ifx #1\expandafter \@firstoftwo
 \else \expandafter \@secondoftwo
 \fi
}%
\providecommand \natexlab [1]{#1}%
\providecommand \enquote  [1]{``#1''}%
\providecommand \bibnamefont  [1]{#1}%
\providecommand \bibfnamefont [1]{#1}%
\providecommand \citenamefont [1]{#1}%
\providecommand \href@noop [0]{\@secondoftwo}%
\providecommand \href [0]{\begingroup \@sanitize@url \@href}%
\providecommand \@href[1]{\@@startlink{#1}\@@href}%
\providecommand \@@href[1]{\endgroup#1\@@endlink}%
\providecommand \@sanitize@url [0]{\catcode `\\12\catcode `\$12\catcode
  `\&12\catcode `\#12\catcode `\^12\catcode `\_12\catcode `\%12\relax}%
\providecommand \@@startlink[1]{}%
\providecommand \@@endlink[0]{}%
\providecommand \url  [0]{\begingroup\@sanitize@url \@url }%
\providecommand \@url [1]{\endgroup\@href {#1}{\urlprefix }}%
\providecommand \urlprefix  [0]{URL }%
\providecommand \Eprint [0]{\href }%
\providecommand \doibase [0]{https://doi.org/}%
\providecommand \selectlanguage [0]{\@gobble}%
\providecommand \bibinfo  [0]{\@secondoftwo}%
\providecommand \bibfield  [0]{\@secondoftwo}%
\providecommand \translation [1]{[#1]}%
\providecommand \BibitemOpen [0]{}%
\providecommand \bibitemStop [0]{}%
\providecommand \bibitemNoStop [0]{.\EOS\space}%
\providecommand \EOS [0]{\spacefactor3000\relax}%
\providecommand \BibitemShut  [1]{\csname bibitem#1\endcsname}%
\let\auto@bib@innerbib\@empty
\bibitem [{\citenamefont {Kunkel}\ and\ \citenamefont
  {Bebenek}(2000)}]{Kunkel2000}%
  \BibitemOpen
  \bibfield  {author} {\bibinfo {author} {\bibfnamefont {T.~A.}\ \bibnamefont
  {Kunkel}}\ and\ \bibinfo {author} {\bibfnamefont {K.}~\bibnamefont
  {Bebenek}},\ }\bibfield  {title} {\bibinfo {title} {{DNA replication
  fidelity}},\ }\href {https://doi.org/10.1146/annurev.biochem.69.1.497}
  {\bibfield  {journal} {\bibinfo  {journal} {Annu. Rev. Biochem.}\ }\textbf
  {\bibinfo {volume} {69}},\ \bibinfo {pages} {497} (\bibinfo {year}
  {2000})}\BibitemShut {NoStop}%
\bibitem [{\citenamefont {Sydow}\ and\ \citenamefont
  {Cramer}(2009)}]{Sydow2009}%
  \BibitemOpen
  \bibfield  {author} {\bibinfo {author} {\bibfnamefont {J.~F.}\ \bibnamefont
  {Sydow}}\ and\ \bibinfo {author} {\bibfnamefont {P.}~\bibnamefont {Cramer}},\
  }\bibfield  {title} {\bibinfo {title} {{RNA polymerase fidelity and
  transcriptional proofreading}},\ }\href
  {https://doi.org/10.1016/j.sbi.2009.10.009} {\bibfield  {journal} {\bibinfo
  {journal} {Curr. Opin. Struct. Biol.}\ }\textbf {\bibinfo {volume} {19}},\
  \bibinfo {pages} {732} (\bibinfo {year} {2009})}\BibitemShut {NoStop}%
\bibitem [{\citenamefont {Rodnina}\ and\ \citenamefont
  {Wintermeyer}(2001)}]{Rodnina2001}%
  \BibitemOpen
  \bibfield  {author} {\bibinfo {author} {\bibfnamefont {M.~V.}\ \bibnamefont
  {Rodnina}}\ and\ \bibinfo {author} {\bibfnamefont {W.}~\bibnamefont
  {Wintermeyer}},\ }\bibfield  {title} {\bibinfo {title} {{Fidelity of
  aminoacyl-tRNA selection on the ribosome: kinetic and structural
  mechanisms}},\ }\href {https://doi.org/10.1146/annurev.biochem.70.1.415}
  {\bibfield  {journal} {\bibinfo  {journal} {Annu. Rev. Biochem.}\ }\textbf
  {\bibinfo {volume} {70}},\ \bibinfo {pages} {415} (\bibinfo {year}
  {2001})}\BibitemShut {NoStop}%
\bibitem [{\citenamefont {Zaher}\ and\ \citenamefont
  {Green}(2009)}]{Zaher2009}%
  \BibitemOpen
  \bibfield  {author} {\bibinfo {author} {\bibfnamefont {H.~S.}\ \bibnamefont
  {Zaher}}\ and\ \bibinfo {author} {\bibfnamefont {R.}~\bibnamefont {Green}},\
  }\bibfield  {title} {\bibinfo {title} {{Fidelity at the molecular level:
  Lessons from protein synthesis}},\ }\href
  {https://doi.org/10.1016/j.cell.2009.01.036} {\bibfield  {journal} {\bibinfo
  {journal} {Cell}\ }\textbf {\bibinfo {volume} {136}},\ \bibinfo {pages} {746}
  (\bibinfo {year} {2009})}\BibitemShut {NoStop}%
\bibitem [{\citenamefont {Hopfield}(1974)}]{Hopfield1974}%
  \BibitemOpen
  \bibfield  {author} {\bibinfo {author} {\bibfnamefont {J.~J.}\ \bibnamefont
  {Hopfield}},\ }\bibfield  {title} {\bibinfo {title} {{Kinetic proofreading: A
  new mechanism for reducing errors in biosynthetic processes Requiring High
  Specificity}},\ }\href {https://doi.org/10.1073/pnas.71.10.4135} {\bibfield
  {journal} {\bibinfo  {journal} {Proc. Natl. Acad. Sci.}\ }\textbf {\bibinfo
  {volume} {71}},\ \bibinfo {pages} {4135} (\bibinfo {year}
  {1974})}\BibitemShut {NoStop}%
\bibitem [{\citenamefont {Ninio}(1975)}]{Ninio1975}%
  \BibitemOpen
  \bibfield  {author} {\bibinfo {author} {\bibfnamefont {J.}~\bibnamefont
  {Ninio}},\ }\bibfield  {title} {\bibinfo {title} {{Kinetic amplification of
  enzyme discrimination}},\ }\href
  {https://doi.org/10.1016/S0300-9084(75)80139-8} {\bibfield  {journal}
  {\bibinfo  {journal} {Biochimie}\ }\textbf {\bibinfo {volume} {57}},\
  \bibinfo {pages} {587} (\bibinfo {year} {1975})}\BibitemShut {NoStop}%
\bibitem [{\citenamefont {Bennett}(1979)}]{Bennett1979}%
  \BibitemOpen
  \bibfield  {author} {\bibinfo {author} {\bibfnamefont {C.~H.}\ \bibnamefont
  {Bennett}},\ }\bibfield  {title} {\bibinfo {title} {Dissipation-error
  tradeoff in proofreading},\ }\href
  {https://doi.org/https://doi.org/10.1016/0303-2647(79)90003-0} {\bibfield
  {journal} {\bibinfo  {journal} {Biosystems}\ }\textbf {\bibinfo {volume}
  {11}},\ \bibinfo {pages} {85 } (\bibinfo {year} {1979})}\BibitemShut
  {NoStop}%
\bibitem [{\citenamefont {Banerjee}\ \emph {et~al.}(2017)\citenamefont
  {Banerjee}, \citenamefont {Kolomeisky},\ and\ \citenamefont
  {Igoshin}}]{Banerjee2017PNAS}%
  \BibitemOpen
  \bibfield  {author} {\bibinfo {author} {\bibfnamefont {K.}~\bibnamefont
  {Banerjee}}, \bibinfo {author} {\bibfnamefont {A.~B.}\ \bibnamefont
  {Kolomeisky}},\ and\ \bibinfo {author} {\bibfnamefont {O.~A.}\ \bibnamefont
  {Igoshin}},\ }\bibfield  {title} {\bibinfo {title} {{Elucidating interplay of
  speed and accuracy in biological error correction}},\ }\href
  {https://doi.org/10.1073/pnas.1614838114} {\bibfield  {journal} {\bibinfo
  {journal} {Proc. Natl. Acad. Sci. U. S. A.}\ }\textbf {\bibinfo {volume}
  {114}},\ \bibinfo {pages} {5183} (\bibinfo {year} {2017})}\BibitemShut
  {NoStop}%
\bibitem [{\citenamefont {Mallory}\ \emph {et~al.}(2019)\citenamefont
  {Mallory}, \citenamefont {Kolomeisky},\ and\ \citenamefont
  {Igoshin}}]{Mallory2019}%
  \BibitemOpen
  \bibfield  {author} {\bibinfo {author} {\bibfnamefont {J.~D.}\ \bibnamefont
  {Mallory}}, \bibinfo {author} {\bibfnamefont {A.~B.}\ \bibnamefont
  {Kolomeisky}},\ and\ \bibinfo {author} {\bibfnamefont {O.~A.}\ \bibnamefont
  {Igoshin}},\ }\bibfield  {title} {\bibinfo {title} {{Trade-offs between
  error, speed, noise, and energy dissipation in biological processes with
  proofreading}},\ }\href {https://doi.org/10.1021/acs.jpcb.9b03757} {\bibfield
   {journal} {\bibinfo  {journal} {J. Phys. Chem. B}\ }\textbf {\bibinfo
  {volume} {123}},\ \bibinfo {pages} {4718} (\bibinfo {year}
  {2019})}\BibitemShut {NoStop}%
\bibitem [{\citenamefont {Wong}\ \emph {et~al.}(1991)\citenamefont {Wong},
  \citenamefont {Patel},\ and\ \citenamefont {Johnson}}]{Wong1991}%
  \BibitemOpen
  \bibfield  {author} {\bibinfo {author} {\bibfnamefont {I.}~\bibnamefont
  {Wong}}, \bibinfo {author} {\bibfnamefont {S.~S.}\ \bibnamefont {Patel}},\
  and\ \bibinfo {author} {\bibfnamefont {K.~A.}\ \bibnamefont {Johnson}},\
  }\bibfield  {title} {\bibinfo {title} {{An induced-fit kinetic mechanism for
  DNA replication fidelity: direct measurement by single-turnover kinetics}},\
  }\href {https://doi.org/10.1021/bi00216a030} {\bibfield  {journal} {\bibinfo
  {journal} {Biochemistry}\ }\textbf {\bibinfo {volume} {30}},\ \bibinfo
  {pages} {526} (\bibinfo {year} {1991})}\BibitemShut {NoStop}%
\bibitem [{\citenamefont {Gromadski}\ and\ \citenamefont
  {Rodnina}(2004)}]{Gromadski2004}%
  \BibitemOpen
  \bibfield  {author} {\bibinfo {author} {\bibfnamefont {K.~B.}\ \bibnamefont
  {Gromadski}}\ and\ \bibinfo {author} {\bibfnamefont {M.~V.}\ \bibnamefont
  {Rodnina}},\ }\bibfield  {title} {\bibinfo {title} {{Kinetic determinants of
  high-Fidelity tRNA discrimination on the ribosome}},\ }\href
  {https://doi.org/10.1016/S1097-2765(04)00005-X} {\bibfield  {journal}
  {\bibinfo  {journal} {Mol. Cell}\ }\textbf {\bibinfo {volume} {13}},\
  \bibinfo {pages} {191} (\bibinfo {year} {2004})}\BibitemShut {NoStop}%
\bibitem [{\citenamefont {Savageau}\ and\ \citenamefont
  {Freter}(1979)}]{Savageau1979Biochemistry}%
  \BibitemOpen
  \bibfield  {author} {\bibinfo {author} {\bibfnamefont {M.~A.}\ \bibnamefont
  {Savageau}}\ and\ \bibinfo {author} {\bibfnamefont {R.~R.}\ \bibnamefont
  {Freter}},\ }\bibfield  {title} {\bibinfo {title} {{Energy cost of
  proofreading to increase fidelity of transfer ribonucleic acid
  aminoacylation}},\ }\href {https://doi.org/10.1021/bi00583a008} {\bibfield
  {journal} {\bibinfo  {journal} {Biochemistry}\ }\textbf {\bibinfo {volume}
  {18}},\ \bibinfo {pages} {3486} (\bibinfo {year} {1979})}\BibitemShut
  {NoStop}%
\bibitem [{\citenamefont {Freter}\ and\ \citenamefont
  {Savageau}(1980)}]{Freter1980}%
  \BibitemOpen
  \bibfield  {author} {\bibinfo {author} {\bibfnamefont {R.~R.}\ \bibnamefont
  {Freter}}\ and\ \bibinfo {author} {\bibfnamefont {M.~A.}\ \bibnamefont
  {Savageau}},\ }\bibfield  {title} {\bibinfo {title} {{Proofreading systems of
  multiple stages for improved accuracy of biological discrimination}},\ }\href
  {https://doi.org/10.1016/0022-5193(80)90284-2} {\bibfield  {journal}
  {\bibinfo  {journal} {J. Theor. Biol.}\ }\textbf {\bibinfo {volume} {85}},\
  \bibinfo {pages} {99} (\bibinfo {year} {1980})}\BibitemShut {NoStop}%
\bibitem [{\citenamefont {Savageau}\ and\ \citenamefont
  {Lapointe}(1981)}]{Savageau1981}%
  \BibitemOpen
  \bibfield  {author} {\bibinfo {author} {\bibfnamefont {M.~A.}\ \bibnamefont
  {Savageau}}\ and\ \bibinfo {author} {\bibfnamefont {D.~S.}\ \bibnamefont
  {Lapointe}},\ }\bibfield  {title} {\bibinfo {title} {{Optimization of kinetic
  proofreading: A general method for derivation of the constraint relations and
  an exploration of a specific case}},\ }\href
  {https://doi.org/10.1016/0022-5193(81)90062-X} {\bibfield  {journal}
  {\bibinfo  {journal} {J. Theor. Biol.}\ }\textbf {\bibinfo {volume} {93}},\
  \bibinfo {pages} {157} (\bibinfo {year} {1981})}\BibitemShut {NoStop}%
\bibitem [{\citenamefont {Ehrenberg}\ and\ \citenamefont
  {Blomberg}(1980)}]{Ehrenberg1980}%
  \BibitemOpen
  \bibfield  {author} {\bibinfo {author} {\bibfnamefont {M.}~\bibnamefont
  {Ehrenberg}}\ and\ \bibinfo {author} {\bibfnamefont {C.}~\bibnamefont
  {Blomberg}},\ }\bibfield  {title} {\bibinfo {title} {{Thermodynamic
  constraints on kinetic proofreading in biosynthetic pathways}},\ }\href
  {https://doi.org/10.1016/S0006-3495(80)85063-6} {\bibfield  {journal}
  {\bibinfo  {journal} {Biophys. J.}\ }\textbf {\bibinfo {volume} {31}},\
  \bibinfo {pages} {333} (\bibinfo {year} {1980})}\BibitemShut {NoStop}%
\bibitem [{\citenamefont {Blomberg}\ and\ \citenamefont
  {Ehrenberg}(1981)}]{Blomberg1981}%
  \BibitemOpen
  \bibfield  {author} {\bibinfo {author} {\bibfnamefont {C.}~\bibnamefont
  {Blomberg}}\ and\ \bibinfo {author} {\bibfnamefont {M.}~\bibnamefont
  {Ehrenberg}},\ }\bibfield  {title} {\bibinfo {title} {{Energy considerations
  for kinetic proofreading in biosynthesis}},\ }\href
  {https://doi.org/10.1016/0022-5193(81)90242-3} {\bibfield  {journal}
  {\bibinfo  {journal} {J. Theor. Biol.}\ }\textbf {\bibinfo {volume} {88}},\
  \bibinfo {pages} {631} (\bibinfo {year} {1981})}\BibitemShut {NoStop}%
\bibitem [{\citenamefont {Murugan}\ \emph {et~al.}(2012)\citenamefont
  {Murugan}, \citenamefont {Huse},\ and\ \citenamefont
  {Leibler}}]{Murugan2012}%
  \BibitemOpen
  \bibfield  {author} {\bibinfo {author} {\bibfnamefont {A.}~\bibnamefont
  {Murugan}}, \bibinfo {author} {\bibfnamefont {D.~A.}\ \bibnamefont {Huse}},\
  and\ \bibinfo {author} {\bibfnamefont {S.}~\bibnamefont {Leibler}},\
  }\bibfield  {title} {\bibinfo {title} {{Speed, dissipation, and error in
  kinetic proofreading}},\ }\href {https://doi.org/10.1073/pnas.1119911109}
  {\bibfield  {journal} {\bibinfo  {journal} {Proc. Natl. Acad. Sci. U. S. A.}\
  }\textbf {\bibinfo {volume} {109}},\ \bibinfo {pages} {12034} (\bibinfo
  {year} {2012})}\BibitemShut {NoStop}%
\bibitem [{\citenamefont {Murugan}\ \emph {et~al.}(2014)\citenamefont
  {Murugan}, \citenamefont {Huse},\ and\ \citenamefont
  {Leibler}}]{Murugan2014}%
  \BibitemOpen
  \bibfield  {author} {\bibinfo {author} {\bibfnamefont {A.}~\bibnamefont
  {Murugan}}, \bibinfo {author} {\bibfnamefont {D.~A.}\ \bibnamefont {Huse}},\
  and\ \bibinfo {author} {\bibfnamefont {S.}~\bibnamefont {Leibler}},\
  }\bibfield  {title} {\bibinfo {title} {{Discriminatory proofreading regimes
  in nonequilibrium systems}},\ }\href
  {https://doi.org/10.1103/PhysRevX.4.021016} {\bibfield  {journal} {\bibinfo
  {journal} {Phys. Rev. X}\ }\textbf {\bibinfo {volume} {4}},\ \bibinfo {pages}
  {021016} (\bibinfo {year} {2014})}\BibitemShut {NoStop}%
\bibitem [{\citenamefont {Wong}\ \emph {et~al.}(2018)\citenamefont {Wong},
  \citenamefont {Amir},\ and\ \citenamefont {Gunawardena}}]{Wong2018}%
  \BibitemOpen
  \bibfield  {author} {\bibinfo {author} {\bibfnamefont {F.}~\bibnamefont
  {Wong}}, \bibinfo {author} {\bibfnamefont {A.}~\bibnamefont {Amir}},\ and\
  \bibinfo {author} {\bibfnamefont {J.}~\bibnamefont {Gunawardena}},\
  }\bibfield  {title} {\bibinfo {title} {{Energy-speed-accuracy relation in
  complex networks for biological discrimination}},\ }\href
  {https://doi.org/10.1103/PhysRevE.98.012420} {\bibfield  {journal} {\bibinfo
  {journal} {Phys. Rev. E}\ }\textbf {\bibinfo {volume} {98}},\ \bibinfo
  {pages} {012420} (\bibinfo {year} {2018})}\BibitemShut {NoStop}%
\bibitem [{\citenamefont {Sartori}\ and\ \citenamefont
  {Pigolotti}(2013)}]{Sartori2013}%
  \BibitemOpen
  \bibfield  {author} {\bibinfo {author} {\bibfnamefont {P.}~\bibnamefont
  {Sartori}}\ and\ \bibinfo {author} {\bibfnamefont {S.}~\bibnamefont
  {Pigolotti}},\ }\bibfield  {title} {\bibinfo {title} {{Kinetic versus
  energetic discrimination in biological copying}},\ }\href
  {https://doi.org/10.1103/PhysRevLett.110.188101} {\bibfield  {journal}
  {\bibinfo  {journal} {Phys. Rev. Lett.}\ }\textbf {\bibinfo {volume} {110}},\
  \bibinfo {pages} {188101} (\bibinfo {year} {2013})}\BibitemShut {NoStop}%
\bibitem [{\citenamefont {Sartori}\ and\ \citenamefont
  {Pigolotti}(2015)}]{Sartori2015}%
  \BibitemOpen
  \bibfield  {author} {\bibinfo {author} {\bibfnamefont {P.}~\bibnamefont
  {Sartori}}\ and\ \bibinfo {author} {\bibfnamefont {S.}~\bibnamefont
  {Pigolotti}},\ }\bibfield  {title} {\bibinfo {title} {{Thermodynamics of
  error correction}},\ }\href {https://doi.org/10.1103/PhysRevX.5.041039}
  {\bibfield  {journal} {\bibinfo  {journal} {Phys. Rev. X}\ }\textbf {\bibinfo
  {volume} {5}},\ \bibinfo {pages} {041039} (\bibinfo {year}
  {2015})}\BibitemShut {NoStop}%
\bibitem [{\citenamefont {Yu}\ \emph {et~al.}(2020)\citenamefont {Yu},
  \citenamefont {Mallory}, \citenamefont {Kolomeisky}, \citenamefont {Ling},\
  and\ \citenamefont {Igoshin}}]{Yu2020}%
  \BibitemOpen
  \bibfield  {author} {\bibinfo {author} {\bibfnamefont {Q.}~\bibnamefont
  {Yu}}, \bibinfo {author} {\bibfnamefont {J.~D.}\ \bibnamefont {Mallory}},
  \bibinfo {author} {\bibfnamefont {A.~B.}\ \bibnamefont {Kolomeisky}},
  \bibinfo {author} {\bibfnamefont {J.}~\bibnamefont {Ling}},\ and\ \bibinfo
  {author} {\bibfnamefont {O.~A.}\ \bibnamefont {Igoshin}},\ }\bibfield
  {title} {\bibinfo {title} {{Trade-offs between speed, accuracy, and
  dissipation in tRNA\textsuperscript{Ile} aminoacylation}},\ }\href
  {https://doi.org/10.1021/acs.jpclett.0c01073} {\bibfield  {journal} {\bibinfo
   {journal} {J. Phys. Chem. Lett.}\ }\textbf {\bibinfo {volume} {11}},\
  \bibinfo {pages} {4001} (\bibinfo {year} {2020})}\BibitemShut {NoStop}%
\bibitem [{\citenamefont {Mallory}\ \emph
  {et~al.}(2020{\natexlab{a}})\citenamefont {Mallory}, \citenamefont
  {Kolomeisky},\ and\ \citenamefont {Igoshin}}]{Mallory2020PNAS}%
  \BibitemOpen
  \bibfield  {author} {\bibinfo {author} {\bibfnamefont {J.~D.}\ \bibnamefont
  {Mallory}}, \bibinfo {author} {\bibfnamefont {A.~B.}\ \bibnamefont
  {Kolomeisky}},\ and\ \bibinfo {author} {\bibfnamefont {O.~A.}\ \bibnamefont
  {Igoshin}},\ }\bibfield  {title} {\bibinfo {title} {{Kinetic control of
  stationary flux ratios for a wide range of biochemical processes}},\ }\href
  {https://doi.org/10.1073/pnas.1920873117} {\bibfield  {journal} {\bibinfo
  {journal} {Proc. Natl. Acad. Sci. U. S. A.}\ }\textbf {\bibinfo {volume}
  {117}},\ \bibinfo {pages} {8884} (\bibinfo {year}
  {2020}{\natexlab{a}})}\BibitemShut {NoStop}%
\bibitem [{\citenamefont {Mallory}\ \emph
  {et~al.}(2020{\natexlab{b}})\citenamefont {Mallory}, \citenamefont
  {Igoshin},\ and\ \citenamefont {Kolomeisky}}]{Mallory2020JPCB}%
  \BibitemOpen
  \bibfield  {author} {\bibinfo {author} {\bibfnamefont {J.~D.}\ \bibnamefont
  {Mallory}}, \bibinfo {author} {\bibfnamefont {O.~A.}\ \bibnamefont
  {Igoshin}},\ and\ \bibinfo {author} {\bibfnamefont {A.~B.}\ \bibnamefont
  {Kolomeisky}},\ }\bibfield  {title} {\bibinfo {title} {{Do we understand the
  mechanisms used by biological systems to correct their errors?}},\ }\href
  {https://doi.org/10.1021/acs.jpcb.0c06180} {\bibfield  {journal} {\bibinfo
  {journal} {J. Phys. Chem. B}\ }\textbf {\bibinfo {volume} {124}},\ \bibinfo
  {pages} {9289} (\bibinfo {year} {2020}{\natexlab{b}})}\BibitemShut {NoStop}%
\bibitem [{\citenamefont {Hartich}\ \emph {et~al.}(2015)\citenamefont
  {Hartich}, \citenamefont {Barato},\ and\ \citenamefont
  {Seifert}}]{Hartich2015}%
  \BibitemOpen
  \bibfield  {author} {\bibinfo {author} {\bibfnamefont {D.}~\bibnamefont
  {Hartich}}, \bibinfo {author} {\bibfnamefont {A.~C.}\ \bibnamefont
  {Barato}},\ and\ \bibinfo {author} {\bibfnamefont {U.}~\bibnamefont
  {Seifert}},\ }\bibfield  {title} {\bibinfo {title} {{Nonequilibrium sensing
  and its analogy to kinetic proofreading}},\ }\bibfield  {journal} {\bibinfo
  {journal} {New J. Phys.}\ }\textbf {\bibinfo {volume} {17}},\ \href
  {https://doi.org/10.1088/1367-2630/17/5/055026}
  {10.1088/1367-2630/17/5/055026} (\bibinfo {year} {2015}),\ \Eprint
  {https://arxiv.org/abs/1502.02594} {1502.02594} \BibitemShut {NoStop}%
\bibitem [{\citenamefont {Galstyan}\ \emph {et~al.}(2020)\citenamefont
  {Galstyan}, \citenamefont {Husain}, \citenamefont {Xiao}, \citenamefont
  {Murugan},\ and\ \citenamefont {Phillips}}]{Galstyan2020}%
  \BibitemOpen
  \bibfield  {author} {\bibinfo {author} {\bibfnamefont {V.}~\bibnamefont
  {Galstyan}}, \bibinfo {author} {\bibfnamefont {K.}~\bibnamefont {Husain}},
  \bibinfo {author} {\bibfnamefont {F.}~\bibnamefont {Xiao}}, \bibinfo {author}
  {\bibfnamefont {A.}~\bibnamefont {Murugan}},\ and\ \bibinfo {author}
  {\bibfnamefont {R.}~\bibnamefont {Phillips}},\ }\bibfield  {title} {\bibinfo
  {title} {{Proofreading through spatial gradients}},\ }\href
  {https://doi.org/10.7554/eLife.60415} {\bibfield  {journal} {\bibinfo
  {journal} {Elife}\ }\textbf {\bibinfo {volume} {9}},\ \bibinfo {pages} {1}
  (\bibinfo {year} {2020})}\BibitemShut {NoStop}%
\bibitem [{\citenamefont {Mallory}\ \emph {et~al.}(2021)\citenamefont
  {Mallory}, \citenamefont {Mallory}, \citenamefont {Kolomeisky},\ and\
  \citenamefont {Igoshin}}]{Mallory2021}%
  \BibitemOpen
  \bibfield  {author} {\bibinfo {author} {\bibfnamefont {J.~D.}\ \bibnamefont
  {Mallory}}, \bibinfo {author} {\bibfnamefont {X.~F.}\ \bibnamefont
  {Mallory}}, \bibinfo {author} {\bibfnamefont {A.~B.}\ \bibnamefont
  {Kolomeisky}},\ and\ \bibinfo {author} {\bibfnamefont {O.~A.}\ \bibnamefont
  {Igoshin}},\ }\bibfield  {title} {\bibinfo {title} {{Theoretical analysis
  reveals the cost and benefit of proofreading in coronavirus genome
  replication}},\ }\href {https://doi.org/10.1021/acs.jpclett.1c00190}
  {\bibfield  {journal} {\bibinfo  {journal} {J. Phys. Chem. Lett.}\ }\textbf
  {\bibinfo {volume} {12}},\ \bibinfo {pages} {2691} (\bibinfo {year}
  {2021})}\BibitemShut {NoStop}%
\bibitem [{\citenamefont {Shoval}\ \emph {et~al.}(2012)\citenamefont {Shoval},
  \citenamefont {Sheftel}, \citenamefont {Shinar}, \citenamefont {Hart},
  \citenamefont {Ramote}, \citenamefont {Mayo}, \citenamefont {Dekel},
  \citenamefont {Kavanagh},\ and\ \citenamefont {Alon}}]{Shoval2012}%
  \BibitemOpen
  \bibfield  {author} {\bibinfo {author} {\bibfnamefont {O.}~\bibnamefont
  {Shoval}}, \bibinfo {author} {\bibfnamefont {H.}~\bibnamefont {Sheftel}},
  \bibinfo {author} {\bibfnamefont {G.}~\bibnamefont {Shinar}}, \bibinfo
  {author} {\bibfnamefont {Y.}~\bibnamefont {Hart}}, \bibinfo {author}
  {\bibfnamefont {O.}~\bibnamefont {Ramote}}, \bibinfo {author} {\bibfnamefont
  {A.}~\bibnamefont {Mayo}}, \bibinfo {author} {\bibfnamefont {E.}~\bibnamefont
  {Dekel}}, \bibinfo {author} {\bibfnamefont {K.}~\bibnamefont {Kavanagh}},\
  and\ \bibinfo {author} {\bibfnamefont {U.}~\bibnamefont {Alon}},\ }\bibfield
  {title} {\bibinfo {title} {{Evolutionary trade-offs, pareto optimality, and
  the geometry of phenotype space}},\ }\href
  {https://doi.org/10.1126/science.1217405} {\bibfield  {journal} {\bibinfo
  {journal} {Science}\ }\textbf {\bibinfo {volume} {336}},\ \bibinfo {pages}
  {1157} (\bibinfo {year} {2012})}\BibitemShut {NoStop}%
\bibitem [{\citenamefont {Pi{\~{n}}eros}\ and\ \citenamefont
  {Tlusty}(2020)}]{Pineros2020}%
  \BibitemOpen
  \bibfield  {author} {\bibinfo {author} {\bibfnamefont {W.~D.}\ \bibnamefont
  {Pi{\~{n}}eros}}\ and\ \bibinfo {author} {\bibfnamefont {T.}~\bibnamefont
  {Tlusty}},\ }\bibfield  {title} {\bibinfo {title} {{Kinetic proofreading and
  the limits of thermodynamic uncertainty}},\ }\href
  {https://doi.org/10.1103/PhysRevE.101.022415} {\bibfield  {journal} {\bibinfo
   {journal} {Phys. Rev. E}\ }\textbf {\bibinfo {volume} {101}},\ \bibinfo
  {pages} {1} (\bibinfo {year} {2020})}\BibitemShut {NoStop}%
\bibitem [{\citenamefont {Johnson}(1993)}]{Johnson1993}%
  \BibitemOpen
  \bibfield  {author} {\bibinfo {author} {\bibfnamefont {K.~A.}\ \bibnamefont
  {Johnson}},\ }\bibfield  {title} {\bibinfo {title} {{Conformational coupling
  in DNA polymerase fidelity}},\ }\href
  {https://doi.org/10.1146/annurev.bi.62.070193.003345} {\bibfield  {journal}
  {\bibinfo  {journal} {Annu. Rev. Biochem.}\ }\textbf {\bibinfo {volume}
  {62}},\ \bibinfo {pages} {685} (\bibinfo {year} {1993})}\BibitemShut
  {NoStop}%
\bibitem [{\citenamefont {Hill}(1977)}]{Hill1977}%
  \BibitemOpen
  \bibfield  {author} {\bibinfo {author} {\bibfnamefont {T.~L.}\ \bibnamefont
  {Hill}},\ }\href@noop {} {\emph {\bibinfo {title} {{Free energy transduction
  in biology}}}}\ (\bibinfo  {publisher} {Academic Press},\ \bibinfo {year}
  {1977})\BibitemShut {NoStop}%
\bibitem [{\citenamefont {Qian}(2006)}]{Qian2006}%
  \BibitemOpen
  \bibfield  {author} {\bibinfo {author} {\bibfnamefont {H.}~\bibnamefont
  {Qian}},\ }\bibfield  {title} {\bibinfo {title} {{Open-System nonequilibrium
  steady state: Statistical thermodynamics, fluctuations, and chemical
  oscillations}},\ }\href {https://doi.org/10.1021/jp061858z} {\bibfield
  {journal} {\bibinfo  {journal} {J. Phys. Chem. B}\ }\textbf {\bibinfo
  {volume} {110}},\ \bibinfo {pages} {15063} (\bibinfo {year}
  {2006})}\BibitemShut {NoStop}%
\bibitem [{\citenamefont {Estrada}\ \emph {et~al.}(2016)\citenamefont
  {Estrada}, \citenamefont {Wong}, \citenamefont {DePace},\ and\ \citenamefont
  {Gunawardena}}]{Estrada2016}%
  \BibitemOpen
  \bibfield  {author} {\bibinfo {author} {\bibfnamefont {J.}~\bibnamefont
  {Estrada}}, \bibinfo {author} {\bibfnamefont {F.}~\bibnamefont {Wong}},
  \bibinfo {author} {\bibfnamefont {A.}~\bibnamefont {DePace}},\ and\ \bibinfo
  {author} {\bibfnamefont {J.}~\bibnamefont {Gunawardena}},\ }\bibfield
  {title} {\bibinfo {title} {{Information integration and energy expenditure in
  gene regulation}},\ }\href {https://doi.org/10.1016/j.cell.2016.06.012}
  {\bibfield  {journal} {\bibinfo  {journal} {Cell}\ }\textbf {\bibinfo
  {volume} {166}},\ \bibinfo {pages} {234} (\bibinfo {year}
  {2016})}\BibitemShut {NoStop}%
\bibitem [{\citenamefont {Ling}\ \emph {et~al.}(2009)\citenamefont {Ling},
  \citenamefont {Reynolds},\ and\ \citenamefont {Ibba}}]{Ling2009}%
  \BibitemOpen
  \bibfield  {author} {\bibinfo {author} {\bibfnamefont {J.}~\bibnamefont
  {Ling}}, \bibinfo {author} {\bibfnamefont {N.}~\bibnamefont {Reynolds}},\
  and\ \bibinfo {author} {\bibfnamefont {M.}~\bibnamefont {Ibba}},\ }\bibfield
  {title} {\bibinfo {title} {{Aminoacyl-tRNA synthesis and translational
  quality control}},\ }\href
  {https://doi.org/10.1146/annurev.micro.091208.073210} {\bibfield  {journal}
  {\bibinfo  {journal} {Annu. Rev. Microbiol.}\ }\textbf {\bibinfo {volume}
  {63}},\ \bibinfo {pages} {61} (\bibinfo {year} {2009})}\BibitemShut {NoStop}%
\bibitem [{\citenamefont {Cvetesic}\ \emph {et~al.}(2012)\citenamefont
  {Cvetesic}, \citenamefont {Perona},\ and\ \citenamefont
  {Gruic-Sovulj}}]{Cvetesic2012}%
  \BibitemOpen
  \bibfield  {author} {\bibinfo {author} {\bibfnamefont {N.}~\bibnamefont
  {Cvetesic}}, \bibinfo {author} {\bibfnamefont {J.~J.}\ \bibnamefont
  {Perona}},\ and\ \bibinfo {author} {\bibfnamefont {I.}~\bibnamefont
  {Gruic-Sovulj}},\ }\bibfield  {title} {\bibinfo {title} {{Kinetic
  partitioning between synthetic and editing pathways in class I aminoacyl-tRNA
  synthetases occurs at both pre-transfer and post-transfer hydrolytic
  steps}},\ }\href {https://doi.org/10.1074/jbc.M112.372151} {\bibfield
  {journal} {\bibinfo  {journal} {J. Biol. Chem.}\ }\textbf {\bibinfo {volume}
  {287}},\ \bibinfo {pages} {25381} (\bibinfo {year} {2012})}\BibitemShut
  {NoStop}%
\bibitem [{\citenamefont {Dulic}\ \emph {et~al.}(2014)\citenamefont {Dulic},
  \citenamefont {Perona}, \citenamefont {Gruic-Sovulj}, \citenamefont {SI},
  \citenamefont {Dulic}, \citenamefont {Perona},\ and\ \citenamefont
  {Gruic-Sovulj}}]{Dulic2014}%
  \BibitemOpen
  \bibfield  {author} {\bibinfo {author} {\bibfnamefont {M.}~\bibnamefont
  {Dulic}}, \bibinfo {author} {\bibfnamefont {J.~J.}\ \bibnamefont {Perona}},
  \bibinfo {author} {\bibfnamefont {I.}~\bibnamefont {Gruic-Sovulj}}, \bibinfo
  {author} {\bibnamefont {SI}}, \bibinfo {author} {\bibfnamefont
  {M.}~\bibnamefont {Dulic}}, \bibinfo {author} {\bibfnamefont {J.~J.}\
  \bibnamefont {Perona}},\ and\ \bibinfo {author} {\bibfnamefont
  {I.}~\bibnamefont {Gruic-Sovulj}},\ }\bibfield  {title} {\bibinfo {title}
  {{Determinants for tRNA-dependent pretransfer editing in the synthetic site
  of isoleucyl-tRNA synthetase}},\ }\href {https://doi.org/10.1021/bi5007699}
  {\bibfield  {journal} {\bibinfo  {journal} {Biochemistry}\ }\textbf {\bibinfo
  {volume} {53}},\ \bibinfo {pages} {6189} (\bibinfo {year}
  {2014})}\BibitemShut {NoStop}%
\bibitem [{\citenamefont {Cvetesic}\ \emph {et~al.}(2015)\citenamefont
  {Cvetesic}, \citenamefont {Bilus},\ and\ \citenamefont
  {Gruic-Sovulj}}]{Cvetesic2015}%
  \BibitemOpen
  \bibfield  {author} {\bibinfo {author} {\bibfnamefont {N.}~\bibnamefont
  {Cvetesic}}, \bibinfo {author} {\bibfnamefont {M.}~\bibnamefont {Bilus}},\
  and\ \bibinfo {author} {\bibfnamefont {I.}~\bibnamefont {Gruic-Sovulj}},\
  }\bibfield  {title} {\bibinfo {title} {{The tRNA A76 hydroxyl groups control
  partitioning of the tRNA-dependent pre- and post-transfer editing pathways in
  class I tRNA synthetase}},\ }\href {https://doi.org/10.1074/jbc.M115.648568}
  {\bibfield  {journal} {\bibinfo  {journal} {J. Biol. Chem.}\ }\textbf
  {\bibinfo {volume} {290}},\ \bibinfo {pages} {13981} (\bibinfo {year}
  {2015})}\BibitemShut {NoStop}%
\bibitem [{\citenamefont {Mckeithan}(1995)}]{McKeithan1995}%
  \BibitemOpen
  \bibfield  {author} {\bibinfo {author} {\bibfnamefont {T.~W.}\ \bibnamefont
  {Mckeithan}},\ }\bibfield  {title} {\bibinfo {title} {{Kinetic proofreading
  in T-cell receptor signal transduction}},\ }\href
  {https://doi.org/10.1073/pnas.92.11.5042} {\bibfield  {journal} {\bibinfo
  {journal} {Proc. Natl. Acad. Sci. U. S. A.}\ }\textbf {\bibinfo {volume}
  {92}},\ \bibinfo {pages} {5042} (\bibinfo {year} {1995})}\BibitemShut
  {NoStop}%
\bibitem [{\citenamefont {Kirsch}(1972)}]{Kirsch1972}%
  \BibitemOpen
  \bibfield  {author} {\bibinfo {author} {\bibfnamefont {J.~F.}\ \bibnamefont
  {Kirsch}},\ }\bibinfo {title} {Linear free energy relationships in
  enzymology},\ in\ \href {https://doi.org/10.1007/978-1-4615-8660-9_8} {\emph
  {\bibinfo {booktitle} {Advances in linear free energy relationships}}},\
  \bibinfo {editor} {edited by\ \bibinfo {editor} {\bibfnamefont {N.~B.}\
  \bibnamefont {Chapman}}\ and\ \bibinfo {editor} {\bibfnamefont
  {J.}~\bibnamefont {Shorter}}}\ (\bibinfo  {publisher} {Springer US},\
  \bibinfo {address} {Boston, MA},\ \bibinfo {year} {1972})\ pp.\ \bibinfo
  {pages} {369--400}\BibitemShut {NoStop}%
\bibitem [{\citenamefont {Hammett}(1970)}]{Hammett1970}%
  \BibitemOpen
  \bibfield  {author} {\bibinfo {author} {\bibfnamefont {L.~P.}\ \bibnamefont
  {Hammett}},\ }\href@noop {} {\emph {\bibinfo {title} {{Physical organic
  chemistry}}}}\ (\bibinfo  {publisher} {McGraw-Hill},\ \bibinfo {year}
  {1970})\BibitemShut {NoStop}%
\bibitem [{\citenamefont {{\c{C}}etiner}\ and\ \citenamefont
  {Gunawardena}(2020)}]{Cetiner2020}%
  \BibitemOpen
  \bibfield  {author} {\bibinfo {author} {\bibfnamefont {U.}~\bibnamefont
  {{\c{C}}etiner}}\ and\ \bibinfo {author} {\bibfnamefont {J.}~\bibnamefont
  {Gunawardena}},\ }\bibfield  {title} {\bibinfo {title} {{Reformulating
  non-equilibrium steady-states and generalised Hopfield discrimination}},\
  }\href@noop {} {\bibfield  {journal} {\bibinfo  {journal} {arXiv}\ ,\
  \bibinfo {pages} {1}} (\bibinfo {year} {2020})},\ \Eprint
  {https://arxiv.org/abs/2011.14994} {arXiv:2011.14994} \BibitemShut {NoStop}%
\bibitem [{\citenamefont {Laidler}(1987)}]{Laidler1987}%
  \BibitemOpen
  \bibfield  {author} {\bibinfo {author} {\bibfnamefont {K.~J.}\ \bibnamefont
  {Laidler}},\ }\href@noop {} {\emph {\bibinfo {title} {{Chemical kinetics}}}}\
  (\bibinfo  {publisher} {Prentice Hall},\ \bibinfo {year} {1987})\BibitemShut
  {NoStop}%
\bibitem [{\citenamefont {Zaher}\ and\ \citenamefont
  {Green}(2010)}]{Zaher2010}%
  \BibitemOpen
  \bibfield  {author} {\bibinfo {author} {\bibfnamefont {H.~S.}\ \bibnamefont
  {Zaher}}\ and\ \bibinfo {author} {\bibfnamefont {R.}~\bibnamefont {Green}},\
  }\bibfield  {title} {\bibinfo {title} {{Hyperaccurate and error-prone
  ribosomes exploit distinct mechanisms during tRNA selection}},\ }\href
  {https://doi.org/10.1016/j.molcel.2010.06.009} {\bibfield  {journal}
  {\bibinfo  {journal} {Mol. Cell}\ }\textbf {\bibinfo {volume} {39}},\
  \bibinfo {pages} {110} (\bibinfo {year} {2010})}\BibitemShut {NoStop}%
\bibitem [{\citenamefont {Wohlgemuth}\ \emph {et~al.}(2011)\citenamefont
  {Wohlgemuth}, \citenamefont {Pohl}, \citenamefont {Mittelstaet},
  \citenamefont {Konevega},\ and\ \citenamefont {Rodnina}}]{Wohlgemuth2011}%
  \BibitemOpen
  \bibfield  {author} {\bibinfo {author} {\bibfnamefont {I.}~\bibnamefont
  {Wohlgemuth}}, \bibinfo {author} {\bibfnamefont {C.}~\bibnamefont {Pohl}},
  \bibinfo {author} {\bibfnamefont {J.}~\bibnamefont {Mittelstaet}}, \bibinfo
  {author} {\bibfnamefont {A.~L.}\ \bibnamefont {Konevega}},\ and\ \bibinfo
  {author} {\bibfnamefont {M.~V.}\ \bibnamefont {Rodnina}},\ }\bibfield
  {title} {\bibinfo {title} {{Evolutionary optimization of speed and accuracy
  of decoding on the ribosome}},\ }\href
  {https://doi.org/10.1098/rstb.2011.0138} {\bibfield  {journal} {\bibinfo
  {journal} {Philos. Trans. R. Soc. Lond., B, Biol. Sci.}\ }\textbf {\bibinfo
  {volume} {366}},\ \bibinfo {pages} {2979} (\bibinfo {year}
  {2011})}\BibitemShut {NoStop}%
\bibitem [{\citenamefont {Belliveau}\ \emph {et~al.}(2021)\citenamefont
  {Belliveau}, \citenamefont {Chure}, \citenamefont {Hueschen}, \citenamefont
  {Garcia}, \citenamefont {Kondev}, \citenamefont {Fisher}, \citenamefont
  {Theriot},\ and\ \citenamefont {Phillips}}]{Belliveau2021}%
  \BibitemOpen
  \bibfield  {author} {\bibinfo {author} {\bibfnamefont {N.~M.}\ \bibnamefont
  {Belliveau}}, \bibinfo {author} {\bibfnamefont {G.}~\bibnamefont {Chure}},
  \bibinfo {author} {\bibfnamefont {C.~L.}\ \bibnamefont {Hueschen}}, \bibinfo
  {author} {\bibfnamefont {H.~G.}\ \bibnamefont {Garcia}}, \bibinfo {author}
  {\bibfnamefont {J.}~\bibnamefont {Kondev}}, \bibinfo {author} {\bibfnamefont
  {D.~S.}\ \bibnamefont {Fisher}}, \bibinfo {author} {\bibfnamefont {J.~A.}\
  \bibnamefont {Theriot}},\ and\ \bibinfo {author} {\bibfnamefont
  {R.}~\bibnamefont {Phillips}},\ }\bibfield  {title} {\bibinfo {title}
  {{Fundamental limits on the rate of bacterial growth and their influence on
  proteomic composition}},\ }\href {https://doi.org/10.1016/j.cels.2021.06.002}
  {\bibfield  {journal} {\bibinfo  {journal} {Cell Syst.}\ }\textbf {\bibinfo
  {volume} {12}},\ \bibinfo {pages} {924} (\bibinfo {year} {2021})}\BibitemShut
  {NoStop}%
\bibitem [{\citenamefont {Klumpp}\ \emph {et~al.}(2013)\citenamefont {Klumpp},
  \citenamefont {Scott}, \citenamefont {Pedersen},\ and\ \citenamefont
  {Hwa}}]{Klumpp2013}%
  \BibitemOpen
  \bibfield  {author} {\bibinfo {author} {\bibfnamefont {S.}~\bibnamefont
  {Klumpp}}, \bibinfo {author} {\bibfnamefont {M.}~\bibnamefont {Scott}},
  \bibinfo {author} {\bibfnamefont {S.}~\bibnamefont {Pedersen}},\ and\
  \bibinfo {author} {\bibfnamefont {T.}~\bibnamefont {Hwa}},\ }\bibfield
  {title} {\bibinfo {title} {{Molecular crowding limits translation and cell
  growth}},\ }\href {https://doi.org/10.1073/pnas.1310377110} {\bibfield
  {journal} {\bibinfo  {journal} {Proc. Natl. Acad. Sci. U. S. A.}\ }\textbf
  {\bibinfo {volume} {110}},\ \bibinfo {pages} {16754} (\bibinfo {year}
  {2013})}\BibitemShut {NoStop}%
\bibitem [{\citenamefont {Bullwinkle}\ and\ \citenamefont
  {Ibba}(2014)}]{Bullwinkle2014}%
  \BibitemOpen
  \bibfield  {author} {\bibinfo {author} {\bibfnamefont {T.~J.}\ \bibnamefont
  {Bullwinkle}}\ and\ \bibinfo {author} {\bibfnamefont {M.}~\bibnamefont
  {Ibba}},\ }\bibinfo {title} {Emergence and evolution},\ in\ \href
  {https://doi.org/10.1007/128_2013_423} {\emph {\bibinfo {booktitle}
  {Aminoacyl-tRNA synthetases in biology and medicine}}},\ \bibinfo {editor}
  {edited by\ \bibinfo {editor} {\bibfnamefont {S.}~\bibnamefont {Kim}}}\
  (\bibinfo  {publisher} {Springer Netherlands},\ \bibinfo {address}
  {Dordrecht},\ \bibinfo {year} {2014})\ pp.\ \bibinfo {pages}
  {43--87}\BibitemShut {NoStop}%
\bibitem [{\citenamefont {Nureki}\ \emph {et~al.}(1998)\citenamefont {Nureki},
  \citenamefont {Vassylyev}, \citenamefont {Tateno}, \citenamefont {Shimada},
  \citenamefont {Nakama}, \citenamefont {Fukai}, \citenamefont {Konno},
  \citenamefont {Hendrickson}, \citenamefont {Schimmel},\ and\ \citenamefont
  {Yokoyama}}]{Nureki1998}%
  \BibitemOpen
  \bibfield  {author} {\bibinfo {author} {\bibfnamefont {O.}~\bibnamefont
  {Nureki}}, \bibinfo {author} {\bibfnamefont {D.~G.}\ \bibnamefont
  {Vassylyev}}, \bibinfo {author} {\bibfnamefont {M.}~\bibnamefont {Tateno}},
  \bibinfo {author} {\bibfnamefont {A.}~\bibnamefont {Shimada}}, \bibinfo
  {author} {\bibfnamefont {T.}~\bibnamefont {Nakama}}, \bibinfo {author}
  {\bibfnamefont {S.}~\bibnamefont {Fukai}}, \bibinfo {author} {\bibfnamefont
  {M.}~\bibnamefont {Konno}}, \bibinfo {author} {\bibfnamefont {T.~L.}\
  \bibnamefont {Hendrickson}}, \bibinfo {author} {\bibfnamefont
  {P.}~\bibnamefont {Schimmel}},\ and\ \bibinfo {author} {\bibfnamefont
  {S.}~\bibnamefont {Yokoyama}},\ }\bibfield  {title} {\bibinfo {title} {Enzyme
  structure with two catalytic sites for double-sieve selection of substrate},\
  }\href {https://doi.org/10.1126/science.280.5363.578} {\bibfield  {journal}
  {\bibinfo  {journal} {Science}\ }\textbf {\bibinfo {volume} {280}},\ \bibinfo
  {pages} {578} (\bibinfo {year} {1998})}\BibitemShut {NoStop}%
\end{thebibliography}%


\begin{thebibliography}{5}%
\makeatletter
\providecommand \@ifxundefined [1]{%
 \@ifx{#1\undefined}
}%
\providecommand \@ifnum [1]{%
 \ifnum #1\expandafter \@firstoftwo
 \else \expandafter \@secondoftwo
 \fi
}%
\providecommand \@ifx [1]{%
 \ifx #1\expandafter \@firstoftwo
 \else \expandafter \@secondoftwo
 \fi
}%
\providecommand \natexlab [1]{#1}%
\providecommand \enquote  [1]{``#1''}%
\providecommand \bibnamefont  [1]{#1}%
\providecommand \bibfnamefont [1]{#1}%
\providecommand \citenamefont [1]{#1}%
\providecommand \href@noop [0]{\@secondoftwo}%
\providecommand \href [0]{\begingroup \@sanitize@url \@href}%
\providecommand \@href[1]{\@@startlink{#1}\@@href}%
\providecommand \@@href[1]{\endgroup#1\@@endlink}%
\providecommand \@sanitize@url [0]{\catcode `\\12\catcode `\$12\catcode
  `\&12\catcode `\#12\catcode `\^12\catcode `\_12\catcode `\%12\relax}%
\providecommand \@@startlink[1]{}%
\providecommand \@@endlink[0]{}%
\providecommand \url  [0]{\begingroup\@sanitize@url \@url }%
\providecommand \@url [1]{\endgroup\@href {#1}{\urlprefix }}%
\providecommand \urlprefix  [0]{URL }%
\providecommand \Eprint [0]{\href }%
\providecommand \doibase [0]{https://doi.org/}%
\providecommand \selectlanguage [0]{\@gobble}%
\providecommand \bibinfo  [0]{\@secondoftwo}%
\providecommand \bibfield  [0]{\@secondoftwo}%
\providecommand \translation [1]{[#1]}%
\providecommand \BibitemOpen [0]{}%
\providecommand \bibitemStop [0]{}%
\providecommand \bibitemNoStop [0]{.\EOS\space}%
\providecommand \EOS [0]{\spacefactor3000\relax}%
\providecommand \BibitemShut  [1]{\csname bibitem#1\endcsname}%
\let\auto@bib@innerbib\@empty
\bibitem [{\citenamefont {Banerjee}\ \emph {et~al.}(2017)\citenamefont
  {Banerjee}, \citenamefont {Kolomeisky},\ and\ \citenamefont
  {Igoshin}}]{Banerjee2017PNAS}%
  \BibitemOpen
  \bibfield  {author} {\bibinfo {author} {\bibfnamefont {K.}~\bibnamefont
  {Banerjee}}, \bibinfo {author} {\bibfnamefont {A.~B.}\ \bibnamefont
  {Kolomeisky}},\ and\ \bibinfo {author} {\bibfnamefont {O.~A.}\ \bibnamefont
  {Igoshin}},\ }\bibfield  {title} {\bibinfo {title} {{Elucidating interplay of
  speed and accuracy in biological error correction}},\ }\href
  {https://doi.org/10.1073/pnas.1614838114} {\bibfield  {journal} {\bibinfo
  {journal} {Proc. Natl. Acad. Sci. U. S. A.}\ }\textbf {\bibinfo {volume}
  {114}},\ \bibinfo {pages} {5183} (\bibinfo {year} {2017})}\BibitemShut
  {NoStop}%
\bibitem [{\citenamefont {Mallory}\ \emph {et~al.}(2019)\citenamefont
  {Mallory}, \citenamefont {Kolomeisky},\ and\ \citenamefont
  {Igoshin}}]{Mallory2019}%
  \BibitemOpen
  \bibfield  {author} {\bibinfo {author} {\bibfnamefont {J.~D.}\ \bibnamefont
  {Mallory}}, \bibinfo {author} {\bibfnamefont {A.~B.}\ \bibnamefont
  {Kolomeisky}},\ and\ \bibinfo {author} {\bibfnamefont {O.~A.}\ \bibnamefont
  {Igoshin}},\ }\bibfield  {title} {\bibinfo {title} {{Trade-offs between
  error, speed, noise, and energy dissipation in biological processes with
  proofreading}},\ }\href {https://doi.org/10.1021/acs.jpcb.9b03757} {\bibfield
   {journal} {\bibinfo  {journal} {J. Phys. Chem. B}\ }\textbf {\bibinfo
  {volume} {123}},\ \bibinfo {pages} {4718} (\bibinfo {year}
  {2019})}\BibitemShut {NoStop}%
\bibitem [{\citenamefont {Yu}\ \emph {et~al.}(2020)\citenamefont {Yu},
  \citenamefont {Mallory}, \citenamefont {Kolomeisky}, \citenamefont {Ling},\
  and\ \citenamefont {Igoshin}}]{Yu2020}%
  \BibitemOpen
  \bibfield  {author} {\bibinfo {author} {\bibfnamefont {Q.}~\bibnamefont
  {Yu}}, \bibinfo {author} {\bibfnamefont {J.~D.}\ \bibnamefont {Mallory}},
  \bibinfo {author} {\bibfnamefont {A.~B.}\ \bibnamefont {Kolomeisky}},
  \bibinfo {author} {\bibfnamefont {J.}~\bibnamefont {Ling}},\ and\ \bibinfo
  {author} {\bibfnamefont {O.~A.}\ \bibnamefont {Igoshin}},\ }\bibfield
  {title} {\bibinfo {title} {{Trade-offs between speed, accuracy, and
  dissipation in tRNA\textsuperscript{Ile} aminoacylation}},\ }\href
  {https://doi.org/10.1021/acs.jpclett.0c01073} {\bibfield  {journal} {\bibinfo
   {journal} {J. Phys. Chem. Lett.}\ }\textbf {\bibinfo {volume} {11}},\
  \bibinfo {pages} {4001} (\bibinfo {year} {2020})}\BibitemShut {NoStop}%
\bibitem [{\citenamefont {Wohlgemuth}\ \emph {et~al.}(2011)\citenamefont
  {Wohlgemuth}, \citenamefont {Pohl}, \citenamefont {Mittelstaet},
  \citenamefont {Konevega},\ and\ \citenamefont {Rodnina}}]{Wohlgemuth2011}%
  \BibitemOpen
  \bibfield  {author} {\bibinfo {author} {\bibfnamefont {I.}~\bibnamefont
  {Wohlgemuth}}, \bibinfo {author} {\bibfnamefont {C.}~\bibnamefont {Pohl}},
  \bibinfo {author} {\bibfnamefont {J.}~\bibnamefont {Mittelstaet}}, \bibinfo
  {author} {\bibfnamefont {A.~L.}\ \bibnamefont {Konevega}},\ and\ \bibinfo
  {author} {\bibfnamefont {M.~V.}\ \bibnamefont {Rodnina}},\ }\bibfield
  {title} {\bibinfo {title} {{Evolutionary optimization of speed and accuracy
  of decoding on the ribosome}},\ }\href
  {https://doi.org/10.1098/rstb.2011.0138} {\bibfield  {journal} {\bibinfo
  {journal} {Philos. Trans. R. Soc. Lond., B, Biol. Sci.}\ }\textbf {\bibinfo
  {volume} {366}},\ \bibinfo {pages} {2979} (\bibinfo {year}
  {2011})}\BibitemShut {NoStop}%
\bibitem [{\citenamefont {Zaher}\ and\ \citenamefont
  {Green}(2010)}]{Zaher2010}%
  \BibitemOpen
  \bibfield  {author} {\bibinfo {author} {\bibfnamefont {H.~S.}\ \bibnamefont
  {Zaher}}\ and\ \bibinfo {author} {\bibfnamefont {R.}~\bibnamefont {Green}},\
  }\bibfield  {title} {\bibinfo {title} {{Hyperaccurate and error-prone
  ribosomes exploit distinct mechanisms during tRNA selection}},\ }\href
  {https://doi.org/10.1016/j.molcel.2010.06.009} {\bibfield  {journal}
  {\bibinfo  {journal} {Mol. Cell}\ }\textbf {\bibinfo {volume} {39}},\
  \bibinfo {pages} {110} (\bibinfo {year} {2010})}\BibitemShut {NoStop}%
\end{thebibliography}%

\end{document}


\title{Supporting Information: The energy cost and optimal design of networks for biological discrimination} 
\author{Qiwei Yu}
\affiliation{Center for Theoretical Biological Physics, Rice University, Houston, TX 77005}
\affiliation{Lewis-Sigler Institute for Integrative Genomics, Princeton University, Princeton, NJ 08544}

\author{Anatoly B.~Kolomeisky} 
\affiliation{Center for Theoretical Biological Physics, Rice University, Houston, TX 77005}
\affiliation{Department of Chemistry, Rice University, Houston, TX 77005}
\affiliation{Department of Chemical and Biomolecular Engineering, Rice University, Houston, TX 77005}
\affiliation{Department of Physics and Astronomy, Rice University, Houston, TX 77005}

\author{Oleg A.~Igoshin}
\email[Corresponding author: ]{igoshin@rice.edu}
\affiliation{Center for Theoretical Biological Physics, Rice University, Houston, TX 77005}
\affiliation{Department of Chemistry, Rice University, Houston, TX 77005}
\affiliation{Department of Bioengineering, Rice University, Houston, TX 77005}
\affiliation{Department of Biosciences, Rice University, Houston, TX 77005}
\date{\today}

\maketitle
\tableofcontents
\newpage
\section{Flux-based formalism of the original Hopfield scheme}
\label{SuppSec:Original Hopfield Proof}
This section provides the detailed derivation of the results for the original Hopfield scheme (Fig.~1A in main text), including the flux-based formalism (Fig.~2A) and the error-cost bound. 
We also consider how finite $\gamma$ introduces a small increase to the minimum energy cost.

\subsection{Deriving the normalized steady-state fluxes for the flux-based formalism}
As shown in Fig.~2A in the main text, the normalized fluxes in the correct half of the network are denoted by $j_{\pm 1}$, $j_{\pm 2}$, and $\beta_{\pm 1}$. The normalized fluxes in the incorrect half of the network are denoted by their primed counterparts, which we shall derive in terms of the correct fluxes.

First, the (normalized) fluxes originating from the free enzyme E are given by
\begin{equation}
	j_1' = \frac{k_1 P_\mathrm{E}}{J_R} = j_1,\quad 
	\beta_{-1}' = \frac{k_{-3}P_\mathrm{E}}{J_R} = \beta_{-1},
\end{equation}
which are exactly equal to their counterparts in the correct half of the network. 
As mentioned in the main text, an additional error rate $\eta_0$ is defined as the ratio of the flux from EW to EW\textsuperscript{${*}$} to the flux from ER to ER\textsuperscript{${*}$}:
\begin{equation}
	\eta_0 = \frac{k_{2}P_\mathrm{EW}}{k_{2}P_\mathrm{ER}} = \frac{k_{2}P_\mathrm{EW}}{J_R} \frac{J_R}{k_{2}P_\mathrm{ER}} = \frac{j_2'}{j_2}.
\end{equation}
Hence we have $j_{2}'=j_2\eta_0$. 
In addition, $j_{-1}'$ can be related to $j_{-1}$ through $j_{2}'$:
\begin{equation}
	\frac{j_{-1}'}{j_{2}'} = \frac{ f k_{-1}}{k_2} =  f  \frac{j_{-1}}{j_2} \Longrightarrow j_{-1}' =  f j_{-1} \frac{j_{2}'}{j_2} =  f \eta_0 j_{-1}.
\end{equation}
Following the same line of thinking, the (normalized) fluxes originating from the activated state EW\textsuperscript{${*}$} are given by:
\begin{align}
	j_p' &= \frac{k_pP_\mathrm{EW^{*}}}{J_R} = \frac{J_W}{J_R}=\eta,\\
	j_{-2}' &= \frac{k_{-2}P_\mathrm{EW^{*}}}{J_R} = \frac{k_{-2}}{k_p} j_p' = j_p' \frac{k_{-2}P_\mathrm{ER^{*}}}{k_pP_\mathrm{ER^{*}}}  = \eta j_{-2},\\
	\beta_1' &= \frac{ f{}  k_{3}P_\mathrm{EW^{*}}}{J_R} =  f{}  \frac{k_{3}}{k_p} y_p' =  f{}   y_p' \frac{k_{3}P_\mathrm{ER^{*}}}{k_pP_\mathrm{ER^{*}}} =  f{}  \eta \beta_1,
\end{align}
where $j_p'=\eta$ was the normalized flux for incorrect formation directly given in Fig.~2A.
Thus, all expressions given in the box in Fig.~2A have been derived.

\subsection{Deriving the error-cost bound}
The stationary conditions for the fluxes for states ER, EW, ER\textsuperscript{*}, and EW\textsuperscript{*} are:
\begin{align}
	j_{1} + j_{-2} &= j_{-1} + j_{2},\\
	j_{2}+\beta_{-1} &= j_{-2}+\beta_1 +1, \\
	j_{1} + \eta j_{-2} & = f\eta_0 j_{-1} + \eta_0 j_{2},\\
	\eta_0 j_2 + \beta_{-1} &= \eta j_{-2} + f\eta \beta_1 + \eta .
\end{align}
The stationary condition for E is guaranteed if the stationary conditions for all other states are satisfied. 
From the first two equations, we eliminate $j_{\pm2}$ and find $j_1-j_{-1} = \beta_1 - \beta_{-1}+1$. From the last two equations, we find $j_1-f\eta_0 j_{-1} = \eta f \beta_1 - \beta_{-1}+\eta$. Subtracting these two relations yields
\begin{equation}
	\qty(f\eta_0-1) j_{-1} = (1-\eta) + (1-\eta f) \beta_{1}.
\end{equation}
Since the normalized fluxes are positive by definition and the error we consider falls within the range $\eta<f^{-1}<1$ (error rates larger than $f^{-1}$ can be achieved without any proofreading), 
the right hand side (RHS) must be positive. Thus, the left hand side (LHS) is also positive, leading to $\eta_0 > f^{-1}$. Indeed, $\eta_0$ only approaches its minimum $f^{-1}$ in the limit $j_{-1}\to+\infty$.
$j_{1}$ would also diverge to infinity in this limit, which corresponds to the fast equilibrium condition in the $j_{\pm 1}$ step.

Recall the cost (Eq.~9 in main text):
\begin{equation}
	C = \frac{(1+\eta f) \beta_1 - 2\beta_{-1}}{1+\eta}.
\end{equation}
From the second and fourth stationary condition:
\begin{equation}
	\beta_1 = j_2-j_{-2} + \beta_{-1} -1 = \frac{1}{f\eta} \qty[\eta_0 j_2 - \eta j_{-2} + \beta_{-1} - \eta] \Rightarrow j_2 = \frac{(f-1)\eta(1+j_{-2}) + (1-\eta f)\beta_{-1}}{\eta f - \eta_0}.
\end{equation}
Thus, $\beta_1$ can be eliminated from the expression for the cost:
\begin{equation}
	\begin{aligned}
		C & = \frac{(1+\eta f) \beta_1 - 2\beta_{-1}}{1+\eta} \\
		& = \frac{(1+\eta_0)j_2 - (1+\eta)(1+j_{-2})}{1+\eta} \\
		& = \frac{1+\eta_0}{1+\eta} \frac{(f-1)\eta(1+j_{-2}) + (1-\eta f)\beta_{-1}}{\eta f - \eta_0} - \qty(1+j_{-2}) \\
		& =  \frac{(\eta_0-\eta)(1+\eta f)}{(1+\eta)(\eta f-\eta_0)} \qty(1+j_{-2})
		+ \frac{(1+\eta_0)(1-\eta f)}{(1+\eta)(\eta f-\eta_0)} \beta_{-1},
	\end{aligned}
\end{equation}
which only depends on $\eta_0$, $j_{-2}$, and $\beta_{-1}$. Since $\eta < f^{-1}<\eta_0$, the coefficients $\frac{(\eta_0-\eta)(1+\eta f)}{(1+\eta)(\eta f-\eta_0)} $ and $\frac{(1+\eta_0)(1-\eta f)}{(1+\eta)(\eta f-\eta_0)}$ are both positive. The cost decreases monotonically with $\eta_0$, $j_{-2}$ and $\beta_{-1}$. The cost is minimized in the limit:
\begin{equation}
	\eta_0 \to f^{-1}, \quad j_{-2} \to 0,\quad \beta_{-1} \to 0.
\end{equation}
The minimum cost reads
\begin{equation}
	C_\text{min} =  \frac{(f_{-1}-\eta)(1+\eta f)}{(1+\eta)(\eta f-f^{-1})}= \frac{1-\eta^2f^2}{(1+\eta)(\eta f^2-1)},
\end{equation}
which gives Eq.~10 in the main text. 
In the optimal system, other fluxes are given by the stationary condition:
\begin{equation}
	j_2 = \frac{\eta f ( f -1)}{\eta f{} ^2-1},\quad \beta_1 = \frac{1-\eta f{} }{\eta f{} ^2-1},\quad j_{\pm 1}\to +\infty.
\end{equation}

\subsection{Effect of the thermodynamic constraint}
\label{SuppSec:OriginalHopfieldThermodynamicConstraint}
Reaching the minimum cost derived above requires vanishing $j_{-2}$ and $\beta_{-1}$. Namely, these two reactions need to be irreversible. However, complete irreversibility is impossible due to the thermodynamic constraint:
\begin{equation}
	\gamma = e^{\beta \Delta \mu_\text{futile}} = \frac{k_1k_2k_3}{k_{-1}k_{-2}k_{-3}} = \frac{j_1j_2\beta_1}{j_{-1}j_{-2}\beta_{-1}},
\end{equation}
where $\Delta \mu_\text{futile}$ is the chemical potential difference for the futile cycle.
For finite $\gamma$, the fluxes $j_{-2}$ and $\beta_{-1}$ are positive, which would cause the minimum cost to increase (i.e. introduce a positive correction term).
Since the bound becomes exact at infinite $\gamma$, we shall calculate the positive correction term to the first order in the large $\gamma$ limit. 
This is also motivated by the fact that $\gamma$ is usually sufficiently large in real biological proofreading networks due to the hydrolysis of energy-rich molecules coupled to the futile cycle.

In the optimal network derived before, the fast equilibrium in the $j_{\pm 1}$ step leads to $j_1/j_{-1}\to 1$. Thus, the thermodynamic constraint reduces to $j_{-2}\beta_{-1} = \gamma^{-1} j_{2}\beta_1$. The energy cost is
\begin{align}
	C & =  \frac{(\eta_0-\eta)(1+\eta f)}{(1+\eta)(\eta f-\eta_0)} \qty(1+j_{-2} +\frac{(1-\eta f)(1+\eta_0)}{(1+\eta f)(\eta_0-\eta)} \beta_{-1})
	\geq \frac{(\eta_0-\eta)(1+\eta f)}{(1+\eta)(\eta f-\eta_0)} \qty(1+2\sqrt{\frac{(1-\eta f)(1+\eta_0)}{(1+\eta f)(\eta_0-\eta)}} \sqrt{j_{-2}\beta_{-1}})
\end{align}
To obtain the first order correction, we substitute with $\eta_0=f^{-1}$ and
\begin{equation}
	j_{-2}\beta_{-1} = \gamma^{-1} j_{2}\beta_1 \approx \gamma^{-1} \cdot \frac{\eta f ( f -1)}{\eta f{} ^2-1} \cdot  \frac{1-\eta f{} }{\eta f{} ^2-1} = \frac{\eta f(f-1)(1-\eta f)}{\qty(\eta f^2-1)^2} \gamma^{-1},
\end{equation}
where $j_2$ and $\beta_1$ are evaluated at the infinite $\gamma$ limit. The cost reads
\begin{align}
	C_\text{min} = \frac{1-\eta^2f^2}{(1+\eta)(\eta f^2-1)} \qty(1 + 2 \sqrt{\frac{\eta f(1-\eta f)(f^2-1)}{(1+\eta f)(\eta f^2-1)^2}} \gamma^{-1/2} + O(\gamma^{-1}))
\end{align}
Therefore, the thermodynamic constraint introduces a correction term of order $O(\gamma^{-1/2})$,  which is negligible in realistic cases where $\gamma \sim e^{20}$.

\section{Flux-based formalism of the $n$-stage dissociation-based-discrimination scheme}
\label{SuppSec:n-stage Hopfield derivation}
In this section, we establish the flux-based formalism and derive the error-cost bound for the $n$-stage dissociation-based-discrimination (DBD) scheme.
The reaction scheme is illustrated in Fig.~1B of the main text with discrimination factors given in Eq.~7 and related text. The notation for the flux-based formalism is given in Fig.~2B of the main text.

\subsection{Deriving the normalized steady-state fluxes for the wrong half of the network}
First, we recall the definition of intermediate error rates $\eta_m$ as the forward flux ratio going from EW\textsubscript{m} (ER\textsubscript{m}) to EW\textsubscript{m+1} (ER\textsubscript{m+1}):
\begin{align}
	\eta_0 = \frac{f_2 k_2 P_{\text{EW}_0}}{k_2 P_{\text{ER}_0}} =  \frac{j_2'}{j_2},\quad
	\eta_m = \frac{f_{2m+2} k_{2m+2} P_{\text{EW}_m}}{k_{2m+2} P_{\text{ER}_{m}}}
	= \frac{\alpha_m'}{\alpha_m}\ (m=1,2,\dots,n).
\end{align}
Since the rate discrimination only appears in dissociation steps, we have $f_2=f_4=\cdots =f_{2m+2} = 1$.
Following the derivation in the original Hopfield scheme, the (normalized) fluxes in the first two steps (i.e. E$\leftrightarrow$EW\textsubscript{0}$\leftrightarrow$EW\textsubscript{1}) are
\begin{align}
	j_1' = j_1,& \quad j_{2}' = \eta_0 j_2,\\
	j_{-1}' = f \eta_0 j_{-1},& \quad j_{-2}' = \eta_1 j_{-2}.
\end{align}
Next, we consider the fluxes associated with the $m$-th intermediate state ER\textsubscript{m}/EW\textsubscript{m}, which are given by
\begin{align}
	\alpha_m' &= \eta_m \alpha_m, \\ 
	\alpha_{-(m-1)}' &= \frac{k_{-2m}P_\mathrm{EW_{m}}}{k_{2m+2}P_\mathrm{EW_{m}}} \alpha_m' =  \frac{\alpha_{-(m-1)}}{\alpha_m} \alpha_m'  = \eta_m \alpha_{-(m-1)}, \\
	\beta_m' & = \frac{ f{}  k_{2m+1}P_\mathrm{EW_{m}}}{k_{2m+2}P_\mathrm{EW_{m}}} \alpha_m' =  f{}  \frac{\beta_m}{\alpha_m} \alpha_m' =  f{} \eta_m \beta_m, \\
	 \quad 
	\beta_{-m}' &= \beta_{-m}.
\end{align}
For the product forming steps, we have $\alpha_n = 1$ and $\alpha_n' = \eta_n=\eta$. 
Thus, we have derived the (normalized) fluxes presented in Fig.~2B for the $n$-stage DBD scheme. 

\subsection{Deriving the error-cost bound}
The cost in the $n$-stage DBD scheme is given by
\begin{equation}
	C_n = \frac{1}{1+\eta}\sum_{m=1}^{n} \qty(\beta_m + \beta_m' - \beta_{-m} - \beta_{-m}')
	= \frac{1}{1+\eta}\sum_{m=1}^{n} \qty[\qty(1+\eta_m f{} )\beta_m - 2\beta_{-m}],
\end{equation}
where $\{ \beta \}$ are the (normalized) stationary fluxes. 
Due to the stationary conditions, summing up the net proofreading fluxes is equivalent to calculating the difference of the total fluxes coming out of the E (free enzyme) state and the total fluxes that lead to products:
\begin{equation}
	C_n = \frac{\qty(1+\eta_0) j_2 - \qty(1+\eta_1)j_{-2}}{1+\eta} - 1. 
\end{equation}
To derive the lower bound of $\qty[\qty(1+\eta_0) j_2 - \qty(1+\eta_1)j_{-2}]$, we first prove the following recursive relation:
\begin{equation}
	\alpha_{m} > \frac{\eta_{m+1}( f{} -1)}{\eta_{m+1} f{} -\eta_{m}}\qty[ \pi_{m+2}+ \alpha_{-m}],\quad (m = 1,2,\dots,n-1)
	\label{Eq: am ansatz}
\end{equation}
where $\pi_{m} = \prod_{k=m}^{n} \frac{\eta_{k}( f{} -1)}{\eta_{k} f{} -\eta_{k-1}}$ and $\pi_{n+1}=1$.
The equality condition for Eq.~\ref{Eq: am ansatz} is $\alpha_{-k}=0$ for $k\geq m+1$ and $\beta_{-k}=0$ for $k\geq m$. We will also prove the following relation for the error rates:
\begin{equation}
	\eta_{m+1} f{}  > \eta_m,\quad (m = 1,2,\dots,n-1)
	\label{Eq: etam ansatz}
\end{equation}

The relations Eq.~\ref{Eq: am ansatz} and Eq.~\ref{Eq: etam ansatz} are derived inductively via the following steps:
\begin{itemize}
	\item {\bf Step 1.} For $m=n-1$, the stationary conditions for states ER\textsubscript{n} and EW\textsubscript{n} read
	\begin{align}
		\alpha_{n-1} - \alpha_{-(n-1)} & = 1  + \beta_n - \beta_{-n}, \\
		\eta_{n-1}\alpha_{n-1} - \eta_{n}\alpha_{-(n-1)} & = \eta_{n} + \eta_n f{} \beta_n - \beta_{-n}
	\end{align}
	where $\eta_n=\eta$ is the final error. 
	Elimination of $\beta_n$ yields 
	\begin{equation}
		\qty(\eta_n  f{}  - \eta_{n-1}) \alpha_{n-1} = \eta_n( f{} -1)\qty(1+\alpha_{-(n-1)}) + \qty(1-\eta_n f{} ) \beta_{-n}.
	\end{equation}
	The coefficient $\qty(1-\eta_n f{} )$ is positive since we are considering error $\eta<\eta_\mathrm{eq} =  f{} ^{-1}$. Therefore, RHS is positive. On the other hand, $\alpha_{n-1}$ is positive. For LHS to also be positive, we must have
	\begin{equation}
		\eta_{n-1} < \eta_n  f,
	\end{equation}
	which recovers Eq.~\ref{Eq: etam ansatz} for $m=n-1$.
	Since $\beta_{-n}>0$, we have
	\begin{equation}
			\qty(\eta_n  f{}  - \eta_{n-1}) \alpha_{n-1} > \eta_n( f{} -1)\qty(1+\alpha_{-(n-1)}) 
		\Rightarrow 
	\alpha_{n-1} > \frac{\eta_n( f{} -1)}{\eta_n f{} -\eta_{n-1}} (1+ \alpha_{-(n-1)}), 
	\end{equation}
	which recovers Eq.~\ref{Eq: am ansatz} for $m=n-1$ (since $\pi_{n+1}=1$). The equality condition is $\beta_{-n}=0$. 

	\item {\bf Step 2.} For any $m=1,2,\dots, n-2$,
	we prove Eq.~\ref{Eq: am ansatz} and Eq.~\ref{Eq: etam ansatz} given the condition that they 
	both hold for $l= m+1$, i.e.
	\begin{equation}
		\eta_{m+2} f{} >\eta_{m+1},\quad \alpha_{m+1} \geq \frac{\eta_{m+2}( f{} -1)}{\eta_{m+2} f{} -\eta_{m+1}}\qty[ \pi_{m+3}+ \alpha_{-(m+1)}],\quad \text{where\ } \pi_{m+3} = \prod_{k=m+3}^{n} \frac{\eta_{k}( f{} -1)}{\eta_{k} f{} -\eta_{k-1}}.
		\label{Eq: am induction asumption}
	\end{equation}
	Consider the stationary conditions for states ER\textsubscript{m+1} and EW\textsubscript{m+1}
	\begin{align}
		\alpha_{m} - \alpha_{-m} & = \alpha_{m+1} - \alpha_{-(m+1)} + \beta_{m+1}- \beta_{-(m+1)}, \\
		\eta_{m}\alpha_{m} - \eta_{m+1}\alpha_{-m} & = \eta_{m+1}\alpha_{m+1} -  \eta_{m+2}\alpha_{-(m+1)}+ \eta_{m+1} f{} \beta_{m+1} - \beta_{-(m+1)}.
	\end{align}
	Eliminating $\beta_{m+1}$, we have
	\begin{equation}
		\qty(\eta_{m+1} f{} -\eta_m) \alpha_m = \eta_{m+1}\qty( f{}  -1)\qty(\alpha_{-m}+\alpha_{m+1}) + \qty(\eta_{m+2}- \eta_{m+1} f{} )\alpha_{-(m+1)} + (1-\eta_{m+1} f{} )\beta_{-(m+1)}.
	\end{equation}
	Since $\beta_{-(m+1)}>0$ and $1-\eta_{m+1} f{} >0$, we obtain the lower bound:
	\begin{equation}
	\qty(\eta_{m+1} f{} -\eta_m) \alpha_m > \eta_{m+1}\qty( f{}  -1)\qty(\alpha_{-m}+\alpha_{m+1}) + \qty(\eta_{m+2}- \eta_{m+1} f{} )\alpha_{-(m+1)}.
	\end{equation}
	Plugging in the lower bound for $\alpha_{m+1}$ given in Eq.~\ref{Eq: am induction asumption}:
	\begin{equation}
	\begin{aligned}
		\qty(\eta_{m+1} f{} -\eta_m) \alpha_m &> \eta_{m+1}\qty( f{}  -1)\qty(\alpha_{-m}+\frac{\eta_{m+2}( f{} -1)}{\eta_{m+2} f{} -\eta_{m+1}}\qty[ \pi_{m+3}+ \alpha_{-(m+1)}]) + \qty(\eta_{m+2}- \eta_{m+1} f{} )\alpha_{-(m+1)}  \\
		& = \eta_{m+1}\qty( f{}  -1)\qty(\alpha_{-m}+ \pi_{m+2} + \frac{\eta_{m+2}( f{} -1)}{\eta_{m+2} f{} -\eta_{m+1}}\alpha_{-(m+1)}) + \qty(\eta_{m+2}- \eta_{m+1} f{} )\alpha_{-(m+1)} \\
		& = \eta_{m+1}\qty( f{}  -1)\qty(\alpha_{-m}+ \pi_{m+2}) + \frac{\qty(\eta_{m+1}-\eta_{m+2})^2 f{} }{\eta_{m+2} f{} -\eta_{m+1}}\alpha_{-(m+1)} \\
		& > \eta_{m+1}\qty( f{}  -1)\qty(\alpha_{-m}+ \pi_{m+2}). 
	\end{aligned}
	\end{equation}
	Since RHS is positive, LHS must also be positive. Thus we have
	\begin{equation}
		\eta_{m+1} f{} >\eta_m.
	\end{equation}
	We can divide both sides by $\qty(\eta_{m+1} f{} -\eta_m)$ which has been shown to be positive. This leads to 
	\begin{equation}
		\alpha_m>  \frac{\eta_{m+1}\qty( f{}  -1)}{\eta_{m+1} f{} -\eta_m}\qty(\alpha_{-m} + \pi_{m+2}). 
	\end{equation}
\end{itemize}
As a result of the mathematical induction, Eq.~\ref{Eq: am ansatz} and Eq.~\ref{Eq: etam ansatz} holds for $m=1,2,\cdots,n-1$. Specifically, the relation for $m=1$ is 
\begin{equation}
	\eta_1 <  f{} ^{n-1} \eta_n,\quad
	\alpha_{1} > \frac{\eta_{2}( f{} -1)}{\eta_{2} f{} -\eta_{1}}\qty[ \pi_{3}+ \alpha_{-1}].
\end{equation}
We repeat the same derivation for states EW\textsubscript{0} and ER\textsubscript{0}. The only difference from repeating the above derivation for $m=0$ is the notation: $\alpha_{\pm 0}$ is now replaced by $j_{\pm 2}$. This gives us
\begin{equation}
	\eta_0 < f \eta_1 < f^n \eta,\quad j_2 > \frac{\eta_1 (f-1)}{\eta_1 f-\eta_0}\qty[\pi_2 + j_{-2}].
\end{equation}
The total cost is
\begin{equation}
	\begin{aligned}
		C_n &= \frac{\qty(1+\eta_0) j_2 - \qty(1+\eta_1)j_{-2}}{1+\eta} - 1 \\
		& > \frac{1+\eta_0}{1+\eta} \pi_1 + \frac{1}{1+\eta}\qty(\frac{\eta_1(f-1)(1+\eta_0)}{\eta_1 f-\eta_0} - (1+\eta_1))j_{-2} -1\\
		& =  \frac{1+\eta_0}{1+\eta} \pi_1 + \frac{\qty(\eta_0-\eta_1)\qty(1+\eta_1 f)}{\qty(1+\eta)\qty(\eta_1 f-\eta_0)}j_{-2} -1. \\
	\end{aligned}
\end{equation}
The coefficient $\frac{\qty(\eta_0-\eta_1)\qty(1+\eta_1 f)}{\qty(1+\eta)\qty(\eta_1 f-\eta_0)}$ is positive since $f^{-1} \eta_0 <\eta_1 <\eta_0$. We also recall that error before proofreading $\eta_0>\eta_\mathrm{eq} = f^{-1}$, with the lower bound reached in the limit of fast equilibrium. 
Therefore, the minimum cost for the $n$-stage DBD scheme for given intermediate error rates $\{\eta_m\}$ is 
\begin{equation}
	C_n > \bar{C}_{n} = \frac{1+\eta_\mathrm{eq}}{1+\eta} \pi_1 - 1 
	= \frac{\qty(1+f^{-1})(f-1)^n}{1+\eta} \prod_{m=1}^n \frac{\eta_m}{\eta_m f- \eta_{m-1}}-1.
	\label{Eq: n stage DBD cost of etas}
\end{equation}
The minimum cost $\bar{C}_{n}$ is reached in the limit
\begin{equation}
	\alpha_{-m}\to 0,\ \beta_{-m}\to 0,  (m=1,2,\dots n);\quad j_{-2} \to 0;\quad \eta_0\to  \eta_\mathrm{eq} = f{} ^{-1}.
\end{equation}
The last condition implies $j_1/j_{-1}\to 1$ and $j_{\pm 1} \to +\infty$. 
These were results reported in Eq.~12 of the main text.

Next, the minimum cost $\bar{C}_n$ can be further optimized with respective to the intermediate error rates $\{ \eta_m \} $. 
In the main text, a symmetry argument is used to illustrate that $\{\eta_m\}$ must form a geometric series for the dissipation to be optimized. Here, we provide the mathematical proof that it is indeed the unique minimum of the energy cost. From Eq.~\ref{Eq: n stage DBD cost of etas}, we define
\begin{equation}
	\tilde{C}_n = \ln\qty[\frac{1+\eta}{\qty(1+f^{-1})(f-1)^n}\qty(\bar{C}_n+1)] = \ln\qty[\prod_{m=1}^n \frac{\eta_m}{\eta_m f- \eta_{m-1}}] = \sum_{m=1}^n \qty( \ln \eta_m - \ln\qty(\eta_m f- \eta_{m-1})).
\end{equation}
For any fixed error rate $\eta$, $\tilde{C}_n$ is apparently a monotonically increasing function of $\bar{C}_n$. Hence, finding the minimum energy cost is equivalent to minimizing $\tilde{C}_n$ with respect to variables $\eta_m$ ($m=1,2,\dots, n-1$), which is done by simply taking the derivative:
\begin{equation}
	\pdv{\tilde{C}_n}{\eta_m} = \frac{1}{\eta_m} - \frac{f}{\eta_m f - \eta_{m-1}} + \frac{1}{\eta_{m+1}f - \eta_m} = \frac{f\qty(\eta_m^2-\eta_{m+1}\eta_{m-1})}{\eta_m \qty(\eta_{m+1}f-\eta_m)\qty(\eta_m f-\eta_{m-1})}.
\end{equation}
Setting the first derivative to zero, we get $\eta_m^2=\eta_{m+1}\eta_{m-1}$, i.e. the intermediate error rates indeed form a geometric series. With the first term $\eta_0 = f^{-1}$ and the last term $\eta_n = \eta$, all the other error rates can be determined as
\begin{equation}
	\eta_m = f^{-1} \qty(\eta f)^{m/n}.
\end{equation}
It can be verified that the optimal error rates satisfy $\eta_m \in \qty(f^{-1} \eta_{m-1}, \eta_{m-1})$, which is consistent with Eq.~\ref{Eq: etam ansatz}. 
To verify that this solution indeed correspond to a minimum of the cost, we calculate the second derivative:
\begin{equation}
\begin{aligned}
	\Eval{\pdv[2]{\tilde{C}_n}{\eta_m}}{\eta_m = f^{-1} \qty(\eta f)^{m/n}}{} &= 
	\Eval{\qty(- \frac{1}{\eta_m^2} + \frac{f^2}{\qty(\eta_m f - \eta_{m-1})^2} + \frac{1}{\qty(\eta_{m+1}f - \eta_m)^2})}{\eta_m = f^{-1} \qty(\eta f)^{m/n}}{} \\
	& = \Eval{\eta_m^{-2}\cdot \qty(-1 + \frac{1}{\qty(1-\frac{\eta_{m-1}}{\eta_m f})^2} + \frac{1}{\qty(\frac{\eta_{m+1}f}{\eta_m}-1)^2})
	}
	{\eta_m = f^{-1} \qty(\eta f)^{m/n}}{} \\
	& = \qty[f^{-1} \qty(\eta f)^{m/n}]^{-2} \cdot \qty(-1 + \frac{1}{\qty(1-f^{-1}\qty(\eta f)^{-1/n})^2} + \frac{1}{\qty(f \qty(\eta f)^{1/n}-1)^2}) \\
	& = f^2 \qty(\eta f)^{-2m/n} \frac{2f \qty(\eta f)^{1/n}}{\qty(f \qty(\eta f)^{1/n}-1)^2} >0.
\end{aligned}
\end{equation}
Thus, the solution found above is a  minimum of the energy cost. Moreover, it is a global minimum. The minimum cost is given by 
\begin{equation}
	\begin{aligned}
		C_{n, \text{min}} = \Eval{\bar{C}_n} {\eta_m = f^{-1} \qty(\eta f)^{m/n}}{} & = \frac{\qty(1+f^{-1})(f-1)^n}{1+\eta} \prod_{m=1}^n \frac{ f^{-1} (\eta f)^{m/n}}{(\eta f)^{m/n}- f^{-1} (\eta f)^{(m-1)/n}}-1\\
		& = \frac{(1+f)(f-1)^n \eta}{(1+\eta)\qty(f(\eta f)^{1/n}-1)^n}-1.
		\label{Eq: n stage DBD cost final}
	\end{aligned}
\end{equation}
This is the minimum energy cost reported in Eq.~13 in main text.

\subsection{Analysing the minimum cost}
The minimum cost (Eq.~\ref{Eq: n stage DBD cost final}) vanishes in the limit $\eta \to \eta_\mathrm{eq} = f^{-1}$ but diverges in the limit $\eta \to \eta_\mathrm{min} = f^{-(n+1)}$. Here, we analyse how the minimum cost depends on the discrimination factor $f$.
\begin{equation}
	\begin{aligned}
		C_{n, \mathrm{min}} &= \frac{(f-1)^n(f+1)\eta}{\qty[f^{1+\frac{1}{n}}\eta^{\frac{1}{n}}-1]^{n}\qty(1+\eta)} -1 = \frac{\qty(1-f^{-1})^n\qty(1+f^{-1})}{\qty[1-\qty(f^{n+1}\eta)^{-1/n}]^{n}\qty(1+\eta)} -1 \\
		& = \frac{\qty(1-f^{-1})^n\qty(1+f^{-1})-\qty[1-\qty(f^{n+1}\eta)^{-1/n}]^{n}\qty(1+\eta)}{\qty[1-\qty(f^{n+1}\eta)^{-1/n}]^{n}\qty(1+\eta)} \\
		& =\frac{\qty[1-(n-1)f^{-1}+O\qty(f^{-2})]-\qty[1+\eta-n\qty(f^{n+1}\eta)^{-1/n} + O\qty[\qty(\eta^{1/n}f^{(n+1)/n})^{-2}]]}{\qty[1-\qty(f^{n+1}\eta)^{-1/n}]^{n}\qty(1+\eta)} \\
		& = \frac{n\qty(f^{n+1}\eta)^{-1/n}-(n-1)f^{-1} + h.o.t.}{\qty[1-\qty(f^{n+1}\eta)^{-1/n}]^{n}\qty(1+\eta)}.
	\end{aligned}
\end{equation}
In the intermediate error range $f^{-(n+1)}\ll \eta \ll   f^{-1}$, the numerator is dominated by the first term which is proportional to $f^{-(n+1)/n}$, and the denominator is approximately $1$. Therefore, the minimum cost decreases with $f$ following a power law:
\begin{equation}
	C_\mathrm{min} \propto f^{-\frac{n+1}{n}}, \quad 
	\qty( \eta^{-1} \ll f \ll \eta^{-1/(n+1)}).
\end{equation}
The power-law exponent $\frac{n+1}{n}$ is verified in Fig.~3A (main text) for $n=3$. More importantly, the power law relation between the minimum cost and the discrimination factor indicates that increasing $f$ leads to a non-diminishing benefit in cost reduction (see main text for detailed discussion). 

\subsection{Partition between proofreading and catalytic fluxes}
The derivation of the minimum energy cost in the above section suggests that in the energetically optimal system, the normalized fluxes in the right half of the network are given by
\begin{equation}
\alpha_m = \pi_{m+1},\quad \beta_m = \alpha_{m-1}-\alpha_m = \frac{\eta_{m-1}-\eta_m}{\eta_m f{} -\eta_{m-1}} \pi_{m+1},
\end{equation}
where $\pi_{m} = \prod_{k=m}^{n} \frac{\eta_{k}( f{} -1)}{\eta_{k} f{} -\eta_{k-1}}$. 
On the other hand, these fluxes are related to the steady-state probability $P_{\mathrm{ER_m}}$, reaction rates $k_{2m+1}$, $k_{2m+2}$, and the correct product formation flux $J_R$ by
\begin{equation}
	\alpha_m = \frac{k_{2m+2}P_{\mathrm{ER_m}}}{J_R}, \quad \beta_m =\frac{k_{2m+1}P_{\mathrm{ER_m}}}{J_R}.
\end{equation}
The ratio of these two fluxes is
\begin{equation}
	\frac{\beta_m}{\alpha_m} = \frac{k_{2m+1}}{k_{2m+2}} =  \frac{\eta_{m-1}-\eta_m}{\eta_m f{} -\eta_{m-1}} = \frac{1-(\eta f{} )^{1/n}}{ f{} (\eta f{} )^{1/n}-1},
\end{equation}
which has taken into account the optimal error rates $\eta_m =  f{} ^{-1}\qty(\eta f{} )^{m/n}$.  This is the partition ratio given in Eq.~14 in the main text. 
The reaction rates $k_{2m+1}$, $k_{2m+2}$ can be expressed in terms of the energy levels of the discrete states and the energy barriers:
\begin{equation}
	k_{2m+1} = k_{2m+1}^0 \exp\qty(\epsilon_m - \epsilon_{m,p}^\dagger),\quad k_{2m+2} = k_{2m+2}^0 \exp\qty(\epsilon_m - \epsilon_{m,m+1}^\dagger).
\end{equation}
$ k_{2m+1}^0 $ and $k_{2m+2}^0$ are prefactors independent of the energy levels. $\epsilon_m$ is the energy level of ER\textsubscript{m}. $\epsilon_{m,p}^\dagger$ and $\epsilon_{m,m+1}^\dagger$ is the energy level of the transition state (energy barrier) between ER\textsubscript{m} and ER\textsubscript{m+1}.
Therefore, the ratio $\beta_m/\alpha_m$ is actually only related to the difference between the energy level of the two transition states:
\begin{equation}
	\frac{\beta_m}{\alpha_m} = \frac{k_{2m+1}}{k_{2m+2}} = \frac{1-(\eta f{} )^{1/n}}{ f{} (\eta f{} )^{1/n}-1} \propto \exp\qty(\epsilon_{m,m+1}^\dagger-\epsilon_{m,p}^\dagger).
\end{equation}
As discussed in the main text, this is a manifestation of how the error-cost relation is kinetically controlled.

\subsection{Effect of the thermodynamic constraints}
Similar to the case of the original Hopfield scheme, the thermodynamic constraints prevent any reaction to be completely irreversible and introduces a correction term to the minimum energy cost (Eq.~\ref{Eq: n stage DBD cost final}) in the $n$-stage DBD scheme. Here we calculate the leading order contribution of this correction term.

In the derivation of the error-cost bound, many of the fluxes were set to zero since they only increase the overall cost. These terms must be recovered as we study the effect of the thermodynamic constraints. Fortunately, due to the linearity of the stationary conditions, they contribute to the cost through a linear relation:
\begin{equation}
	C
	 = C\qty(j_{-2}, \qty{\alpha}, \qty{\beta})
	 = C_0 + a_0 j_{-2} + \sum_{m=1}^{n-1} a_m \alpha_{-m} + \sum_{m=1}^{n} b_m \beta_{-m},
\end{equation}
where $C_0$ is the minimum cost in Eq.~\ref{Eq: n stage DBD cost final}. 
The coefficients $a_i$ ($i=0,1,2,\dots,n-1$) and $b_i$ ($i=1,2,3,\dots,n$) are positive functions of $f{}$ and $\eta_m$ ($m=1,2,\dots, n$).
Following the inductive method used to derive the bound, we find the following coefficients:
\begin{align}
	a_0 &= \frac{\qty(\eta_0-\eta_1)(1+\eta_1 f{})}{(1+\eta)(\eta_1f-\eta_0)}\\
	a_i & = \frac{1+\eta_0}{1+\eta} \frac{\pi_1}{\pi_{i+2}} \frac{f{} \qty(\eta_i-\eta_{i+1})^2}{\eta_i\eta_{i+1}(f{}-1)^2},\quad i=1,2,\dots, n-1\\
	b_i & = \frac{1+\eta_0}{1+\eta} \frac{\pi_1}{\pi_{i+1}} \frac{1-f{}\eta_i}{\eta_i(f{}-1)},\quad i=1,2,\dots, n,
\end{align}
where $\eta_0 = f{}^{-1}$, $\pi_{m} = \prod_{k=m}^{n} \frac{\eta_{k}( f{} -1)}{\eta_{k} f{} -\eta_{k-1}}$, and $\pi_{n+1}=1$. 

The thermodynamic constraints are:
\begin{equation}
\begin{aligned}
	\gamma  & = \frac{j_2}{j_{-2}} \cdot \frac{\beta_1}{\beta_{-1}}
	= \frac{j_2}{j_{-2}} \cdot \frac{\alpha_1\beta_2}{\alpha_{-1}\beta_{-2}}
	= \cdots 
	= \frac{j_2}{j_{-2}} \cdot \prod_{k=1}^{m-1} \frac{\alpha_k}{\alpha_{-k}} \frac{\beta_m}{\beta_{-m}}, \quad (m=1,2,\dots n)
\end{aligned}
\end{equation}
For any futile cycle, the thermodynamic correction to the energy cost is of the order $\gamma^{-1/L}$, where $L$ is the number of reactions needed to be driven strongly forward in this cycle. 
This is because the cost always depends on the reverse reaction fluxes, which should vanish without the thermodynamic constraint, in a linear fashion. 
Thus, in the presence of the thermodynamic constraint, the cost is minimized when those reverse fluxes are of the same order of magnitude, i.e. of order $\gamma^{-1/L}$. 
Hence, the first order contribution in $\gamma$ comes from the largest futile cycle, which has $(n+1)$ reaction steps that need to be driven forward. 
The thermodynamic constraint for this cycle can be reorganized to:
\begin{equation}
	j_{-2}\beta_{-n} \prod_{m=1}^{n-1} \alpha_{-m} = \frac{j_2\beta_n}{\gamma}   \prod_{m=1}^{n-1} \alpha_{m}.
\end{equation}
Therefore, the first correction to the cost is calculated as follows:
\begin{equation}
	\begin{aligned}
		C &= C_0 + \qty(a_0 j_{-2} + \sum_{m=1}^{n-1} a_m \alpha_{-m} + b_n \beta_{-n}) + \sum_{m=1}^{n-1} b_m \beta_{-m}\\
		&\geq C_0 + (n+1) \qty(a_0j_{-2}\cdot \prod_{m=1}^{n-1} a_m \alpha_{-m} \cdot b_n \beta_{-n})^{1/(n+1)}+ \sum_{m=1}^{n-1} b_m \beta_{-m} \\
		& = C_0 + (n+1) \qty(a_0j_{2}\cdot b_n \beta_{n}   \prod_{m=1}^{n-1} a_m\alpha_{m} )^{1/(n+1)} \gamma^{-1/(n+1)} + O(\gamma^{-2/(n+1)})\\
		& = C_0 + C_1\gamma^{-1/(n+1)} + O(\gamma^{-1/n}).
	\end{aligned}
\end{equation}
The $O(\gamma^{-1/n})$ term is due to the second largest futile cycle which has length $n$.
The coefficient $C_1$ is given by
\begin{equation}
	\begin{aligned}
	C_1 &=  \qty(n+1) \qty(a_0j_{2}\cdot \prod_{m=1}^{n-1} a_m \alpha_{m} \cdot b_n \beta_{n})^{1/(n+1)},
	\end{aligned}
\end{equation}
where the coefficients ($a_0$, $a_m$, $b_n$) and fluxes ($j_{2}$, $\alpha_m$, $\beta_n$) are evaluated in the optimal scheme, i.e. as if the thermodynamic constraints are not present. 
Thus, the correction is of the order $\gamma^{-1/(n+1)}$.
Although the correction term becomes increasingly significant as $n$ is increased, the number of proofreading pathways in real biological systems is usually limited, so the correction term remains small.
Moreover, note that the correction term due to thermodynamic constraints is always positive, so the original error-cost bound could never be violated. 

\section{A simple kinetic model for $n$-stage proofreading}
\label{SuppSec:MinimalModel}
In this section, we study the $n$-stage proofreading scheme shown in Fig.~4A by directly solving the Chemical Master Equation (CME).
We introduce $P_m(t)$ to denote the probability for state ER\textsubscript{m} at time $t$ and $P_{-m}(t)$ to denote the probability for state EW\textsubscript{m}. 
The probability for the free enzyme state E is denoted by $P_0(t)$. The probabilities are normalized by the condition
\begin{equation}
	\sum_{m=-n}^n P_m(t) = 1, \quad \forall t \in \qty(-\infty, + \infty).
\end{equation}
The CME reads
\begin{align}
	\dv{P_0(t)}{t} &= (1+a) \kappa_n P_n(t) + (1+ f{}  a) \kappa_n P_{-n}(t) + f{}  a  \sum_{m=1}^{n-1} \kappa_m P_{-m}(t) + a \sum_{m=1}^{n-1} \kappa_m P_{m}(t)- \qty(1+f^{-1}) \kappa_0 P_0(t), \\
	\dv{P_m(t)}{t} &= \kappa_{m-1} P_{m-1}(t) - (1+a) \kappa_m P_m(t),\quad m=1,2,\dots n \\
	\dv{P_{-1}(t)}{t} &=  f^{-1} \kappa_0 P_{0}(t) - (1+ f{}  a) \kappa_1 P_{-1}(t), \\
	\dv{P_{-m}(t)}{t} &= \kappa_{m-1} P_{-(m-1)}(t) - (1+ f{}  a) \kappa_m P_{-m}(t),\quad m=2,3,\dots n .
\end{align}
We are interested in the steady-state solution, which satisfies $\dv{P_m}{t}=0$ ($m=-n,-(n-1),\dots,n-1,n$). 
The stationary condition for state $m$ ($m=1,2,3,\dots,n$) leads to:
\begin{equation}
	\dv{P_m(t)}{t} = \kappa_{m-1}P_{m-1}(t) - (1+a) \kappa_{m}P_m(t) =0  \Rightarrow P_m = \frac{1}{(1+a)^m} \frac{\kappa_0}{\kappa_m} P_0.
\end{equation}
Similarly, the stationary condition for state $(-m)$ leads to
\begin{equation}
	P_{-m} = \frac{1}{ f{} (1+ f{}  a)^m} \frac{\kappa_0}{\kappa_m} P_0.
\end{equation}
The error rate $\eta$ is given by
\begin{equation}
	\eta = \frac{J_W}{J_R} = \frac{P_{-n}}{P_n} = f^{-1} \qty(\frac{1+a}{1+ f{}  a})^n.
\end{equation}
Note that the error $\eta$ is always bound between $\eta_\mathrm{min} =  f{} ^{-n-1}$ (in the limit $a\to\infty$) and $\eta_\mathrm{eq} =  f{} ^{-1}$ (in the limit $a\to 0$). From this relation, we can solve for $a$ as a function of $\eta$:
\begin{equation}
	a = \frac{1-\qty(\eta f{} )^{1/n}}{ f{}  \qty(\eta f{} )^{1/n}-1} .
\end{equation}
On the other hand, the energy cost $C$ is given by
\begin{equation}
\begin{aligned}
	C &= \frac{1}{J_R+J_W}\qty(a \sum_{m=1}^n \kappa_m P_m +  f{}  a  \sum_{m=1}^n \kappa_m P_{-m}) \\
	& = \frac{a}{(1+\eta)\kappa_n P_n}\sum_{m=1}^n \kappa_m \qty(P_m +  f{} P_{-m}) \\
	& = \frac{a(1+a)^n}{(1+\eta)\kappa_0 P_0}\sum_{m=1}^n \kappa_0 P_0 \qty( \frac{1}{(1+a)^m} +  \frac{1}{(1+f{}a)^m}) \\
	& = \frac{a(1+a)^n}{1+\eta} \qty[ \frac{1-(1+a)^{-n}}{a}+\frac{1-(1+f{} a)^{-n}}{f{} a}]\\
	& = \frac{(1+a)^n\qty(1+f^{-1})}{1+\eta} - 1. 
\end{aligned}
\end{equation}
Substituting $a$ with $a = \frac{1-\qty(\eta f{} )^{1/n}}{ f{}  \qty(\eta f{} )^{1/n}-1} $, we obtain the full expression for the minimum cost
\begin{equation}
	C = \qty(1+f)\qty(\frac{f-1}{ f{} \qty(\eta f{} )^{1/n}-1})^n\frac{\eta}{1+\eta}-1 = \frac{(1+ f{})\qty( f{} -1)^n}{\qty( f\qty(\eta f{} )^{1/n}-1)^n}\frac{\eta}{1+\eta}-1
	\label{Eq:MinimumModelDissipation}
\end{equation}
which is exactly the dissipation bound for $n$-stage DBD scheme reported in the main text (Eq.~13). 

In this simplified model, both error and energy dissipation are modulated by the partition ratio $a$, which is equivalent to the flux-splitting ratio $\beta_m/\alpha_m$ in the flux-based formalism calculated above. 
When $a\to 0$, the system approaches the non-dissipative, equilibrium discrimination regime with $\eta\to\eta_\mathrm{eq} =  f{} ^{-1}$ and $C\to0$. 
When $a\to\infty$, the system approaches the limit to which error can be reduced by dissipative proofreading, namely $\eta\to\eta_\mathrm{min} = f{} ^{-n-1}$ and $C\to\infty$. 
Moreover, the system is optimized as long as the partition ratio $a$ is uniform for all proofreading pathways. The continuous tuning of $a\in(0,\infty)$ therefore represents a trade-off between error and dissipation, where error can be reduced by increasing $a$ at the cost of more dissipation.

\section{Michaelis-Menten scheme with dissipative resetting}
\label{SuppSec:ArbitraryDiscriminationFactors}

This section provides detailed derivation of the error-cost relation in the MM-with-proofreading scheme reported in Fig.~5 in the main text. 
More complex reaction networks can be considered as combination or generalization of this type of reaction network. 

\begin{figure}[bth]
	\centering
	\includegraphics[width=0.6\linewidth]{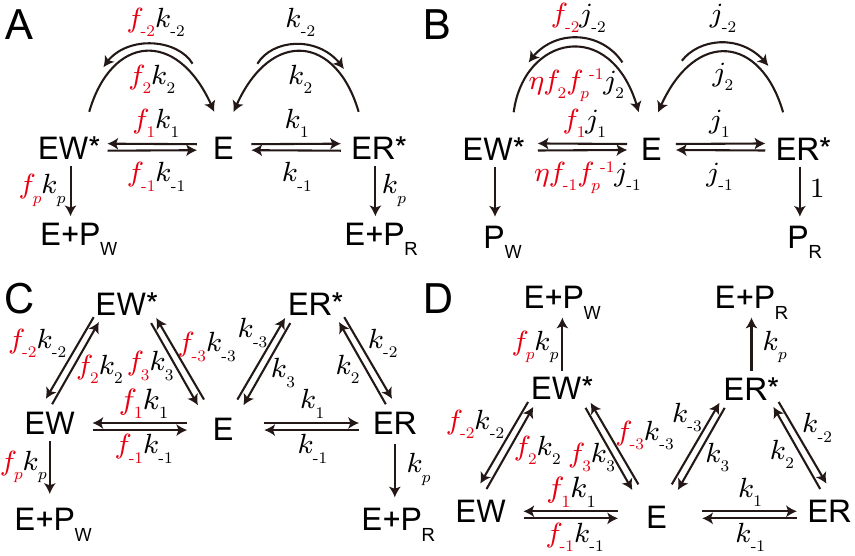}
	\caption{Reaction schemes used in the main text and SI. 
	(A) Michaelis-Menten scheme with dissipative resetting (reproduced from Fig.~5A in main text for comparison with the flux-based formalism).
	(B) Flux-based formalism for the MM-with-proofreading scheme.
	(C) Reaction scheme for T7 DNA polymerase, reproduced from ref.~\cite{Banerjee2017PNAS,Mallory2019}.
	(D) Reaction scheme for \textit{E. coli} ribosome, reproduced from ref.~\cite{Banerjee2017PNAS,Mallory2019}.
	}
	\label{fig:SchemesOnecycle}
\end{figure}

The reaction scheme is presented in Fig.~\ref{fig:SchemesOnecycle}A with notations introduced in the main text. Due to the kinetic control of both error and energy cost, we introduce a set of variables $\xi_i$ to quantify the difference between energy barriers:
\begin{equation}
	\xi_1 = f_1,\quad \xi_2 = \frac{f_1f_2}{f_{-1}} = f_{-2},\quad \xi_p = \frac{f_1f_p}{f_{-1}} = \frac{f_{-2}f_p}{f_2}.
\end{equation}
Their relation with energy barrier differences are reported in Eq.~17 in the main text.

In the absence of proofreading ($k_{\pm 2}=0$), the minimum error is determined by the maximum difference in energy barriers
\begin{equation}
	\eta_\mathrm{eq} = \min\qty(\xi_1,\xi_p) = f_1 \min\qty(1, \frac{f_p}{f_{-1}}).
\end{equation}
The minimum error is achieved by making the step with the largest barrier difference rate-limiting.
Namely, $k_p$ is rate-limiting if $f_p<f_{-1}$, and $k_1$ is rate-limiting if $f_p>f_{-1}$. 

We study the relation between error and energy cost in the parameter regime where the dissipative proofreading mechanism is relevant, i.e. it could achieve some error rate $\eta<\eta_\mathrm{eq}$ which is otherwise inaccessible. 
The condition for the proofreading mechanism to reduce error below $\eta_\mathrm{eq}$ is 
\begin{equation}
	f_2 > \max\qty(f_{-1},f_p).
	\label{Eq:MMproofreading f2 range}
\end{equation}
In the main text, this condition is justified with the heuristic argument that proofreading only improves the accuracy if it creates more bias in the dissociation of incorrect complexes compared to the bias in non-dissipative dissociation (unbinding) or product formation.
In the following, this condition is justified \textit{a posteriori} after the minimum error is derived.

In Fig.~\ref{fig:SchemesOnecycle}B, we present the flux-based formalism for the MM-with-proofreading scheme, where the noncognate fluxes (i.e. $j_{\pm1,\pm2}'$) has already been derived and labeled on the reactions. 
$\eta$ stands for the error. The fluxes are constrained by the stationary conditions for ER and EW:
\begin{equation}
	\begin{aligned}
		j_1 - j_{-1} = 1 + j_2 - j_{-2},\quad 
		f_1 j_1 - \eta\frac{f_{-1}}{f_p} j_{-1} = \eta + \eta\frac{f_2}{f_p} j_2 - f_{-2} j_{-2}. 
	\end{aligned}
\end{equation}
The energy cost $C$ is given by 
\begin{equation}
	C = \frac{1}{1+\eta} \qty(j_2-j_{-2} + \eta \frac{f_2}{f_p}j_2 - f_{-2} j_{-2}) 
	= \frac{1}{1+\eta} \qty[ (1+f_1)j_1 - \qty(1+ \eta \frac{f_{-1}}{{f_p}})j_{-1}]-1.
\end{equation}

Eliminating $j_2$ from the stationary conditions yields:
\begin{equation}
	\qty(\eta \frac{f_2}{f_p} - f_1) j_1 = 
	\eta\frac{f_2-f_{-1}}{f_p} j_{-1} + \eta \frac{f_2-f_p}{f_p} + f_2\qty(\frac{f_1}{f_{-1}}-\frac{1}{f_p}\eta)j_{-2},
\end{equation}
where $f_{-2}$ has been substituted by $\frac{f_1f_2}{f_{-1}}$ due to thermodynamic constraints. We note that the right hand side is positive due to conditions $f_2>\max\qty(f_{-1},f_p)$ and $\eta<\eta_\mathrm{eq} \leq \frac{f_1 f_p}{f_{-1}}$. Thus, the left hand side must also be positive, leading to the minimum error
\begin{equation}
	\eta > \eta_\mathrm{min} = \frac{f_1f_p}{f_2}. 
\end{equation}
The condition for the minimum error in the presence of proofreading to be smaller than the minimum error without proofreading is
\begin{equation}
	\eta_\mathrm{min} < \eta_\mathrm{eq} \Leftrightarrow
	\frac{f_1f_p}{f_2} < f_1 \mathrm{min}\qty(1, \frac{f_p}{f_{-1}}) \Leftrightarrow
	f_2 > \frac{f_p}{\mathrm{min}\qty(1, \frac{f_p}{f_{-1}}) } = \max(f_p, f_{-1}),
\end{equation}
which recovers the condition Eq.~\ref{Eq:MMproofreading f2 range}. 
This is the condition for the nonequilibrium proofreading mechanism to be relevant.
It can be verified that if $f_2$ is smaller than either $f_{-1}$ or $f_p$, the minimum error can always be achieved without proofreading. 

Finally, we consider the energy cost for $\eta\in\qty(\eta_\mathrm{min},\eta_\mathrm{eq})$:
\begin{equation}
	\begin{aligned}
		C &=\frac{1}{1+\eta} \qty[ (1+f_1)j_1 - \qty(1+ \eta \frac{f_{-1}}{{f_p}})j_{-1}]-1\\
		& = \frac{1}{1+\eta} \qty[ (1+f_1)\frac{\eta\frac{f_2-f_{-1}}{f_p} j_{-1} + \eta \frac{f_2-f_p}{f_p} + f_2\qty(\frac{f_1}{f_{-1}}-\frac{1}{f_p}\eta)j_{-2}}{\eta \frac{f_2}{f_p}-f_1} - \qty(1+ \eta \frac{f_{-1}}{{f_p}})j_{-1}]-1\\
		& =  C_0 + a_1 j_{-1} + a_2 j_{-2}.
	\end{aligned}
\end{equation}
The coefficients are given by
\begin{align}
	C_0 & = \frac{(f_1-\eta)\qty(1+\eta \frac{f_2}{f_p})}{(1+\eta)(\eta \frac{f_2}{f_p}-f_1)},\\
	a_1 & = \frac{\qty(f_1-\eta \frac{f_{-1}}{f_p})\qty(1+\eta \frac{f_2}{f_p})}{\qty(1+\eta)\qty(\eta \frac{f_2}{f_p}-f_1)},\\
	a_2 & = \frac{(1+f_1)f_2 \qty(\frac{f_1}{f_{-1}}- \frac{\eta}{f_p})}{(1+\eta)\qty(\eta \frac{f_2}{f_p}-f_1)}.\\
\end{align}
Since $a_{1,2}>0$, the energy cost is minimized when the reverse fluxes $j_{-1,-2}\to 0$. The minimum energy cost is given by
\begin{equation}
	C_\mathrm{min} = C_0 =\frac{(f_1-\eta)\qty(1+\eta \frac{f_2}{f_p})}{(1+\eta)(\eta \frac{f_2}{f_p}-f_1)},
	\label{Eq: MM-with-proofreading bound expression}
\end{equation}
which recovered Eq.~16 of the main text.

The effect of the thermodynamic constraint $\gamma = \frac{k_1k_2}{k_{-1}k_{-2}}$ can be analysed following the method used in the original Hopfield scheme (section~\ref{SuppSec:OriginalHopfieldThermodynamicConstraint}). The correction is of the order $\gamma^{-1/2}$ due to having two reactions driven irreversibly forward in the futile cycle.

\section{Parameters and additional simulation results for the real biological systems}
\label{SuppSec:RealSystems}

In this section, we provide additional details for the three biological examples analysed in the main text (Fig.~6). 

\subsection{T7 DNA polymerase}
The reaction network and parameters for the DNA replication network are obtained from previous works~\cite{Mallory2019,Banerjee2017PNAS}. 
For reference purposes, the reaction network has been reproduced in Fig.~\ref{fig:SchemesOnecycle}C.

\paragraph{Relation between various error rates.} 
We first verify that the native system operates in the regime where dissipative proofreading is necessary.
The error of the native system is $\eta_\mathrm{wt} =  7.39\times 10^{-8}$; the minimum error for discrimination without proofreading is $\eta_\mathrm{eq}=\min\qty(\xi_1,\xi_p) = \xi_1 = 8.00\times 10^{-6}$; the minimum error for the first step is $\eta_0= \xi_1 = f_1 =  8.00\times 10^{-6}$; the overall minimum error is $\eta_\mathrm{min} = \frac{f_p}{f_2} f_1 = 3.34\times10^{-11}$. Therefore, the relation between these error rates is
\begin{equation}
	\eta_0 = \eta_\mathrm{eq} >\eta_\mathrm{wt}>\eta_\mathrm{min}.
\end{equation}
The native system $\eta_\mathrm{wt}$ falls within the non-equilibrium discrimination regime.

\paragraph{Optimal and native proofreading systems}
The only difference between the DNA replication network and the MM-with-proofreading scheme is the addition of intermediate states EW\textsuperscript{*} and ER\textsuperscript{*}. The additional states will not change the error-cost bound since proofreading reaction is driven irreversibly forward in the optimal scheme, as indicated by the derivation in the last section. Hence, the error-cost bound is the same as that derived in the MM-with-proofreading scheme:
\begin{equation}
	C_\mathrm{min} = \frac{(f_1-\eta)\qty(1+\eta \frac{f_2}{f_p})}{(1+\eta)(\eta \frac{f_2}{f_p}-f_1)},
\end{equation}
which is the red line in Fig.~6A of the main text. This bound indeed encapsulates all the systems sampled. 

At the native error rate, the optimal partition ratio is given by:
\begin{equation}
	a_\mathrm{optimal} = \qty(j_2)_\mathrm{optimal} =  \frac{f_1-\eta_\mathrm{wt}}{\eta_\mathrm{wt}\frac{f_2}{f_p}-f_1} = 4.5 \times 10^{-4}. 
\end{equation}
The native partition ratio is
\begin{equation}
	a_\mathrm{wt} = \frac{k_2}{k_p} = 8.0 \times 10^{-4}.
\end{equation}

\subsection{\textit{E. coli} ribosome}
The reaction network and parameters for the protein replication network are also obtained from previous works~\cite{Mallory2019,Banerjee2017PNAS}. 
For reference purposes, the reaction network has been reproduced in Fig.~\ref{fig:SchemesOnecycle}D.

\paragraph{Relation between various error rates.} 
The error of the native system is $\eta_\mathrm{wt} =  8.65\times 10^{-4}$; 
the minimum error for discrimination without proofreading is $\eta_\mathrm{eq}=\min\qty(\xi_1,\xi_2,\xi_p) = \xi_p = 1.45\times 10^{-6}$; 
the minimum error for the first two steps is $\eta_0=\min(\xi_1,\xi_2) =\xi_2 =  3.45\times 10^{-4}$; 
the overall minimum error is $\eta_\mathrm{min} = \frac{f_p}{f_3} \eta_0 = 1.83\times10^{-7}$. Therefore, the relation between these error rates is
\begin{equation}
	\eta_\mathrm{wt}> \eta_0 >\eta_\mathrm{eq} >\eta_\mathrm{min}.
\end{equation}
The native error rate $\eta_\mathrm{wt}$ falls within the equilibrium discrimination regime, which can in principle be achieved without the proofreading step.
As discussed in the main text, achieving $\eta_\mathrm{eq}$ requires the product formation step $k_p$ to be much smaller than the preceding reactions $k_{\pm 1, \pm 2}$, which could not be realized due to speed requirements. 
Similarly, achieving $\eta_0$ without proofreading requires GTP hydrolysis ($k_2$) to be rate-limiting, which is also prevented by speed requirements.

\paragraph{Energy-cost bound in the translation network.}
The network has only one proofreading pathway, and the error-cost bound takes the same form as the bound in the MM-with-proofreading scheme (Eq.~\ref{Eq: MM-with-proofreading bound expression}) with $f_1$ replaced by $\eta_0$ (the minimum error in the first two steps) and $f_2$ replaced by $f_3$ (the discrimination factor for the proofreading step). Therefore, the error-cost bound in the ribosome network is
\begin{equation}
	C_\mathrm{min} = \frac{\qty(\eta_0-\eta)\qty(1+\eta \frac{f_3}{f_p})}{(1+\eta)\qty(\eta \frac{f_3}{f_p}-\eta_0)} 
	= \frac{\qty(\frac{f_1f_2}{f_{-1}}-\eta)\qty(1+\eta \frac{f_3}{f_p})}{(1+\eta)\qty(\eta \frac{f_3}{f_p}-\frac{f_1f_2}{f_{-1}})}.
	\label{Eq:ribosome bound}
\end{equation}
This bound correspond to the red line in Fig.~6B in main text.

\begin{figure}
	\centering
	\includegraphics[width=0.65\linewidth]{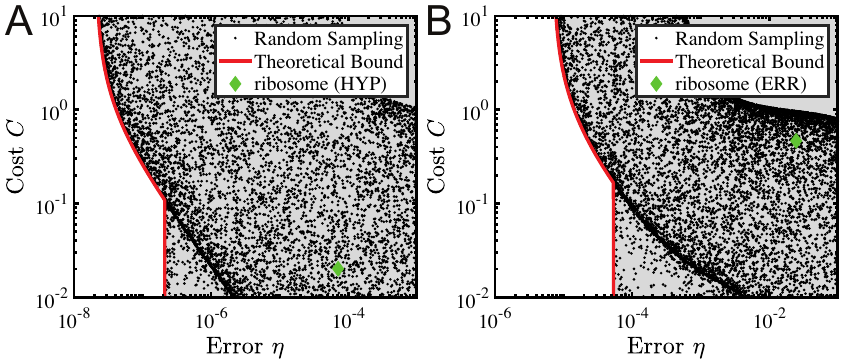}
	\caption{The error-dissipation relations in two mutants of the \textit{E. coli} ribosome.
		Left: mutant \textit{rpsL141}, which is hyperaccurate (HYP).
		Right: mutant \textit{rpsD12}, which is more error-prone than WT (ERR). 
	}
	\label{fig:RibosomeMutantResults}
\end{figure}

\paragraph{Results in mutants.} 
The simulation results for the ERR (error-prone) and HYP (hyperaccurate) mutants are qualitatively similar to the results in WT. 
The parameters for these two mutants are obtained from ref.~\cite{Banerjee2017PNAS}.
The numeric results are presented in Fig.~\ref{fig:RibosomeMutantResults}.

\subsection{\textit{E. coli} isoleucyl-tRNA synthetase (IleRS)}
\begin{figure}[tbh]
	\centering
	\includegraphics[width=0.8\linewidth]{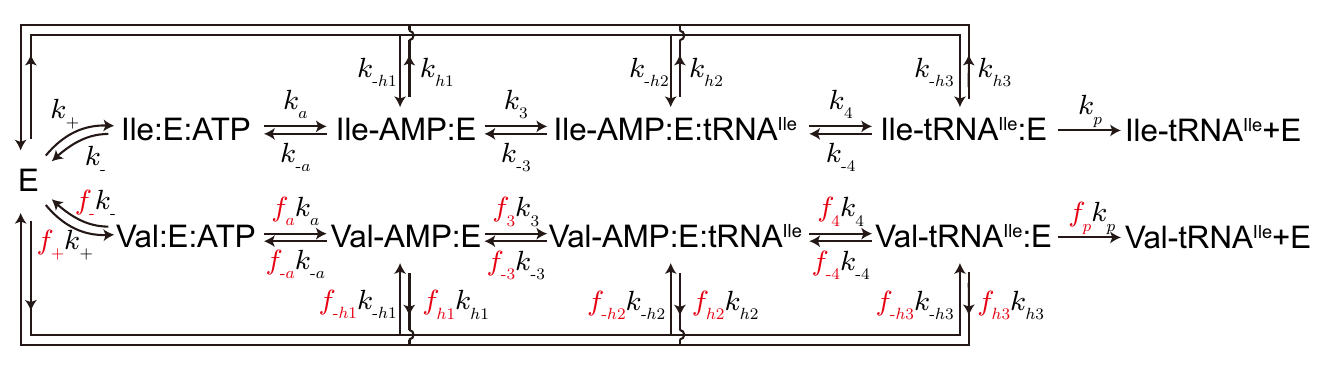}
	\caption{
		Reaction scheme for \textit{E. coli} tRNA\textsuperscript{Ile} aminoacylation, reproduced from ref.~\cite{Yu2020}. 
	}
	\label{fig:IleRSScheme}
\end{figure}

The reaction network and parameters for the IleRS network are obtained from ref.~\cite{Yu2020}. 
The reaction network is presented in Fig.~\ref{fig:IleRSScheme}.
The error and cost of the IleRS network are bounded by the following piecewise function:
\begin{equation}
	C_\mathrm{min} =  \begin{cases}
	\frac{1+\eta_0}{1+\eta} \frac{(b_1-1)(b_2-1)(b_3-1)\eta}{\qty(\qty(b_1b_2b_3\eta)^{1/3}-\eta_0^{1/3})^3} -1 & \frac{\eta_0}{b_1b_2b_3}<\eta<\frac{\eta_0b_1^2}{b_2b_3}\\
	\frac{1+\eta_0}{1+\eta}\frac{(b_2-1)(b_3-1)\eta}{\qty(\qty(b_2b_3\eta)^{1/2}-{\eta_0}^{1/2})^2}  -1 & \frac{\eta_0b_1^2}{b_2b_3} <\eta<\frac{\eta_0b_2}{b_3} \\
	\frac{1+\eta_0}{1+\eta} \frac{\qty(b_3-1)\eta}{b_3\eta-\eta_0} -1& \frac{\eta_0b_2}{b_3} <\eta <\eta_0
	\end{cases}
	\label{Eq:IleRS bound}
\end{equation}
where $b_1 =\frac{f_{h1}}{f_{3}}$, $b_2 = \frac{f_{h2}}{f_{4}}$, and $b_3=\frac{f_{h3}}{f_{p}}$. $\eta_0=\min\qty(f_+,\frac{f_+}{f_-}f_a)$ is the minimum error of the equilibrium discrimination by the first two steps (binding and activation).
Eq.~\ref{Eq:IleRS bound} corresponds to the red line in Fig.~6C in the main text. The three error intervals correspond to the three phases of proofreading in Fig.~6D in the main text.
In this section, we provide detailed derivation for the error-cost bound and the optimal partition ratios.

\paragraph{Derivation of the error-cost bound.}
Here, we show that due to the different discrimination factors in the three proofreading pathways, the optimal partition ratios are no longer uniform, and there will be three proofreading regimes due to the sequential ``shutdown'' of proofreading pathways.

First, we investigate the case where all three proofreading pathways are utilized, which can be considered as three MM-with-proofreading schemes applied in tandem. The minimum cost in the MM-with-proofreading scheme (Eq.~\ref{Eq: MM-with-proofreading bound expression}) gives the ratio of the total input flux (product formation plus proofreading) to the output (product-forming) flux:
\begin{align}
	\frac{J_\text{in}}{J_\text{out}} = \frac{J_\text{hydrolysis}}{J_\text{product}} = 1+C = 1 + \frac{\qty(f_1-\eta)\qty(1+\eta \frac{f_2}{f_p})}{\qty(1+\eta)\qty(\eta \frac{f_2}{f_p}-f_1)} = 
	1 + \frac{\qty(\eta_0-\eta)\qty(1+\eta b)}{\qty(1+\eta)\qty(\eta b-\eta_0)}
	= \frac{\eta(\eta_0+1)(b-1)}{\qty(1+\eta)\qty(\eta b-\eta_0)},
\end{align}
where $b=\frac{f_2}{f_p}$ is the discrimination of the partition ratio, and $\eta_0=f_1$ can be considered as the error of the last proofreading stage. The optimal partition ratio corresponding to this minimum cost is
\begin{eqnarray}
	a = j_2 = \frac{\eta_0-\eta}{\eta b-\eta_0}.
\end{eqnarray}
The minimum proofreading cost can thus be calculated by taking the product of the ratios ${J_\text{in}}/{J_\text{out}}$ in all three proofreading pathways, assuming optimal partition ratios. 
We denote the error at the three proofreading stages as $\eta_{1,2,3}$, respective. $\eta_3=\eta$ is the final error rate. $\eta_0=\min\qty(f_+, \frac{f_+}{f_-}f_a)$ is the minimum error before proofreading. The minimum cost is therefore
\begin{align}
	C &= \frac{J_\text{in}}{J_\text{out}}- 1\\
	& = \frac{\eta_1(\eta_0+1)(b_1-1)}{\qty(1+\eta_1)\qty(\eta_1 b_1-\eta_0)}
	\frac{\eta_2(\eta_1+1)(b_2-1)}{\qty(1+\eta_2)\qty(\eta_2 b_2-\eta_1)}
	\frac{\eta_3(\eta_2+1)(b_3-1)}{\qty(1+\eta_3)\qty(\eta_3 b_3-\eta_2)} -1 \\
	& = \frac{\eta_1\eta_2\eta_3(1+\eta_0)(b_1-1)(b_2-1)(b_3-1)}{\qty(1+\eta_3)\qty(\eta_1 b_1-\eta_0)\qty(\eta_2 b_2-\eta_1)\qty(\eta_3 b_3-\eta_2)} -1 \\
	& = \frac{1+\eta_0}{1+\eta} \frac{\qty(1-b_1^{-1})\qty(1-b_2^{-1})\qty(1-b_3^{-1})}{\qty(1-\frac{\eta_0}{\eta_1b_1})\qty(1-\frac{\eta_1}{\eta_2b_2})\qty(1-\frac{\eta_2}{\eta_3b_3})} -1
\end{align}
The denominator can be maximized with Jensen's inequality. Since $f(x) = \ln(1-e^x)$ ($x\in(0,1)$) is a concave function ($f''(x)<0$), we have
\begin{align}
	&f\qty(\ln\frac{\eta_0}{\eta_1b_1})+f\qty(\ln\frac{\eta_1}{\eta_2b_2})+f\qty(\ln\frac{\eta_2}{\eta_3b_3})\leq 3f\qty(\frac{1}{3}\ln\frac{\eta_0}{b_1b_2b_3\eta_3})
	\\
	\Rightarrow & \qty(1-\frac{\eta_0}{\eta_1b_1})\qty(1-\frac{\eta_1}{\eta_2b_2})\qty(1-\frac{\eta_2}{\eta_3b_3}) \leq \qty[1-\qty(\frac{\eta_0}{b_1b_2b_3\eta_3})^{1/3}]^3.
\end{align}
Hence, we obtain the minimum cost in this regime:
\begin{align}
	C_\mathrm{min} =  \frac{1+\eta_0}{1+\eta} \frac{\qty(1-b_1^{-1})\qty(1-b_2^{-1})\qty(1-b_3^{-1})}{\qty[1-\qty(\frac{\eta_0}{b_1b_2b_3\eta_3})^{1/3}]^3} -1
	= \frac{1+\eta_0}{1+\eta} \frac{(b_1-1)(b_2-1)(b_3-1)\eta}{\qty(\qty(b_1b_2b_3\eta)^{1/3}-\eta_0^{1/3})^3} -1,
\end{align}
where $\eta_3=\eta$. 
The condition for minimizing the cost is 
\begin{align}
	&\frac{\eta_0}{\eta_1 b_1} =\frac{\eta_1}{\eta_2 b_2} = \frac{\eta_2}{\eta b_3} = \qty(\frac{\eta_0}{b_1b_2b_3 \eta})^{1/3} \\
	\Rightarrow &  \eta_1 = \eta_0^{2/3} \eta^{1/3} \qty(\frac{b_2b_3}{b_1^2})^{1/3}, \quad \eta_2 = \eta_0^{1/3} \eta^{2/3} \qty(\frac{b_3^2}{b_1b_2})^{1/3}.
	\label{Eq:IleRS 3stage optimal eta}
\end{align}
Note that the error rates no longer form a geometric series. Instead, their ratios are modulated by factors $b_{1,2,3}$. The optimal partition ratios are:
\begin{equation}
	a_1 = \frac{\eta_0-\eta_1}{\eta_1b_1-\eta_0},\quad 
	a_2 = \frac{\eta_{1}-\eta_2}{\eta_2b_2-\eta_{1}},\quad
	a_3 = \frac{\eta_{2}-\eta}{\eta b_3-\eta_{2}}.
\end{equation}
where $\eta_{1,2}$ take the optimal values indicated in Eq.~\ref{Eq:IleRS 3stage optimal eta}.
All three partition ratios decrease as the error $\eta$ is increased. In the $n$-stage DBD scheme, the partition ratios are equal, and they go to zero simultaneously at $\eta_\mathrm{eq} = f^{-1}$. 
For the IleRS network, however, the three partition ratios are not equal, and one of them vanishes first. This takes place in the proofreading pathway with the least $b$, which is $b_1$ in the IleRS network:
\begin{align}
	a_1 =0 \Leftrightarrow \eta_0=\eta_1 \Leftrightarrow \eta =  \eta_\text{th1}= \eta_0 \frac{b_1^2}{b_2b_3}.
\end{align}
For error rates greater than the threshold $\eta_\text{th1}$, the above calculation leads to a negative partition ratio ($a_1<0$), which must be regularized to zero.

Hence, the three-stage proofreading analysis only applies to $\eta \in \qty(\frac{\eta_0}{b_1b_2b_3}, \frac{\eta_0 b_1^2}{b_2b_3})$. 
For larger error, the first proofreading pathway does not function ($a_1=0$), and we treat the system as two MM-with-proofreading schemes operating in tandem. Similarly, an error-cost bound can be obtained:
\begin{align}
	C_\mathrm{min} = \frac{1+\eta_0}{1+\eta}\frac{(b_2-1)(b_3-1)\eta}{\qty(\qty(b_2b_3\eta)^{1/2}-{\eta_0}^{1/2})^2}  -1 ,\quad  \frac{\eta_0b_1^2}{b_2b_3} <\eta<\frac{\eta_0b_2}{b_3}. 
\end{align}
The maximum error for this two-pathway regime is determined by $a_2=0$, which leads to $\eta=\eta_\text{th2}=\frac{\eta_0b_2}{b_3}$. 
For error larger than this value, we have $a_1=a_2=0$, and the optimal system operates as if there is only one proofreading pathway:
\begin{align}
	C_\mathrm{min} = \frac{1+\eta_0}{1+\eta} \frac{\qty(b_3-1)\eta}{b_3\eta-\eta_0} -1,\quad \frac{\eta_0b_2}{b_3} <\eta <\eta_0.
\end{align}
Therefore, we have derived the piecewise error-cost bound for the IleRS network, which is in agreement with the numeric sampling (Fig.~6C, main text). 

\paragraph{Analysis of the native system.}
In the IleRS network, the error rate thresholds which separate the three proofreading regimes are
\begin{align} 
	\eta_0 = 9.2 \times 10^{-3},\quad 
	\eta_\text{th2} = \eta_0 \frac{b_2}{b_3} = 1.3 \times 10^{-4},\quad  
	\eta_\text{th1} = \eta_0 \frac{b_1^2}{b_2b_3} = 7.7 \times 10^{-5},\quad
	\eta_\mathrm{min} = \frac{\eta_0}{b_1b_2b_3} = 4.5 \times 10^{-8}. 
\end{align}
The native system operates in the one-stage proofreading phase, where the optimal system only utilizes the last (post-transfer) proofreading pathway:
\begin{equation}
	\eta_\mathrm{wt} = 2.2\times 10^{-4} \in \qty(\eta_\text{th2},\  \eta_0).
\end{equation}
The native partition ratios are
\begin{equation}
	a_1 = \frac{f_{h1}}{f_3} = 3.4 \times 10^{-6} ,\quad
	a_2 = \frac{f_{h2}}{f_4} = 3.6 \times 10^{-3},\quad
	a_3 = \frac{f_{h3}}{f_p} = 8.9 \times 10^{-2}.
\end{equation}
The optimal partition ratios are
\begin{equation}
	a_1=a_2=0,\quad 
	a_3 = \frac{\eta_0-\eta_\mathrm{wt}}{\eta_\mathrm{wt}b_3-\eta_0} = 3.9 \times 10^{-2}.
\end{equation}
Hence, it would seem that the first two proofreading pathways are not utilized, consistent with the theory prediction ($a_1=a_2=0$).
The last proofreading pathway is responsible for most of the proofreading, but the third-stage partition ratio in the native system ($8.9\times 10^{-2}$) is more than twice of its optimal value ($3.9\times 10^{-2}$).
The reason for the extra proofreading is that $\eta_0$, which is the minimum error before proofreading, is never realized in the real system. 
It is only achieved if the amino acid activation step $k_a$ is much slower than binding $k_{\pm}$, but such time scale separation is not realized in the native system. 
The error rate before proofreading, which in theory could be as low as $\eta_0 = f_a \frac{f_+}{f_-} = 9.2 \times 10^{-3}$, is actually $\eta_\mathrm{activation} = 2.1 \times 10^{-2}$ in the native system (calculated by taking the ratio of net fluxes in the activation step). 
If we calculate the partition ratio with $\eta_0$ replaced by $\eta_\mathrm{activation}$, the optimal partition ratio becomes $a_3'=9.3\times 10^{-2}$, which is indeed closer to the native system. 
The reason why $\eta_0$ could not be realized is similar to what was discussed in the main text about the ribosome network. 
$\eta_0$ could be approached by either speeding up binding/unbinding reactions or by slowing down the amino acid activation step. 
Reducing the activation rate, however, will slow down the speed of product formation.
One possible interpretation is that while the binding and unbinding reactions are already as fast as possible, the native system chooses not to further decrease the activation rate so as to produce isoleucyl-tRNA\textsuperscript{Ile} sufficiently fast, which necessitates additional proofreading in the post-transfer proofreading pathway. 

Hence, the main conclusion here is that the deviation of the native IleRS from the optimal error-cost bound is due to prioritizing speed in the activation step. This is consistent with the trade-off analysis in the previous work~\cite{Yu2020}, where $k_a$ prefers to optimize speed rather than error or dissipation. If $k_a$ (and the reverse reaction $k_{-a}$) becomes much slower than the binding and unbinding rates $k_{\pm}$, the accuracy before proofreading will be improved, which will lead to a smaller partition ratio $a_3$ and lower cost $C$.

\subsection{Detailed model of the ribosome}
\begin{figure}
	\includegraphics[width=0.8\textwidth]{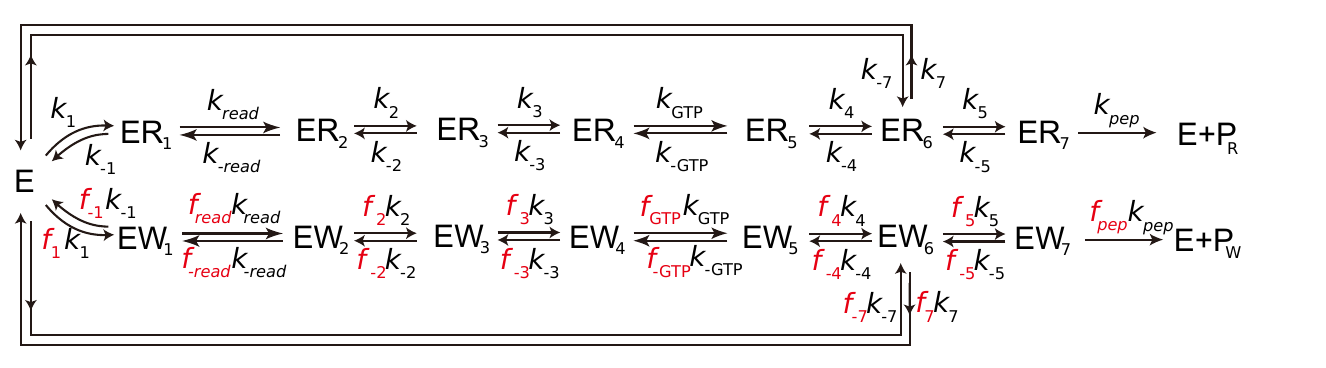}
	\caption{Schematics of the detailed ribosome model~\cite{Wohlgemuth2011}.}
	\label{Fig:RodninaScheme}
\end{figure}
The ribosome model presented in the main text was based on previous theoretical work~\cite{Banerjee2017PNAS} and experimental work~\cite{Zaher2010}. Here, we apply our theoretical framework to study another model of the ribosome, which is based ref.~\cite{Wohlgemuth2011}. The reaction scheme is shown in Fig.~\ref{Fig:RodninaScheme}. Compared to the ribosome model discussed in the main text, this model now includes multiple intermediate states. However, there is still only one proofreading pathway, namely the futile cycle containing $k_7$.
For the sake of generality, we allow for discrimination in all reaction steps, subject to the thermodynamic constraint:
\begin{equation}
	\frac{f_1f_\mathrm{read}f_2f_3f_\mathrm{GTP}f_4f_7}{f_{-1}f_\mathrm{-read}f_{-2}f_{-3}f_\mathrm{-GTP}f_{-4}f_{-7}}=1.
\end{equation}

In the following, we derive the error-cost bound with the flux-based formalism detailed above.

For step $i$, the rate constant is denoted by $k_i$ in the cognate network and $k_i'=k_if_i$ in the noncognate network. The normalized flux is denoted by $j_i$ in the cognate network and $j_i'$ in the noncognate network. 
The fluxes forming products are $j_\mathrm{pep}=1$ and $j_\mathrm{pep}'=\eta$, where $\eta$ is the final error rate. The cost is defined by 
\begin{equation}
	C = \frac{j_7+j_7'-j_{-7}-j_{-7}'}{j_\mathrm{pep}+j_\mathrm{pep}'} = \frac{j_7+j_7'-j_{-7}-j_{-7}'}{1+\eta}.
\end{equation}
Similar to the steps taken to derive the error-cost bound in previous sections of the SI, we establish the relation between fluxes inductively. We define the intermediate error rates:
\begin{equation}
	\eta_\mathrm{read} = \frac{j_{\mathrm{read}}'}{j_{\mathrm{read}}},\quad 
	\eta_{2} = \frac{j_{2}'}{j_{2}},\quad 
	\eta_{3} = \frac{j_{3}'}{j_{3}},\quad 
	\eta_\mathrm{GTP} = \frac{j_{\mathrm{GTP}}'}{j_{\mathrm{GTP}}},\quad 
	\eta_{4} = \frac{j_{4}'}{j_{4}},\quad 
	\eta_{5} = \frac{j_{5}'}{j_{5}}. 
\end{equation}
All the noncognate fluxes $\{j'\}$ can now be expressed in terms of the cognate fluxes $\{j\}$, the discrimination factors $\{f\}$, and the error rates $\{\eta\}$.
The stationary conditions for states ER\textsubscript{7} and EW\textsubscript{7} read
\begin{align}
	j_5 - j_{-5} = 1,\quad j_5' - j_{-5}' = \eta,
\end{align}
where $j_{5}' = \eta_5 j_5$ and $j_{-5}' = \eta \frac{f_{-5}}{f_\mathrm{pep}} j_{-5}$. These equations lead to 
\begin{equation}
	\qty(\eta-\eta_5\frac{f_\mathrm{pep}}{f_{-5}})j_5 = \qty({1-\frac{f_\mathrm{pep}}{f_{-5}}})\eta.
\end{equation}
Since $j_{5}=1+j_{-5}>1$, the intermediate error rate $\eta_5$ satisfies:
\begin{equation}
	\eta_5 <\eta_{5,\max} = \eta \cdot \max\qty(1, \frac{f_{-5}}{f_\mathrm{pep}}).
\end{equation}

The stationary conditions for states ER\textsubscript{6} and EW\textsubscript{6} read
\begin{align}
	j_4-j_{-4} &= j_7-j_{-7} + j_5 - j_{-5}, \\
	 j_4'-j_{-4}' &= j_7'-j_{-7}' + j_5' - j_{-5}',
\end{align}
where $j_4'=\eta_4j_4$, $j_{-4}' = \eta_5f_{-4} f_5^{-1} j_{-4}$, $j_{7}'=\eta_5 f_5^{-1} f_7 j_7$, $j_{-7}'=f_{-7}j_{-7}$. Elimination of the forward proofreading flux $j_7$ yields
\begin{equation}
	\qty(\eta_5 \frac{f_7}{f_5}-\eta_4)j_4 = \eta_5 \frac{f_7-f_{-4}}{f_5}j_{-4} + \qty(f_{-7}-\eta_5 \frac{f_7}{f_5})j_{-7} + \qty(\frac{f_7}{f_5}\eta_5-\eta).
\end{equation}
Based on analysis employed in previous models, error rates $\eta_5>{f_5}\frac{f_{-7}}{f_7}$ could be achieved without any proofreading (by making the $k_5$ step rate-limiting). Hence, we study the cost for error rates $\eta_5< {f_5}\frac{f_{-7}}{f_7}$. The coefficient $\qty(f_{-7}-\eta_5 \frac{f_7}{f_5})$ is positive. 
Proofreading preferentially dissociates noncognate complexes, indicating $f_7>f_{-4}$ and $f_7>f_5$ (which is the case for experimental data). 
Thus, LHS must also be positive, leading to 
\begin{equation}
	\eta_{5}>\eta_4 \frac{f_5}{f_7}.
\end{equation} 
The minimum $j_4$ is 
\begin{equation}
	j_4 \geq \frac{\qty(\frac{f_7}{f_5}\eta_5-\eta) + \eta_5\frac{f_7-f_{-4}}{f_5}j_{-4}}{\eta_5\frac{f_7}{f_5}-\eta_4},
\end{equation}
with equality condition $j_{-7} \to 0$. The cost is
\begin{align}
	C = \frac{j_7+j_7'-j_{-7}-j_{-7}'}{1+\eta} &= \frac{j_4+j_4'-j_{-4}-j_{-4}'}{1+\eta} -1\\
	& = \frac{(1+\eta_4)j_4-\qty(1+\frac{\eta_5f_{-4}}{f_5})j_{-4}}{1+\eta} -1\\
	& \geq \frac{\qty(\eta_4-\eta)\qty(1+\frac{f_7}{f_5}\eta_5)}{\qty(1+\eta)\qty(\frac{f_7}{f_5}\eta_5-\eta_4)} + \frac{\qty(\eta_4-\eta_5\frac{f_{-4}}{f_5})\qty(1+\frac{f_7}{f_5}\eta_5)}{\qty(1+\eta)\qty(\frac{f_7}{f_5}\eta_5-\eta_4)}j_{-4}
\end{align}
For error rates satisfying $\eta_4>\eta_5 f_5^{-1}f_{-4}$, the proofreading pathway is unnecessary, and the minimum cost is zero. For error rates satisfying $\eta_4>\eta_5 f_5^{-1}f_{-4}$, the cost is minimized in the limit $j_{-4} \to 0$:
\begin{equation}
	C \geq \frac{\qty(\eta_4-\eta)\qty(1+\frac{f_7}{f_5}\eta_5)}{\qty(1+\eta)\qty(\frac{f_7}{f_5}\eta_5-\eta_4)}.
\end{equation}
The above minimum cost increases with $\eta_4$ but decreases with $\eta_5$.
The maximum value of $\eta_5$ is given by:
\begin{equation}
	\eta_{5,\max} = \eta \cdot \max\qty(1, \frac{f_{-5}}{f_\mathrm{pep}}). \label{Eq: eta5max}
\end{equation}
The minimum value of $\eta_4$ is determined by the maximal difference in the energy barriers along the chain of reversible reactions from state E to state ER\textsubscript{6}/EW\textsubscript{6}:
\begin{align}
	\eta_4 &> \eta_{4,\min} = e^{-\Delta\mu_\mathrm{max}} \\
	&= 
	\mathrm{min}\qty(f_1,  \frac{f_1f_\mathrm{read}}{f_{-1}},
	\frac{f_1f_\mathrm{read}f_2}{f_{-1}f_\mathrm{-read}},
	\frac{f_1f_\mathrm{read}f_2f_3}{f_{-1}f_\mathrm{-read}f_{-2}},
	\frac{f_1f_\mathrm{read}f_2f_3f_\mathrm{GTP}}{f_{-1}f_\mathrm{-read}f_{-2}f_{-3}},
	\frac{f_1f_\mathrm{read}f_2f_3f_\mathrm{GTP}f_4}{f_{-1}f_\mathrm{-read}f_{-2}f_{-3}f_\mathrm{-GTP}}).\label{Eq:eta4min}
\end{align}
Therefore, we have derived the error-cost bound:
\begin{equation}
	C_\mathrm{min} = \frac{\qty(\eta_{4,\min}-\eta)\qty(1+\frac{f_7}{f_5}\eta_{5,\max})}{\qty(1+\eta)\qty(\frac{f_7}{f_5}\eta_{5,\max}-\eta_{4,\min})},\quad \eta\in\qty(\eta_{\min},\eta_\mathrm{eq}),
	\label{Eq:detailed ribosome bound}
\end{equation}
where $\eta_{4,\min}$ and $\eta_{5,\max}$ are given by Eq.~\ref{Eq:eta4min} and Eq.~\ref{Eq: eta5max}, respectively. The minimum error is
\begin{equation}
	\eta_\mathrm{min} = \frac{\eta_{4,\min}}{f_7}\cdot\min\qty(1,\frac{f_\mathrm{pep}}{f_{-5}}),
\end{equation}
and the minimum error without proofreading is
\begin{equation}
	\eta_\mathrm{eq} = \min\qty(\eta_{4,\min},
	\frac{f_1f_\mathrm{read}f_2f_3f_\mathrm{GTP}f_4f_5}{f_{-1}f_\mathrm{-read}f_{-2}f_{-3}f_\mathrm{-GTP}f_{-4}}
	,
	\frac{f_1f_\mathrm{read}f_2f_3f_\mathrm{GTP}f_4f_5f_\mathrm{pep}}{f_{-1}f_\mathrm{-read}f_{-2}f_{-3}f_\mathrm{-GTP}f_{-4}}).
\end{equation}

Thus, the flux-based formalism could be used to fully determine the fundamental error-cost bound in this detailed kinetic model of the ribosome. The methodology is completely same as that used in for the other models, and the cost-error bound exhibits similar quantitative behavior. In fact, the mathematical form of this bound (Eq.~\ref{Eq:detailed ribosome bound}) is similar to that of the ribosome model studied in the main text (Eq.~\ref{Eq:ribosome bound}).

\bibliography{KPR_ref}